\documentclass[useAMS,usenatbib]{mn2e}
\usepackage{psfig}
\usepackage{epsfig}

\def\kms{$\mathrm{km\;s}^{-1}$}

\def\Msun{{\rm\,M_\odot}}

\def\h3{$h_{3}$}
\def\h4{$h_{4}$}
\def\ang{{\AA}}

\title[Dark matter in early--type galaxies: dynamical modelling  of IC~1459, IC~3370, NGC~3379 and NGC~4105]{Dark matter in early--type galaxies: dynamical modelling  of IC~1459, IC~3370, NGC~3379 and NGC~4105}
\author[S. Samurovi\'c and I.~J. Danziger]
{S. Samurovi\'c$^{1,2,3,4}$\thanks{E-mail:
srdjan@ts.astro.it (SS); danziger@ts.astro.it (IJD)} and I.~J. Danziger$^{2}$\footnotemark[1]
\\
$^{1}$Dipartimento di Astronomia, Universit\`a di Trieste, Via Tiepolo 11, I-34131 Trieste, Italy\\
$^{2}$INAF, Osservatorio Astronomico di Trieste, Via Tiepolo 11, I-34131 Trieste, Italy\\
$^{3}$Astronomical Observatory, Volgina 7, 11160 Belgrade, Serbia and Montenegro\\
$^{4}$Institute Isaac Newton of Chile, Yugoslavia Branch}

\begin{document} 
 
\date{Accepted .......... Received ........; in original form .........}
 
\pagerange{\pageref{firstpage}--\pageref{lastpage}} \pubyear{2005}

\maketitle

\label{firstpage}

\begin{abstract}
We analyse long--slit spectra of four early--type galaxies which extend from
$\sim$ 1 to $\sim 3$ effective radii: IC~1459, IC~3370, NGC~3379 and NGC~4105.
We have extracted the full line--of--sight velocity distribution (in the case of
NGC~3379 we also used data from the literature)  which we model using the
two--integral approach. Using two--integral modelling we find no strong evidence
for dark haloes, but the fits suggest that three--integral modelling is
necessary. We also find that the inferred constant mass--to--light ratio in all
four cases is typical for early--type galaxies. Finally, we also  discuss the
constraints on the mass--to--light ratio which can be obtained using X--ray
haloes in the case of IC~1459, NGC~3379 and NGC~4105 and compare the estimated
values with the predictions from the dynamical modelling.
\end{abstract}


\section{Introduction}

The problem of   dark matter in galaxies remains perhaps the most important astrophysical problem in contemporary cosmology and extragalactic astronomy.
Although its nature is still unknown, general but not unanimous opinion is that it  exists and that it is a  necessary ingredient of
every viable cosmological model (see recent overview of the dark matter problem in galaxies in Binney 2004).
 The existence of the dark matter in spiral galaxies is rather
clear mainly because of existence of cool gas which 
provides a powerful tool for obtaining  rotation curves,
that are, for most spirals, nearly flat thus indicating the presence of dark
mass in their outer parts -- dark haloes (see, e.g., Binney \& Tremaine 1987 Sec 10.1). 
There are problems in the determination of its shape, but observations tend to 
conclude that the dark halo is flattened (see, e.g., Samurovi{\'c},
{\'C}irkovi{\'c} \& Milo{\v s}evi{\'c}--Zdjelar 1999).

However, the problem of dark matter 
in elliptical galaxies (early--type galaxies, in general) is more complicated -- it is more
difficult to confirm the presence of dark haloes around ellipticals.
Since elliptical galaxies  contain little or no cool gas usually 
one cannot use 21--cm observations to trace kinematics of neutral hydrogen out to
large radii, as  is possible in the case of spirals.
 In an attempt to check whether ellipticals have dark haloes one can use
stellar kinematics, but since their outer parts are very faint, it is
usually difficult to obtain spectra to constrain kinematics at large
radii. An additional problem is related to the fact that one does not
{\it a priori} know anything about the orbits of stars in ellipticals.
Current investigations lead to the conclusion that there is less 
unambiguous evidence for the dark matter in ellipticals than in the case of spirals. Moreover, there are hints that in ellipticals the dark matter is not needed at all or, more precisely,
not needed in some early--type galaxies, out to a given  observed distance from the galactic centre (see, for example, Romanowsky et al. 2003).

The dark matter problem in elliptical galaxies 
can be studied using different tracers of mass: as in Danziger (1997) one can split the different methodological approaches into three large groups of tracers: gas, test particles and lensing methods. Without going into details about 
the features of all these groups we can note that the different methodologies provide different claims about the existence of the dark haloes in early--type galaxies.
Here, we will constrain ourselves to the methods that provide data out to large radii (larger than $\sim 2$--$3 R_e$, where $R_e$ is the effective radius).
The approach based on the study of the {\it hot gas} (X--ray haloes with temperature $T\sim 10^7$ K) (see a review in  Mathews \& Brighenti  2003) is based on two assumptions:
(i) the gas obeys the perfect gas law and (ii)  hydrostatic equilibrium 
holds. This method in general provides large mass--to--light ratios in the early--type galaxies (cf. Loewenstein \& White 1999).

Some methods based on  {\it test particles} include  planetary nebulae, 
globular clusters  and integrated stellar light. {\it The planetary nebulae (PNe)} method  seems to be a very promising tool in dark matter research because PNe are detectable even in moderately distant galaxies through their strong emission lines. M\'endez et al. (2001) studied 535 PNe in the flattened elliptical galaxy NGC~4697 and found that, assuming isotropic velocity distribution, this galaxy does not need dark matter interior to $3R_e$.
Using the PNe method one derives mass--to--light ratios which are much lower than those derived using X--ray haloes: for example,  Romanowsky et al. (2003) found for their sample  (NGC~821 observed out to $3.5R_e$, NGC~3379 observed out to $\sim 5.5R_e$ and NGC~4494 observed out to $\sim 3R_e$) that velocity dispersion profiles show declines which indicates ``the presence of little if any dark matter in these galaxies'  haloes''.
Probably for NGC~3379 a better estimate of $R_e=54.8$ arcsec is given by 
Capaccioli et al. (1990) and therefore the radial distance of $\sim 5.5 R_e$ by Romanowsky et al. (based on RC3) should actually be $\sim 3.5R_e$.
Here we adopt the Capaccioli et al. value for the effective radius of NGC~3379, see Sec. 3.3.
This value was also used in the paper of Ciardullo et al. (1993), see below.

We have analyzed NGC~3379 below using long--slit spectra out to 2.2 $R_e$ and found after a detailed dynamical modelling that there is no need to apply the dark matter hypothesis
in order to obtain a successful fit for this galaxy.
Peng, Ford \& Freeman (2004)   recently presented their results of an imaging and spectroscopic survey for PNe in NGC~5128. 
They found that PNe exist at distances out to 80 kpc ($\sim 15 R_e$) making this study the most comprehensive kinematical study of an elliptical galaxy to date, both in the number of velocity tracers and in radial extent.
They found that the dark matter is necessary to explain the observed stellar kinematics, but their value of $M/L_B$ is low: within 80 kpc   the total  
dynamic mass $\sim 5\times 10^{11}{\rm {\rm M_\odot}}$ is found with $M/L_B\sim 13$. The problem with this method is that 
the $h_3$ and $h_4$ parameters which describe the the full line--of--sight velocity distribution (LOSVD) profile (see below) 
are not observationally determined as for the integrated spectra
and an ambiguity about the mass distribution in the outer parts of galaxy may exist: strong tangential anisotropies can mimic the existence of the dark matter (Tonry 1983; Binney \& Merrifield 1998 Sec 11.2). 

The {\it globular clusters (GCs)} methodology can also be used in a search for dark matter.  
C\^ot\'e et al. (2003)  studied M49 ($=$ NGC~4472) galaxy and showed that the  radial velocities and density profiles of globular clusters provide ``unmistakable evidence'' for a massive dark halo interior to $\sim 3R_e$.
Note again the problem of the estimate of the effective radius:
C\^ot\'e et al. determined $R_e=186$ arcsec whereas RC3 gives $R_e=104$ arcsec.
This lower value of $R_e$ would place their last measured point at $\sim 5.5R_e$.
Recently, Bridges et al. (2003) presented their results obtained using Gemini/GMOS spectrograph of several  early--type galaxies. It is important to note that they have  observed 22 globular clusters in the aforementioned galaxy NGC~3379 and found no evidence of dark matter out to $\sim 200$ arcsec: 
in their preliminary analysis they reached the conclusion that the mass--to--light ratio {\sl decreases} (from $\sim$ 8 at $\sim 60 $ arcsec to $\sim$ 4 at $\sim$ 200 arcsec,
 in the $V$--band). 
Richtler et al. (2004)  analyzed the GC system of  the galaxy NGC~1399:
they used 468 radial velocities assuming a pure Gaussian distribution to conclude that the velocity dispersion of this galaxy remains approximately constant between 2 and 9 arcmin  
(which corresponds to approximately 2.86 and 12.86 $R_e$) thus implying the existence of dark matter in this central galaxy of the Fornax cluster.
The problem of the lack of the  full LOSVD mentioned earlier still exists with this methodology. 
 
A large set of dark matter investigations in   early--type galaxies is made through studies of {\it integrated stellar light}. Because this paper deals with the two--integral (2I) approach we   now 
focus    on the papers that use this technique. Binney, Davies \&
Illingworth (1990) in their seminal paper established a 2I axisymmetric modelling based on the photometric observations. They analyzed galaxies NGC~720, NGC~1052, and NGC~4697 and modelled velocities and velocity dispersions out to $\sim$ 1 $R_e$. van der Marel, Binney \& Davis (1990) applied this approach to NGC~3379 (out to $\sim$ 1 $R_e$), NGC~4261 (out to $\sim$ 1 $R_e$), NGC~4278 (out to $\sim$ 1 $R_e$) and NGC~4472 (out to 0.5 $R_e$). Cinzano \& van der Marel (1994) modelled the galaxy NGC~2974 out to 0.5 $R_e$ introducing a new moment -- modelling of the Gauss--Hermite  moments (for definitions see below) defined previously in van der Marel \& Franx (1993). 
Note however, that all these modelling procedures did not take into account dark matter, because they dealt with the regions in which dark matter was not expected to make a significant contribution.  
Saglia et al. (1993) presented kinematical and line strength profiles of NGC~4472, IC~4296 and NGC~7144 and from their dynamical modelling
(quadratic programming)  concluded that there is a strong evidence for dark matter in these galaxies.
Carollo et al. (1995) observed and modelled a set of elliptical galaxies
(NGC~2434, NGC~2663, NGC~3706 and NGC~5018). They used a 2I modelling procedure to model the stellar line--of--sight velocity distribution 
(using velocity dispersion and Gauss--Hermite $h_4$ parameter) out to two effective radii.  They concluded that the massive dark matter haloes must be present in three of the four galaxies, and in the case of NGC~2663 there was no evidence of dark matter.
Rix et al. (1997) used the Schwarzschild (1979) method for construction of axisymmetric and triaxial models of galaxies in equilibrium without explicit knowledge of the integrals of motion. They introduced into the analysis velocity, velocity dispersion and Gauss--Hermite parameters $h_3$ and $h_4$. They used the  galaxy NGC~2434 (from Carollo et al. 1995) to perform a detailed spherical dynamical modelling in order to conclude that this galaxy contains a lot of dark matter: they found that about half of the mass within one effective radius is dark. 

Statler, Smecker--Hane \& Cecil (1996) studied  stellar kinematical fields of the post--merger elliptical galaxy NGC~1700 out to four effective radii. In a 
subsequent paper Statler, Dejonghe \& Smecker--Hane (1999) found, using 2I axisymmetric models as well as three--integral (3I) 
quadratic programming models 
that NGC~1700 must have a radially increasing mass--to--light ratio, and that NGC~1700 ``appears to represent the strongest stellar dynamical evidence to date for dark haloes in elliptical galaxies''. 
Statler \& McNamara (2002) observed this galaxy in the X--ray domain and using gas modelling estimated the gas temperature to be $\sim 0.5$ keV.
Note, however, that these authors found that probably the hypothesis of hydrostatic equilibrium is not applicable in this case which would therefore make difficult the comparison between mass profiles based on the X--ray data and stellar dynamics.
Saglia et al. (2000) modelled the galaxy NGC~1399 using 2I models (major photometric axis only) out to $\sim$ 2.5 $R_e$. They marginally detected 
the influence of the dark component that starts from 1.5 $R_e$. 

Kronawitter et al. (2000) modelled a sample of 21 elliptical galaxies out to 1--2 $R_e$: for three of them (NGC~2434, NGC~7507, NGC~7626) they found that  models based on luminous matter should be ruled out.
De Bruyne et al. (2001) modelled NGC~4649 and NGC~7097 using 3I quadratic programming method  and found that in the case of NGC~4649 
a constant mass--to--light ratio ($M/L_V =9.5$)  fit can provide good agreement with the data and that a marginally better fit can be obtained including 10 per cent of dark matter at 1.2 $R_e$. In the case of NGC~7097 both kinematic and photometric data can be fitted out to 1.6 $R_e$ using a constant mass--to--light ratio $\sim$ 7.2. 
Cretton, Rix \& de Zeeuw (2000) modelled the giant elliptical galaxy NGC~2320 
using the Schwarzschild orbit superposition method and found that the 
models with a radially constant mass--to--light ratio and logarithmic models with dark   matter provide comparably good  fits to the data and have similar dynamical structure (but note that the mass--to--light ratio in the $V$--band is rather large: $\sim$  15 for the mass--follows--light models and $\sim 17$ for the logarithmic models).

Finally, we mention the paper of Cappellari et al. (2002) in which the internal parts of one of the two galaxies which we analyze in this paper, IC~1459, 
were modelled using 3I axisymmetric technique.

From what is given above it is obvious that  concerning the existence of dark matter in early--type galaxies
the situation remains rather unclear: some appear to have dark matter haloes, and the others appear to be devoid of the dark component. One important note may 
be appropriate: recent observations that extend out to large radial distances from the centre ($>5R_e$) (for example, PNe, GCs methodologies), although rare and incomplete (because of the lack of knowledge of the full LOSVD) strongly suggest that the evidence for the dark matter in these outer regions is scarce, and that the calculated mass--to--light ratios are small (either equal or marginally larger than  those estimated in the inner regions).
This is obviously at odds with values claimed from the X--ray observations. 
 This study addresses, what we can call, ``intermediate''   regions out to $\sim 1-3R_e$ taking into account the full LOSVD.

In this paper we analyze in detail using  2I approach four early--type galaxies: IC~1459, IC~3370, NGC~3379 and NGC~4105 for which we have long--slit spectra that extend out to $\sim 1-3R_e$. In Sect. 2 we give some theoretical foundations related to both observational and modelling aspects. 
In Sect. 3 we briefly present the features of these four galaxies together with the photometric observations.
Section 4 provides observational and data reduction information. 2I modelling results are given in Section 5. In Section 6 we compare the results  obtained with the 2I technique with those obtained using X--ray haloes technique (in the case of IC~1459, NGC~3379 and NGC~4105).
Finally, our conclusions are presented in Section 7.

\section{Theoretical foundations}

\subsection{Line of sight velocity distributions}
 
 In the case of all external galaxies, one cannot obtain data necessary for the reconstruction of the distribution function of the stellar system directly: one can observe line--of--sight velocities and angular coordinates.
 Since individual stars cannot be resolved, one has to deal with integrated stellar light that represents the average of the stellar properties of numerous unresolved stars that lie along each line of sight (LOS).
  The first step in the analysis of the shifts and broadenings is to define the line of sight velocity distribution (LOSVD, also called velocity profile, VP): this is a function $F(v_{\rm LOS})$ that defines the fraction of the stars that contribute to the spectrum that have LOS velocities between $v_{\rm LOS}$ and $v_{\rm LOS}+{\rm d}v_{\rm LOS}$ and is given as $F(v_{\rm LOS}){\rm d}v_{\rm LOS}.$ Now, if one assumes that all stars have identical spectra $S(u)$ (where $u$ is the spectral velocity in the galaxy's spectrum), then the intensity that is received from a star with LOS velocity $v_{\rm LOS}$ is $S(u-v_{\rm LOS})$. When one sums over all stars one gets:
\begin{equation}
G(u)\propto \int {\rm d}v_{\rm LOS}F(v_{\rm LOS})S(u-v_{\rm LOS}).
\end{equation}
This relation represents the starting point for a study of stellar kinematics in external galaxies (cf. Binney \& Merrifield 1998 Sec 11.1). The observer obtains $G(u)$ for a LOS through a galaxy by obtaining its spectrum. If the galaxy is made of a certain type of star, one can estimate $S(u)$   using a spectrum of a star from the Milky Way galaxy.

One possible solution is to assume that the LOSVD has the Gaussian form. Sargent et al. (1977) invented the method known as Fourier Quotient Method, that has a problem of large errors for the ratio ${\tilde G(k) \over \tilde S(k)}$ that vary from point to point. The cross--correlation method based on the calculation of the cross--correlation function between the galaxy and the stellar spectra was pioneered by Simkin (1974) and developed further by Tonry and Davis (1979)
and Statler (1995).
 
In this paper we model the LOSVD   as truncated Gauss--Hermite ($F_{\rm TGH}$) series that consists of a Gaussian that is multiplied by a polynomial (van der Marel \& Franx 1993; also Gerhard 1993):
\begin{equation}
F_{\rm TGH}(v_{\rm LOS}) = \Gamma {\alpha (w)\over \sigma} \exp(-{1\over 2}w^2)\left [1+
\sum _{k=3}^nh_kH_k(w)\right ]
\end{equation}
here $\Gamma$ represents the line strength,
 $w\equiv (v_{\rm LOS}-\bar v)/\sigma$, $\alpha\equiv {1\over {\sqrt {2\pi}}} \exp(-w^2/2)$, where $\bar v$ and $\sigma$ are free parameters.  $h_k$ are constant coefficients and $H_k(w)$ is a Gauss--Hermite function, that is a polynomial of order $k$. We will truncate the series at $k=4$ (although higher values are also possible), for which the polynomials are:
\begin{equation}  
H_0(w)=1,
\end{equation}
\begin{equation}  
  H_1(w)=\sqrt{w},
  \end{equation}
  \begin{equation}
  H_2(w)={1\over \sqrt{2}}(2w^2-1),
  \end{equation}
  \begin{equation}
  H_3(w)= {1\over \sqrt{6}} (2\sqrt{2}w^3-3\sqrt{2}w),
  \end{equation}
  and
  
  \begin{equation}
  H_4(w)= {1\over \sqrt{24}} (4w^4 -12w^2+3).
 \end{equation}
  It can be shown (van der Marel \& Franx 1993) that $H_l(w)$ (in this case,
$l=0,\dots,4)$ are orthogonal with respect to the weight function $\alpha^2(w)$.
 
Now the LOSVD can be calculated by varying the values of $\bar v$, $\sigma$, $h_3$ and $h_4$ until the convolution of the function $F_{TGH}(v_{\rm LOS})$ with a template star spectrum  best reproduces the observed galaxy spectrum.
 The optimal fit is then reached using a non--linear least--squares fitting algorithm. If the form of the  LOSVD is close to the Gaussian form, then $\bar v$ and $\sigma$ will be approximately equal to $\overline{ v_{\rm LOS}}$ and $\sigma_{\rm  LOS}$. Parameters $h_3$ and $h_4$ are important because they measure asymmetric and symmetric departures from the Gaussian.
 If one detects a positive (negative) value of the $h_3$ parameter that would mean that the distribution is skewed towards  higher (lower) velocities with respect to the systemic velocity.
On the other hand, if one detects $h_4 >0 $ this means that the distribution is more peaked than the Gaussian at small velocities with more extended high--velocity tails; for $h_4 <0$ the distribution is more flat--topped than the Gaussian. In the study of the dark matter in the early--type galaxies the  value of the $h_4$ parameter plays a crucial role because it is constraining
the level of tangential anisotropy which is extremely important since it is well known that the excess of tangential motions can mimic the existence of the dark matter haloes in these galaxies (Danziger 1997; Gerhard 1993).

\subsection{Two--integral (2I) modelling}

For the  2I modelling procedures we used the modelling technique developed by  Binney,  Davies \& Illingworth (1990, hereafter BDI), and subsequently used by van der Marel, Binney \& Davies (1990)   and Cinzano \& van  der Marel (1994). Only in Cinzano \& van  der Marel (1994)
does  the modelling include $h_3$ and $h_4$ parameters. These $h_l$ models are based on the assumptions of Gaussian bulge and disk components.
Statler et al. (1999) used the modified version of this method to analyze the mass distribution in NGC~1700. Here we  present briefly the assumptions and the modelling steps.

\begin{figure}
\centering
\resizebox{\hsize}{!}{ 
\includegraphics{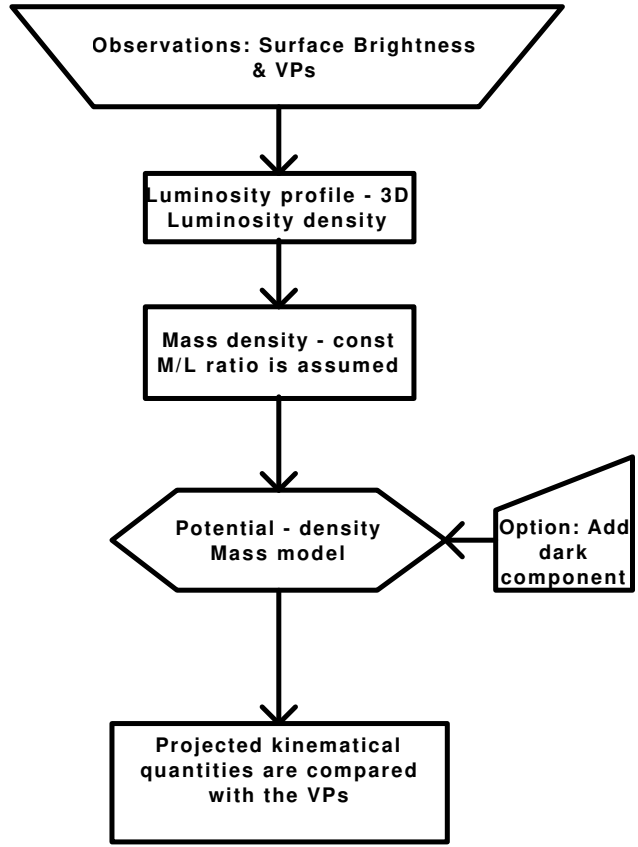}} 
\caption{Flowchart of 2I modelling.}
\label{fig:2I-flow}
\end{figure}

2I modelling is the first step in understanding of the dynamics of the elliptical galaxies, because in cases of small departures from triaxiality (which is far more probable, and very strong in  the case of IC~3370 as  will be shown below), comparison of real systems with the models can provide useful insights. The assumptions of axisymmetry and the fact that the velocity dispersion tensor is everywhere isotropic are the starting points for the procedure that includes the following three steps (cf. BDI): (i)
inversion of the luminosity profiles and obtaining three--dimensional luminosity density that provides the mass density (under the assumption of constant mass--to--light ratio); (ii)  evaluation of the potential and derivation of the velocity dispersion and azimuthal streaming (under the assumptions that the form of the distribution function is $f(E,L_z)$, where $E$ is the energy and $L_z$ is the angular momentum of the individual star about the symmetry axis of the galaxy and  that the velocity dispersion is isotropic as a starting point) and (iii) comparison of 
the projected kinematical quantities from the model with the observed kinematic parameters; optionally, a disc, and/or a dark halo can be included in the modelling procedure.  The flowchart of the modelling procedure is given in Fig. \ref{fig:2I-flow}.

The first step in the modelling procedure involves making a least--squares fit for a flattened Jaffe model (Jaffe 1983, Eqs. (2) and (3))  or a flattened modified Hubble model. The brightness distribution in the case of the modified Hubble 
profile is given as (see Binney \& Tremaine 1987, Eq. (2--39)):
\begin{equation}
\label{hubble}
I(R)={2j_0a\over 1+(R/a)^2},
\end{equation}
 (where  $j_0$ is the central luminosity density and $a$ is the core radius),
and was used in the 2I modelling procedures of the two galaxies in this paper.
Because we were mainly interested in the outer parts of galaxies we did not correct for the effects of seeing that are of importance near the centres of galaxies. 
Six Lucy iterations were used to get a fit of the surface brightness profile to the model. In cases when the disc was taken into account, the surface
brightness of the disc was subtracted  assuming that the disc is exponential.

For the evaluation of the dynamical quantities  one assumes that the spheroid's mass density $\rho(R,z)=\Upsilon _sj(R,z)$ where $\Upsilon_s$ is some constant mass--to--light ratio, and $j(R,z)$ is the luminosity density. 
Here, it would be appropriate to present an estimate of the typical mass--to--light ratio in  elliptical galaxies. van der Marel (1991)
found in his sample of 37 bright ellipticals that the average mass--to--light ratio in the $B$--band is:
$M/L_B=(5.95\pm 0.25)h_{50}$ thus $M/L_B=8.33\pm 0.35$ for $h_0=0.70$. He also found that the mass--to--light ratio is correlated with the total luminosity: $M/L_B= 3.84 h_{50} (L_B/L_{*,B})^{0.35}$,
where $L_{*,B}\equiv 3.3 \times 10^{10}h_{50}^{-2}L_\odot$.

The details related to the 2I modelling procedure which we use are given in BDI  and  Cinzano \& van der Marel (1994) and Fig. \ref{fig:2I-flow}. Below we give some of the most important steps.

Since we assume  in this paper  that the distribution function is of the form $f(E,L_z)$
the second radial velocity moment, $\overline {v_R^2}\equiv \sigma_R^2$, and the second vertical velocity moment, $\overline {v_z^2}\equiv \sigma_z^2$ are everywhere equal and $\overline {v_Rv_z}=0$. Solving the Jeans equations we  search for unknowns $\overline {v_\phi^2}$ and $\sigma_R^2 =\sigma_z^2$. Using a free parameter, $k$, one can, as usual, assign a part of the second azimuthal velocity moment $\overline{v_\phi^2}$ to streaming:
\begin{equation}
\overline {v_\phi} = k \sqrt{\overline{v_\phi^2}-\sigma_R^2}.
\end{equation}
Then we project the dynamical quantities on to the sky to obtain predictions, always taking into account a given inclination angle $i$ (which is a free parameter given at the begining of the 2I procedure): this is an angle measured from the galaxy's rotation axis to the line--of--sight: for edge--on galaxy $i=90^\circ$ and for face--on galaxy  $i=0^\circ$. We always start the modelling with the value $k=1$ which is a natural starting point, because it implies an isotropic dispersion vector.

For the modelling of the stellar kinematics  we used   
 freely available  ``Two--integral Jeans modelling Software'' made by R. van der Marel and J. Binney, 
 again on the x86 GNU/Linux PC platform (see the comment about the software used for the  extraction of the kinematics   above).

\section{General information and photometric observations}

In all the analyses that we are performing we assume that the Hubble
constant is equal to 70 ${\rm km\ s^{-1}  Mpc^{-1}}$.
The standard steps were performed in the data reduction of the photometric observations of IC~1459 and IC~3370: bias subtraction, flat--fielding, 
cleaning of cosmic rays and subtraction of sky background. We used the {\sc IRAF}\footnote {{\sc IRAF} is distributed by NOAO, which is operated by AURA Inc., under contract with the National Science Foundation.}
task {\sc ellipse} to extract the full photometric profiles in both cases.

\subsection{IC~1459}

IC~1459 is a giant E3 elliptical galaxy. Its
  absolute blue magnitude is -20.52,
  heliocentric radial velocity $1691$   \kms \ (taken from NED database and in agreement with our observations).
\  It covers $5.2 \times 3.8$ arcmin on the sky (RC3).
One arcsec in the galaxy corresponds to $\sim 117.16$ pc. The effective radius is 33 arcsec (=3.87 kpc).
One of its most characteristic features is a fast counterrotating stellar core
(Franx \& Illingworth 1988). 
It has other peculiarities: twisted isophotes (Williams \& Schwarzschild 1979),
a dust lane and patches near  the nucleus (Sparks et al. 1986) and 
an ionized gaseous disc at the core that rotates along the major axis in the same direction as the majority of stars in the galaxy -- this is the opposite direction to that of the stellar core (Forbes, Reizel \& Williger 1995).
The nucleus of IC~1459 has a strong (1Jy) compact radio source (Slee et al 1994).
Recently, Fabbiano et al. (2003) observed this galaxy with {\it Chandra} ACIS-S: these observations will be used below, when comparing stellar dynamics and X--ray data.

Verdoes Kleijn et al. (2000) analyzed kinematical observations of the nuclear gas disc, and found a central black hole of mass $M_{\rm BH}=(2-6)\times 10^{8}{\rm M_\odot}$. Cappellari et al. (2002) observed IC~1459 using several slit positions and constructed axisymmetric 
3I models of this galaxy using the Schwarzschild orbit superposition method. They found, using stellar and gas kinematics, that $M_{\rm BH}=(1.1\pm  0.3)\times 10^{9}{\rm M_\odot}$. The mass--to--light ratio
found in this paper was $\sim 8.8$ (when converted to the $B$--band and the distance used in this paper, $D=24.16$ Mpc) and is in agreement with our estimate (we found that $M/L=5-10$).

\subsubsection{Photometric observations}
Photometric observations were made by one of us (IJD)   
during 1997 August 28--30  using the ESO 
NTT and EMMI in the Red Medium Spectroscopy mode in the $V$--band.
We present the results obtained using the aforementioned {\sc IRAF} routine in 
Fig. \ref{fig:IC1459_IC3370_phot} (left)
where surface brightness was transformed to the $B$--band using relation $B-V=0.99$ taken from the LEDA\footnote{\tt http://leda.univ-lyon1.fr/.} database. The photometric profile was compared with that of Franx \& Illingworth (1988) and it was found that it was in a good agreement.

\subsection{IC~3370}

IC~3370 is a bright galaxy, classified as E2--E3 (elliptical) galaxy,  absolute blue magnitude --21.4,
heliocentric radial velocity $2930$   \kms \ (taken from NED database and in agreement with our observations).
It covers $2.9 \times 2.3$ arcmin on the sky (RC3).
However, it is a rather unusual elliptical galaxy and according to Jarvis (1987, hereafter referred to as J87) it should be classified as S0pec (see below). 
One arcsec in the galaxy corresponds to $\sim 203.02$ pc. The effective radius is 35 arcsec (=7.10 kpc).

\subsubsection{Photometric observations}

We used frames kindly provided by O. Hainaut using ESO NTT and EMMI in the RILD mode on July 3--4, 2002 in the $B$--band.
The photometry of IC~3370 is  very interesting and it is given in detail in
J87. We present here some additional elements that are complementary to that study  and are of importance for the analysis that we are undertaking.

\begin{figure*}
\centering
\resizebox{\hsize}{!}{ 
\includegraphics{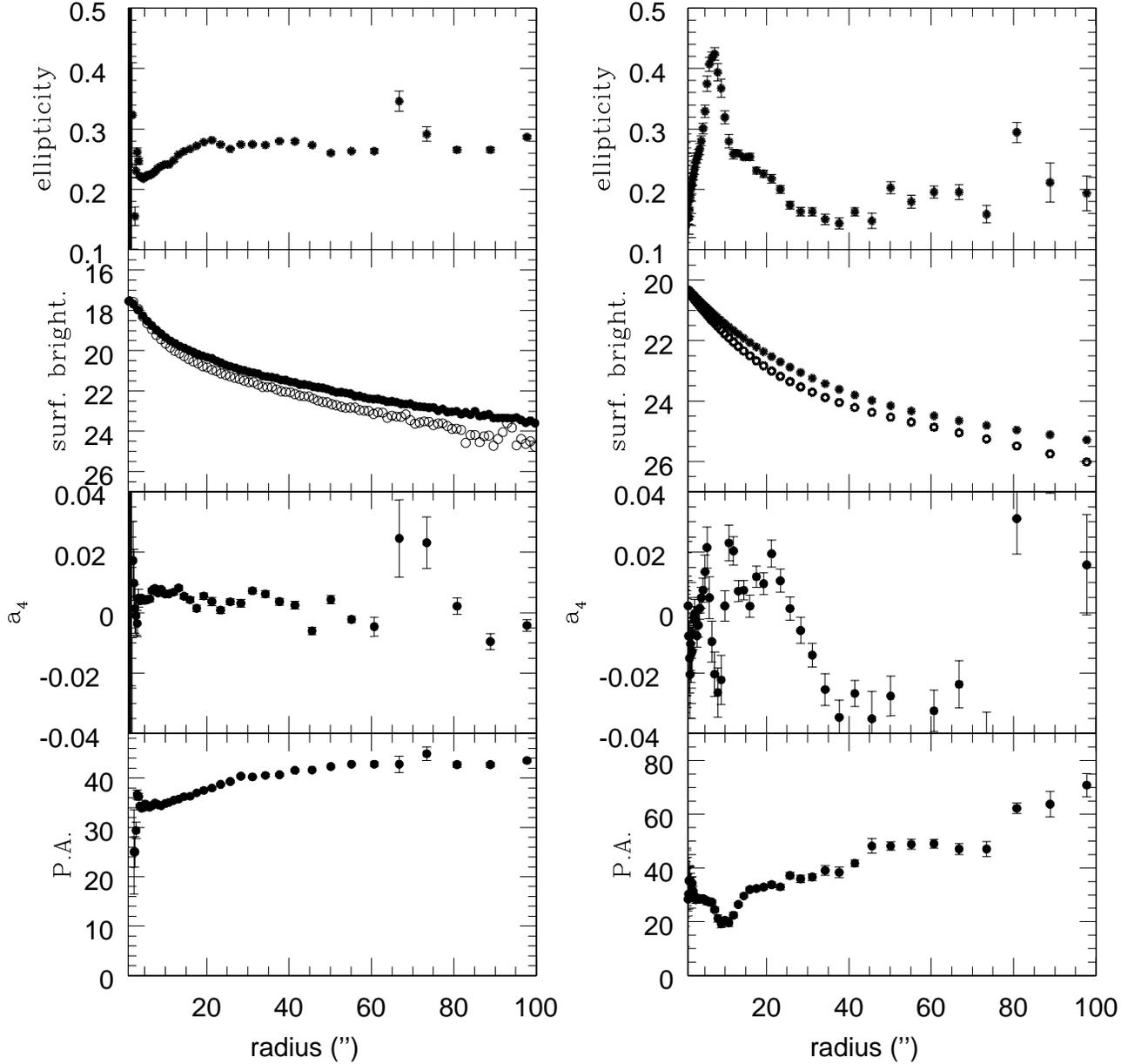}} 
\caption{Photometric profiles of IC~1459 and IC~3370. {\sl Left}: Photometric profiles for IC~1459 (in the $V$--band). From top to bottom: ellipticity, surface brightness for the B filter (see text) in mag arcsec$^{-2}$ (for major axis: full circles; for minor axis: open circles),   $a_4$ parameter and position angle.
{\sl Right}: Photometric profiles for IC~3370 (in the $B$--band). From top to bottom: ellipticity, surface brightness  in mag asrcsec$^{-2}$ (for major axis: full circles; for minor axis: open circles),   $a_4$ parameter and position angle.
}
\label{fig:IC1459_IC3370_phot}
\end{figure*}

One should note that J87 took for the major axis the position angle (PA) of $40^\circ$,
Carollo, Danziger \& Buson (1993) took for the same axis  P.A. of $51^\circ$, while the spectra in this study were taken using P.A. $= 60^\circ$. The reason for these differences lies in a very particular photometry of this galaxy that has strong isophotal twisting as shown in J87 and  in Fig. \ref{fig:IC1459_IC3370_phot}  (see position angle (P.A.) plot). 
This may be evidence for the fact that this galaxy is triaxial, because the isophotes of an axisymmetric system must always be aligned with one another (see, for example, Binney \& Merrifield 1998 Sec 4.2). Fasano \& Bonoli (1989) using a sample of 43 isolated ellipticals found that the twisting observed in these galaxies is intrinsic (triaxiality).
Jarvis has taken the mean position angle of isophotes to be equal to $40\pm 2^\circ$ which is true for the data up to 80 arcsec. However, at larger radii the P.A. tends to increase, so the usage of larger value of $60^\circ$ (and $150^\circ$ for the minor axis) is justified (see Fig.  \ref{fig:IC1459_IC3370_phot}).

In Fig. \ref{fig:IC1459_IC3370_phot} (right) we present relevant photometric data obtained using {\sc IRAF} task {\sc ellipse}: ellipticity,  magnitude in the $B$--band for major axis (filled circles) and minor axis (open circles), $a_4$ parameter
(which measures the fourth Fourier component of the intensity variations along the best--fitting ellipse to an isophote) and the position angles, as a function of distance. The value of $a_4$ is positive  up to one effective radius (for almost all values of radius), thus indicating that the isophotes are discy, while beyond one effective radius, the isophotes become boxy since $a_4$ is negative. Since $a_4$ increases rapidly up to $\sim 5$ arcsec this  can lead to the conclusion of the embedded  disc.   The existence of the stellar disc was shown in J87.

\subsection{NGC~3379}

\begin{figure*}
\centering
\resizebox{\hsize}{!}{ 
\includegraphics{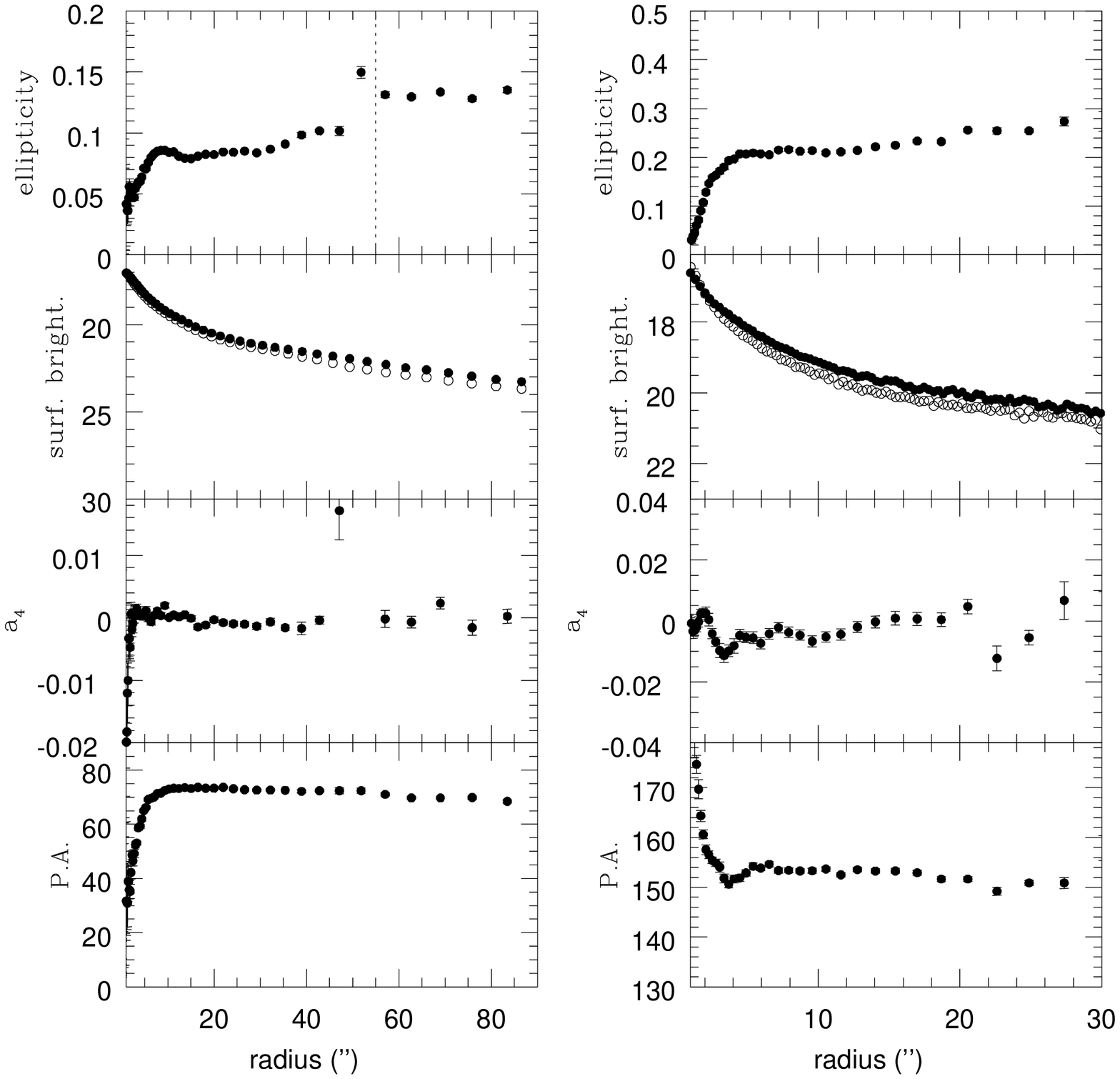}} 
\caption{Photometric profiles of NGC~3379 and NGC~4105.
Vertical dotted line indicates one effective radius (in the case of NGC~3379 only).
 {\sl Left}: Photometric profiles for NGC~3379 in the I-band. From top to bottom: ellipticity, surface brightness (for the R filter from Capaccioli et al. (1990))  in mag arcsec$^{-2}$ (for major axis: full circles; for minor axis: open circles),   $a_4$ parameter (fourth harmonic deviations from ellipse)  and position angle.
{\sl Right}: Photometric profiles for NGC~4105 (in the $R$--band). From top to bottom: ellipticity, surface brightness  in mag asrcsec$^{-2}$ (for major axis: full circles; for minor axis: open circles),   $a_4$ parameter and position angle.
}
\label{fig:NGC3379_NGC4105_phot}
\end{figure*}

NGC~3379 is a bright E1 galaxy (note the ellipticity $\epsilon \approx 0.15$ in  Fig. (\ref{fig:NGC3379_NGC4105_phot}) (left)); there are still some doubts whether this is a bona fide normal elliptical  or a face-on lenticular galaxy (Statler 2001; Gregg et al. 2004 Sec 7), with a heliocentric radial velocity of 911 \kms, (taken from the NED database)  and absolute $B$ magnitude -20.57. One arcsec in the galaxy corresponds to $\sim 63.12$ pc. The effective radius is 54.8 arcsec (=3.46 kpc); see also the discussion in Introduction. Its kinematics  have been the subject of several papers by Statler and collaborators (Statler 1994; Statler \& Smecker-Hane 1999; Statler 2001).

\subsubsection{Photometric observations}
The  surface brightness was taken from the paper of  Capaccioli et al. (1990), whereas  
ellipticity, $a_4$ parameter 
and position angle as function of radius 
were extracted from images from the ESO archive ($I$--band, NTT SUSI)  using standard {\sc IRAF} commands. This galaxy shows small departures of $a_4$ parameter from zero beyond $\sim 2$ arcsec. The
position angle is approximately constant beyond $\sim 15$ arcsec.

\subsection{NGC~4105}
 
NGC4105 is an E2 galaxy, with heliocentric radial velocity of 1918 \kms \  (taken from the NED database), and absolute $B$ magnitude -20.72. One arcsec in the galaxy corresponds to $\sim 134.14$ pc. The effective radius is 35 arcsec (=4.69 kpc).

NGC~4105 belongs to a binary system with a companion galaxy NGC~4106.
which is $\sim$ 1 arcmin away from NGC~4105. 
 Longhetti et al. (1998) claimed to see evidence of the interaction in the velocity profiles  of NGC~4105. In their study they used the position angle of 118 degrees to find asymmetric profiles for both velocity and velocity dispersion. In our study we used position angle of the major photometric axis of 150 degrees and of the minor photometric axis of 60 degrees. For these position angles we notice departures from symmetry, but since the effect is not so strong  out to $\sim 1R_e$ (most notably for the velocity dispersion and the $h_4$ parameter, for the major axis, see Sec. 4.4 below)
in the following discussion  we will neglect the possible gravitational influence of NGC~4106.

 \subsubsection{Photometric observations}
Photometric data were extracted from frames obtained courtesy of M. Carollo \& K. Freeman (who used ESO 2.2 m telescope with EFOSC) using standard {\sc IRAF} routines.
The photometric profiles are given in the $R$--band.
We present the results obtained using the   {\sc IRAF} routines in 
Fig. \ref{fig:NGC3379_NGC4105_phot} (right).
Beyond $\sim 6$ arcsec the position angle is approximately constant.
Photometry of NGC~4105 has been presented in Reduzzi \& Rampazzo (1996):
our photometric profiles are in a good agreement with the results presented
in this paper.

We  present the basic data related to IC~3370, IC~1459, NGC~3379 and NGC~4105 with the data sources in Table 1. Note that for IC~3370 because of the fact that the X--ray was not detected the estimates based on the X--ray data are missing
(see Sec. 6 for details).

\begin{table*}
\caption{Basic properties of IC~1459, IC~3370,  NGC~3379 and NGC~4105}
\begin{center}
\begin{tabular}{rrrrrrrrrr}
\hline
\noalign{\smallskip}
\multicolumn{1}{c}{Name} &
\multicolumn{1}{c}{$r$} &
\multicolumn{1}{c}{$r$} &
\multicolumn{1}{c}{$r/R_e$} &
\multicolumn{1}{c}{B} &
\multicolumn{1}{c}{$\log L_{\rm X}$} &
\multicolumn{1}{c}{T} &
\multicolumn{1}{c}{$D$} &
\multicolumn{1}{c}{M} &
\multicolumn{1}{c}{$({M\over L})_B$} \\
\noalign{\smallskip}
\multicolumn{1}{c}{\ \ } &
\multicolumn{1}{c}{[arcsec]} &
\multicolumn{1}{c}{[kpc]} &
\multicolumn{1}{c}{\ \ } &
\multicolumn{1}{c}{[mag]} &
\multicolumn{1}{c}{[{erg s}${}^{-1}$]} &
\multicolumn{1}{c}{[keV]} &
\multicolumn{1}{c}{[Mpc]} &
\multicolumn{1}{c}{[$10^{11}{\rm M_\odot}$]} &
\multicolumn{1}{c}{\ \ }  \\
\noalign{\smallskip}
\multicolumn{1}{c}{(1)} &
\multicolumn{1}{c}{(2)} &
\multicolumn{1}{c}{(3)} &
\multicolumn{1}{c}{(4)} &
\multicolumn{1}{c}{(5)} &
\multicolumn{1}{c}{(6)} &
\multicolumn{1}{c}{(7)} &
\multicolumn{1}{c}{(8)} &
\multicolumn{1}{c}{(9)} &
\multicolumn{1}{c}{(10)} \\
\noalign{\smallskip}
\hline
\noalign{\smallskip}  
 {IC~1459}  & 100  & 11.7  & 2.86 &11.14 & 41.19 &0.60 & 24.16 &	3.91  &  13.00 \\
 {IC~3370}  &  120 & 24.36  & 3.43 & 11.59 & -- & -- &41.86 & -- &  -- \\
 {NGC~3379}  & 80  & 5.04  & 1.46 &10.10 & 39.78 &0.26 &  13.01 &	0.92  &  7.00 \\
  {NGC~4105}  & 30  & 4.02  & 0.86 &11.27 & 41.94 &0.76 &  27.66 &	1.70  &  10.80 \\
\noalign{\smallskip}
\hline
\noalign{\medskip}
\end{tabular}
\end{center}
\begin{minipage}{18cm}
NOTES -- 
Col. (2):  radius  out to which long--slit spectra  extend (in arcsec).
Col. (3):  same as Col. (2) but in kpc.
Col. (4):  same as Col. (2) but in units of $R_e$.
Col. (5):  total B magnitude (taken from LEDA).
Col. (6):  X--ray luminosity (taken from: Brown \& Bregman 1998 for IC~1459, 
Brown \& Bregman 1998 for NGC~3379 and Fabbiano, Kim \& Trinchieri 1992 for NGC~4105).
Col. (7): temperature (taken from:  Fabbiano et al. 2003 and  Davis \& White 1996 for IC~1459,  Brown \& Bregman 2001 for NGC~3379 and Davis \& White 1996 for NGC~4105)
Col. (8): distance calculated using $H_0=70$ 
$ {\rm km \ s^{-1}\ Mpc^{-1}}$ (using heliocentric radial velocities from the NED archive).
Col. (9): mass in units of $10^{11}{\rm M_\odot}$  estimated using Eq. (\ref{eqn:TOT1}) for the radius given in the Col. (2).
Col. (10):   estimate of the mass--to--light ratio using Eq. (\ref{eqn:ML1}) again for the radius in the Col. (2). See Fig. {\ref{fig:IC1459_xray}} for estimates for IC~1459 based on other temperatures.
           
\end{minipage}
\label{tab:table_1}
\end{table*}

\section{Stellar kinematics}   

Long--slit spectra observations  of IC~1459 and IC~3379 were made by one of us (IJD)   
during 1997 August 28--30  using the ESO 
NTT and EMMI in the Red Medium Spectroscopy mode.
The  central wavelength was chosen to be near the Mg$_2$ feature: $\sim$ 5150 \AA. The range that was covered was $\sim$ 700 \AA. 
The spectra were rebinned at the telescope over 2 pixels giving a scale of 0.56 arcsec pixel$^{-1}$.
The spectroscopic observations of NGC~3379 and NGC~4105 were obtained courtesy of M. Carollo and K. Freeman.
Galaxy NGC~3379 was observed  using the Double Beam Spectrograph attached to the Australian National University 2.3 m telescope at Siding Springs Observatory.    NGC4105 was observed using ESO 2.2 m telescope with EFOSC.
The  central wavelength was  the Mg$_2$ feature: $\sim$ 5150 \AA. 
The range   covered was $\sim$ 1000 \AA \ (for NGC~3379)
and $\sim$ 1200 \AA \ (for NGC~4105).

The standard  data reduction steps were performed for all the galaxies: bias subtraction, flat--fielding, 
cleaning of cosmic rays and subtraction of sky background. 
We used the {\sc MIDAS}\footnote{{\sc MIDAS} is developed and maintained by the European Southern Observatory.} package in which we implemented several 
routines necessary for fast reduction
(wavelength calibration)
 and post--reduction procedures (logarithmic binning of the spectra).
Wavelength calibration was done using the Helium--Argon (for IC~1459, IC~3370, NGC~4105) and Neon--Argon (for NGC~3379) comparison lamps spectra.
Sky subtraction was done by taking an average of 30 rows near the edges of the exposure frames. Finally the spectra were rebinned on a logarithmic scale.
Also, spectra of several template stars  were reduced as described above, continuum divided, and averaged over several rows
 in order to obtain one stellar template spectrum of high signal--to--noise ratio (S/N).

For the extraction of the stellar kinematics  we used  
freely available  ``Gauss--Hermite Fourier Fitting Software'' made by R. van der Marel and M. Franx.
Since this package was written for the Sun FORTRAN compiler for the Sun UNIX platform, initial testing was done using  Sun Sparc (Sun-Blade-100) platform. Later, it was modified and ported to the x86 
GNU/Linux PC platform that uses a GNU FORTRAN compiler. Detailed tests were done, and it was found that the results obtained in two different environments were in the excellent agreement. All the results presented in this paper were obtained in the GNU/Linux environment.

\subsection{IC~1459}

\begin{figure*}
\centering
\resizebox{\hsize}{!}{ 
\includegraphics{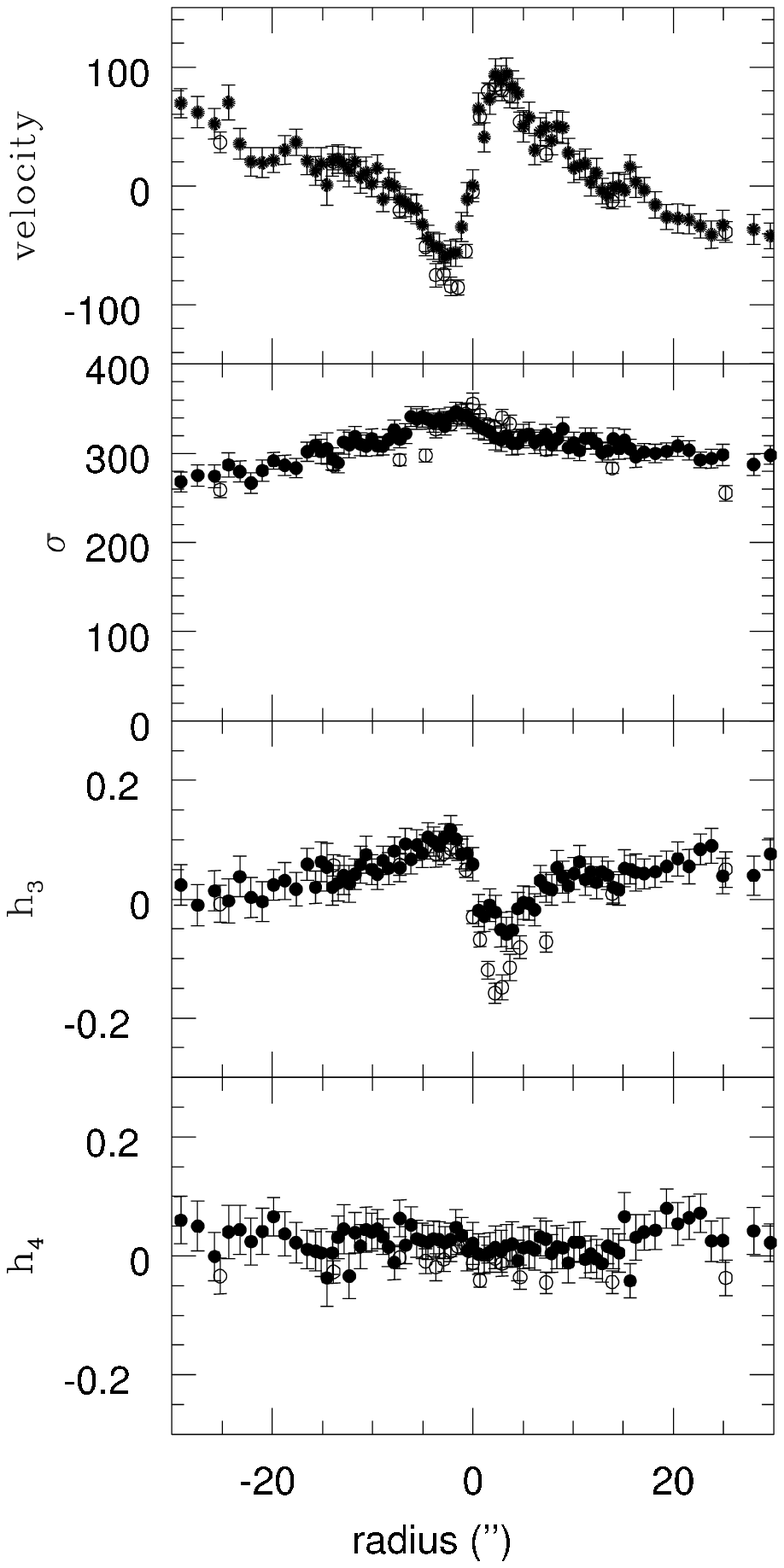}
\includegraphics{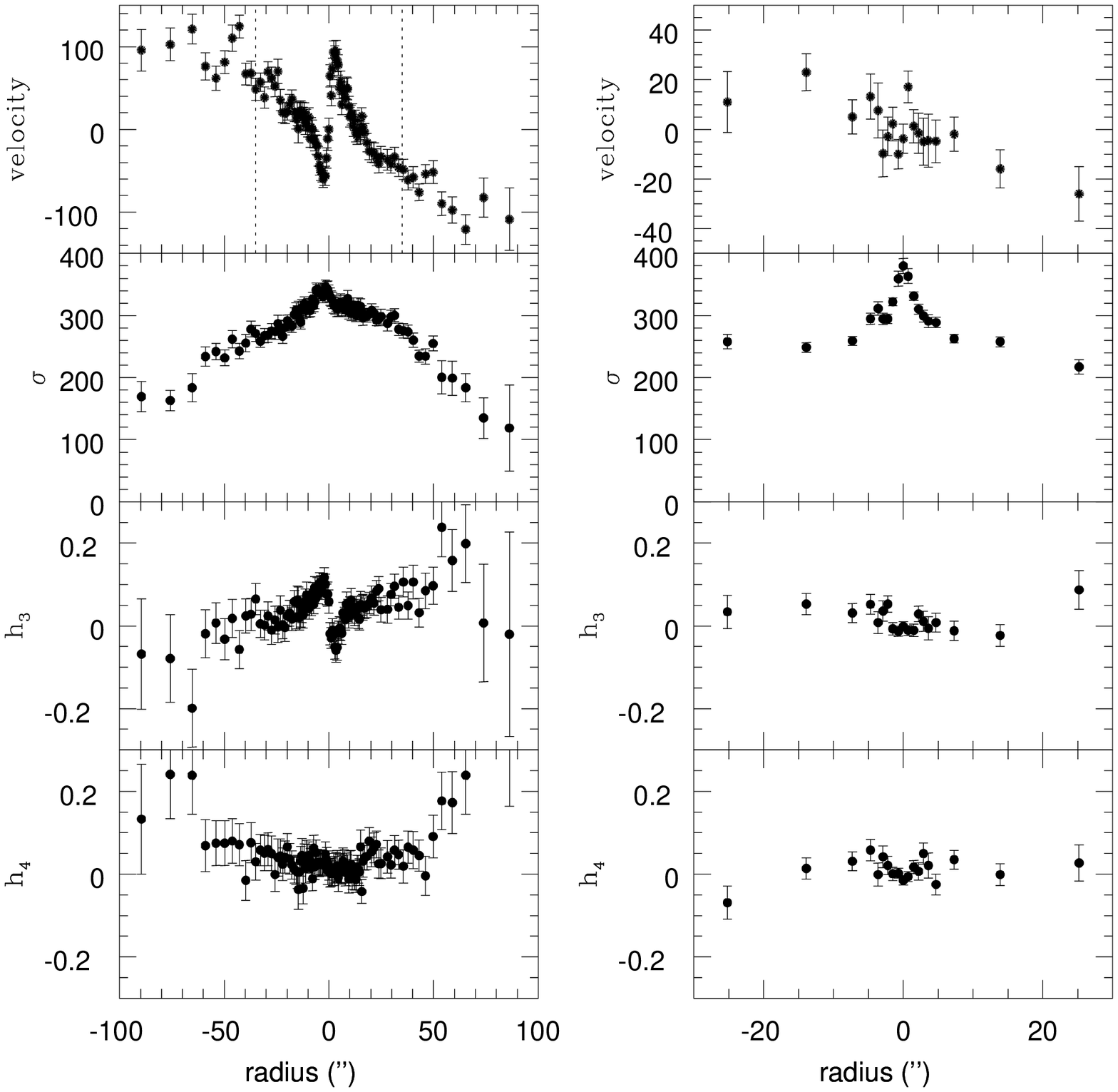}} 
\caption{ Stellar kinematics of IC~1459. {\sl Left}: Comparison of the kinematic profiles for the major axis of IC~1459 (P.A.$=40^\circ$, black circles) and the data taken from Cappellari et al. (2002) (P.A.$=39^\circ$, open circles). 
{\sl Middle}: major axis data.
{\sl Right}: minor axis data (taken from Cappellari et al. 2002).
From top to bottom: velocity, velocity dispersion, $h_3$ and $h_4$ parameters. One effective radius in case of the major axis is plotted using dashed line. 
Note that in case of the minor axis it is out of scale.
}
\label{fig:IC1459_capp_kin}
\end{figure*}

Several exposures were taken for two  different position angles:
for the galactic major axis (P.A. = 40$^\circ$) total exposure of 35,100 s,
and for the  minor axis (P.A. = 130$^\circ$) total exposure of 3,600 s.
Because of the fact that only one  exposure was available for the minor axis,
the removal of the cosmic ray hits  was not successful and we have taken the minor axis stellar kinematics from Cappellari et al. (2002). 
We compared the results 
for the major axis and plot the comparison in Fig. \ref{fig:IC1459_capp_kin} (left panel).
The agreement is good, except for the velocity and $h_3$ parameter near the galactic centre where some discrepancy exists. Note, however,
that Cappellari et al. (2002) used P.A.=39$^\circ$ and observations which we analyze here were made at P.A.=40$^\circ$. In the outer parts agreement is excellent for the whole velocity profile.
  The template star HR~5582 (type K3$^-$) was used. The instrumental dispersion was $\sim$ 3.5  {\AA} ($\sim$ 190 \kms) and was determined using Helium--Argon spectrum
in a region $\sim$ 5000  {\AA}. The slit width was 3 arcsec.

In Fig.  \ref{fig:IC1459_capp_kin} (middle and right panel) we show the major and minor axis kinematic parameters. 
Major axis data show the rapid increase of velocity in the inner $\sim$
3 arcsec: velocity rises to $\sim$ 100 \kms \ (note however a small asymmetry in our determination of velocity). Velocity dispersion is large at the centre: $\sim$ 350 \kms, and decreases rapidly to $\sim$ 240 \kms \ (at $\sim$ 40~arcsec). 
There is a plateau in velocity dispersion 
between $\sim$ 20 arcsec and 30 arcsec after which velocity dispersion decreases.
 The  $h_3$ parameter shows a typical behaviour, i.e. it rises (falls)  when velocity rapidly increases (decreases). In the outer parts it shows small departures from zero. The $h_4$ parameter shows very small departures from zero in the inner parts, and in the outer parts there is an increase of its value, suggesting existence of the radial anisotropy.  Minor axis data provide evidence of small velocities, and larger central velocity dispersion ($\sim$ 380 \kms). Both the $h_3$ and $h_4$ parameters show very small departures from zero throughout the observed parts of the galaxy.

\subsection{IC~3370}
Several exposures were taken for three  different position angles:
for the galactic major axis (P.A. = 60$^\circ$) total exposure of 21,600 s,
for the  minor axis (P.A. = 150$^\circ$) total exposure of 7,200 s. 
Also, the spectra of the intermediate axis were taken (P.A.=20$^\circ$), and the total exposure time was 14,400 s.
The template star spectrum of HR~2701  (type K0III) was used. The instrumental dispersion was $\sim$ 3.5 
 {\AA} ($\sim$ 190 \kms) and was determined using a Helium--Argon spectrum
 in a region $\sim$ 5000  {\AA}. The slit width was 3 arcsec.
 
\begin{figure*}
\centering
\resizebox{\hsize}{!}{ 
\includegraphics{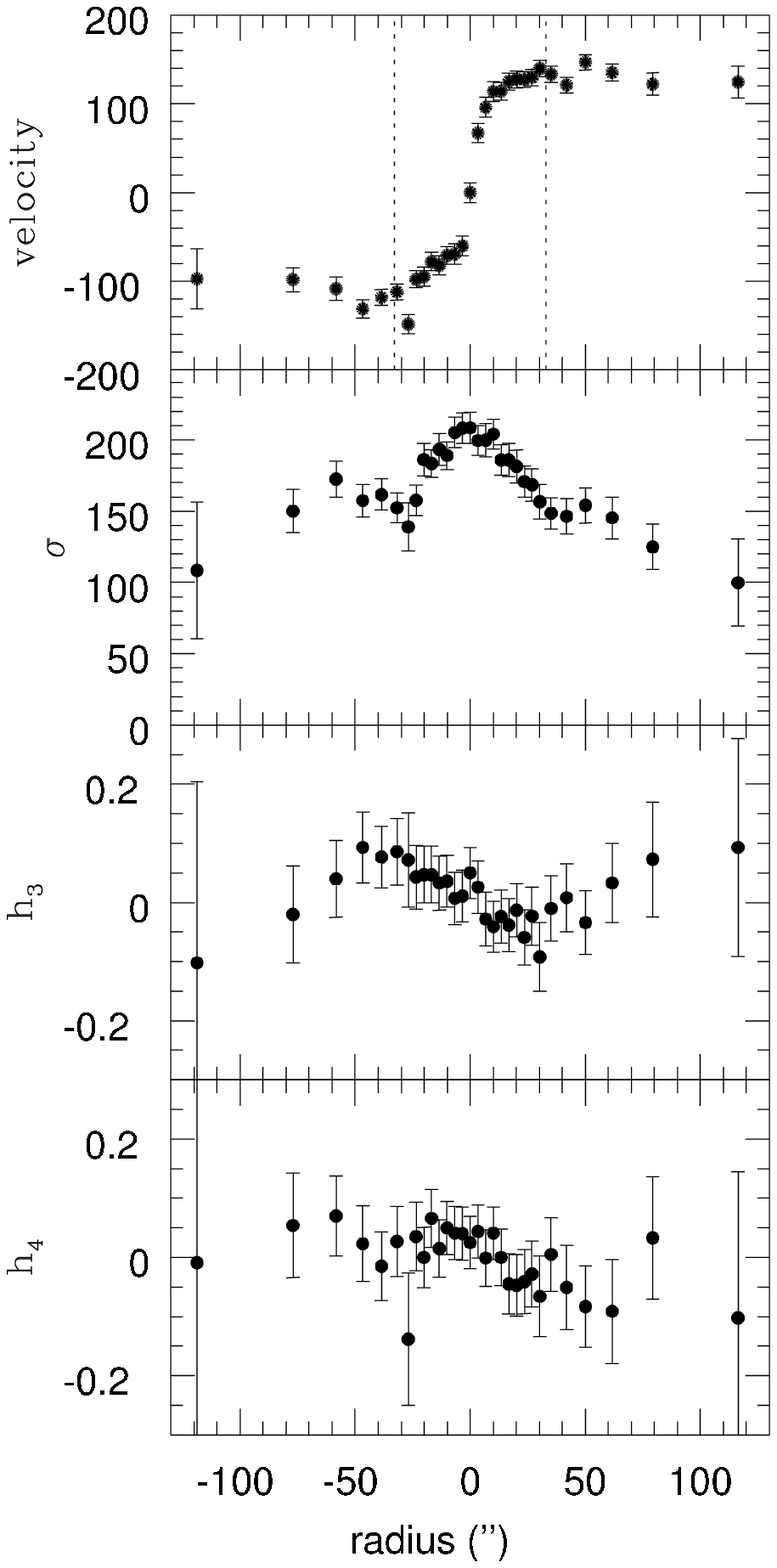}
\includegraphics{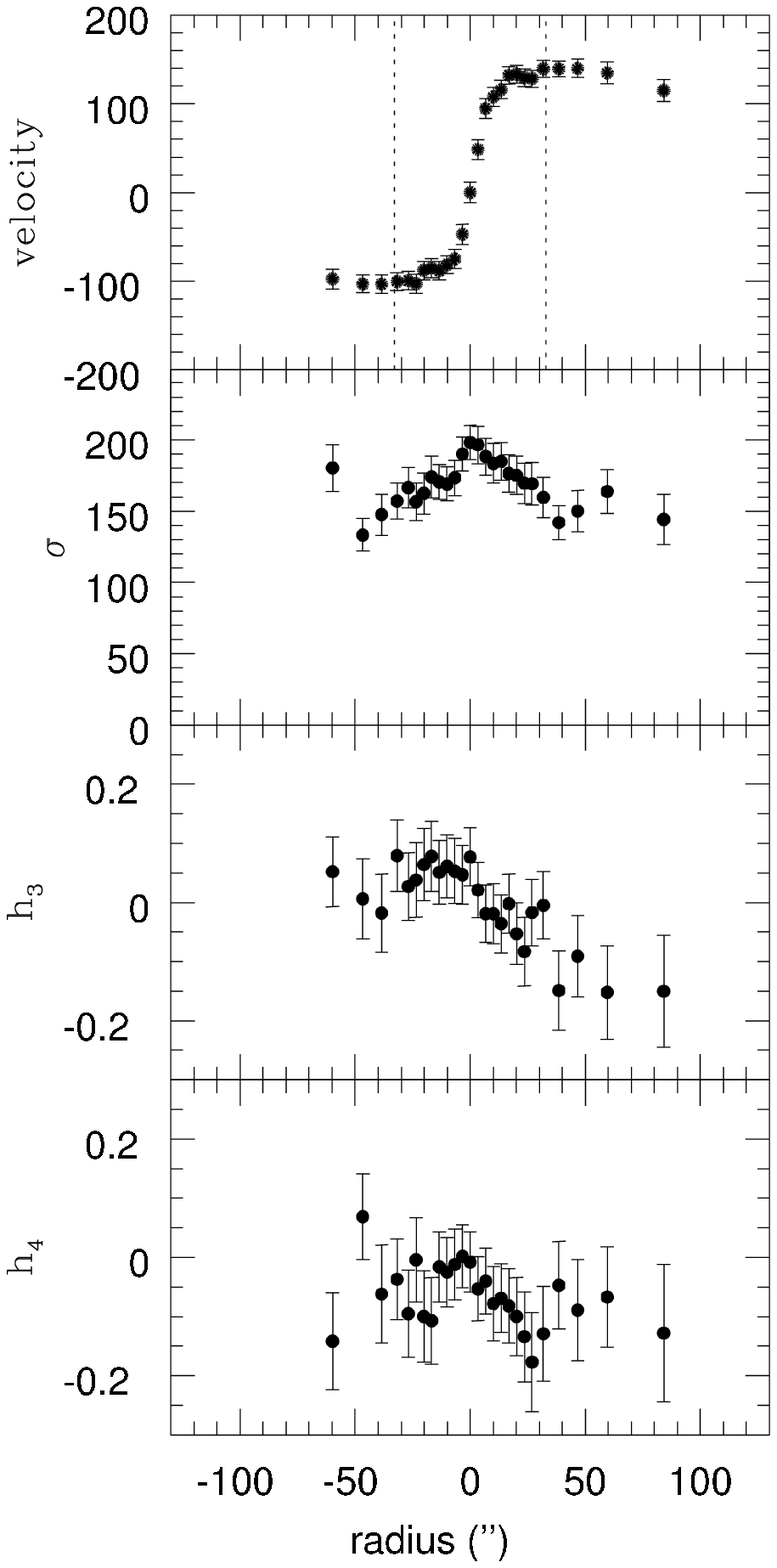}
\includegraphics{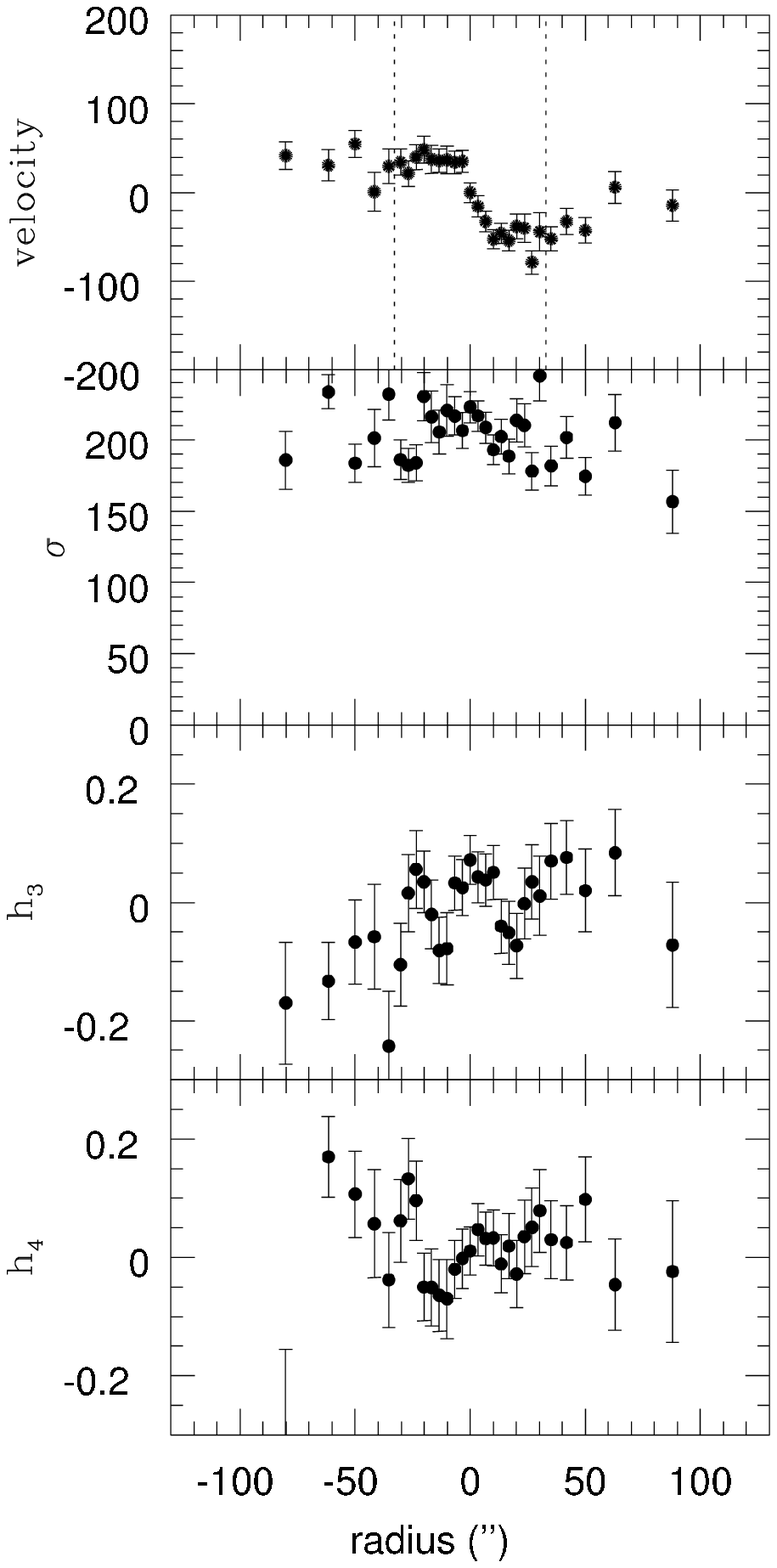}} 
\caption{Kinematic profiles for major axis  (P.A.$=60^\circ$, left), 
the intermediate (P.A.$=150^\circ$, middle) and minor (P.A.$=20^\circ$, right) axes of IC~3370. From top to bottom: velocity, velocity dispersion, $h_3$ and $h_4$ parameters. One effective radius is plotted using dashed lines.}

\label{fig:IC3370_kin}
\end{figure*}

In Fig.  \ref{fig:IC3370_kin} (left panel) we show the major axis kinematic parameters. This galaxy indeed shows behaviour that is characteristic for an S0 galaxy:
for example, its major axis kinematics can be compared to that of NGC~1461, lenticular  galaxy from the Fisher (1997) sample. Note  the usual behaviour of $h_3$ parameter: when the velocity rises, $h_3$ decreases, and vice versa. 
In Fig. \ref{fig:IC3370_kin} we present intermediate (middle panel) and minor
(right panel) axis kinematic profiles:
IC~3370 has  minor axis rotation that provides an additional hint (apart from the isophotal twist) of the triaxiality. Note the small values (consistent with zero)  of $h_3$ and $h_4$ at the large distances from the centre for the major axis and their generally small values in the case of the minor axis -- they provide  evidence of the lack of excessive tangential motions, that may mimic the dark matter in the outer parts of the galaxy (see above). In the case of the intermediate axis beyond 60 arcsec we have a hint of negative values of the $h_4$ parameter which could mimic dark matter in these regions.

\begin{figure*}
\centering
\resizebox{\hsize}{!}{ 
\includegraphics{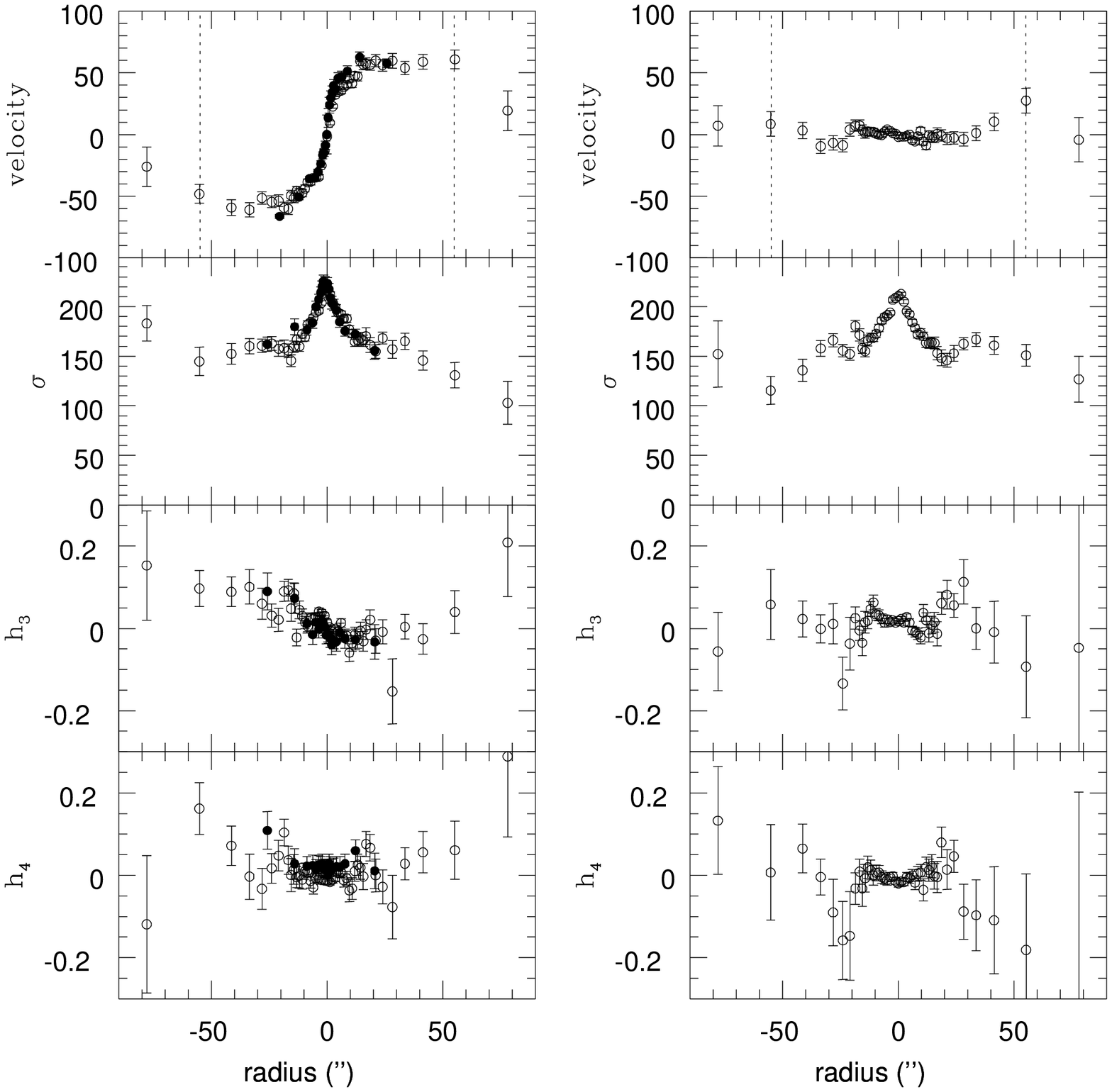}} 
\caption{Open circles: stellar kinematics 
of NGC~3379 (taken from
Statler \& Smecker-Hane (1999)) for major (left) and minor (right) axis.
Black circles: our extraction of the stellar kinematics given for comparison.  
From top to bottom: velocity, velocity dispersion, $h_3$ and $h_4$ parameters.  Dashed line indicates one effective radius. 
}
\label{fig:NGC3379_maj_min}
\end{figure*}

\subsection{NGC~3379}

For  NGC~3379 the long slit spectra of the major axis (P.A.=70$^\circ$) were taken on March 13-14, 1997 and the total exposure time was 6,000 s.  
The scale was 0.59 arcsec pixel $^{-1}$ and the wavelength calibration was done using a Neon-Argon lamp. The template star was K--giant cpd-43. 
The instrumental dispersion was $\sim 2$ \ang \ ($\sim$ 100 \kms) and was determined using a Neon-Argon spectrum in a region $\sim$ 5000 \ang. 
Slit width was 1.5 arcsec.
Since we had only major axis (P.A.=70$^\circ$) data, which we reduced, we have taken data from Statler \& Smecker-Hane (1999) for the major and the minor axis (P.A.=340$^\circ$). 
We compared the results for the inner region which  we have in common for the major axis and found that they are in an excellent agreement (see Fig. (\ref{fig:NGC3379_maj_min} (left)).
The data that we had extend out to $\approx$ 30  arcsec, so in the modelling procedures (see next Section) we will use 
 Statler \& Smecker-Hane (1999) measurements only at all radii because their data extend  to
a larger radius (80 arcsec that is $\approx 1.46$ $R{}_e$)
and are also available for the minor axis.

This galaxy shows a steep increase of velocity:  it rises to $\sim$ 60 \kms \ 
in the inner 20 arcsec. After a plateau between $\sim$ 20 arcsec and $\sim$ 60 arcsec the velocity shows a tendency to decrease. The velocity dispersion peaks at $\sim$ 230 \kms \ and then decreases rapidly. There is a plateau between $\sim$ 20 arcsec and $\sim$ 50 arcsec. One can see that there is an obvious asymmetry at $\sim$ 80 arcsec.
 The $ h_3$ parameter is small out to $\sim 50$  arcsec, but shows departures from zero at $\sim$ 70 arcsec. $h_4$ remains small throughout the whole observed galaxy, except in the outer parts for which there is a hint of departures from zero, but since error bars are large, it is difficult to draw firm conclusions. Minor axis data suggest that   NGC~3379 does not show significant rotation on the minor axis.
The  velocity dispersion profile is similar to that of the major axis.
 The $h_3$ and $h_4$ parameters are small throughout the whole observed galaxy on the minor axis (see Fig. \ref{fig:NGC3379_maj_min}).

\begin{figure*}
\centering
\resizebox{\hsize}{!}{ 
\includegraphics{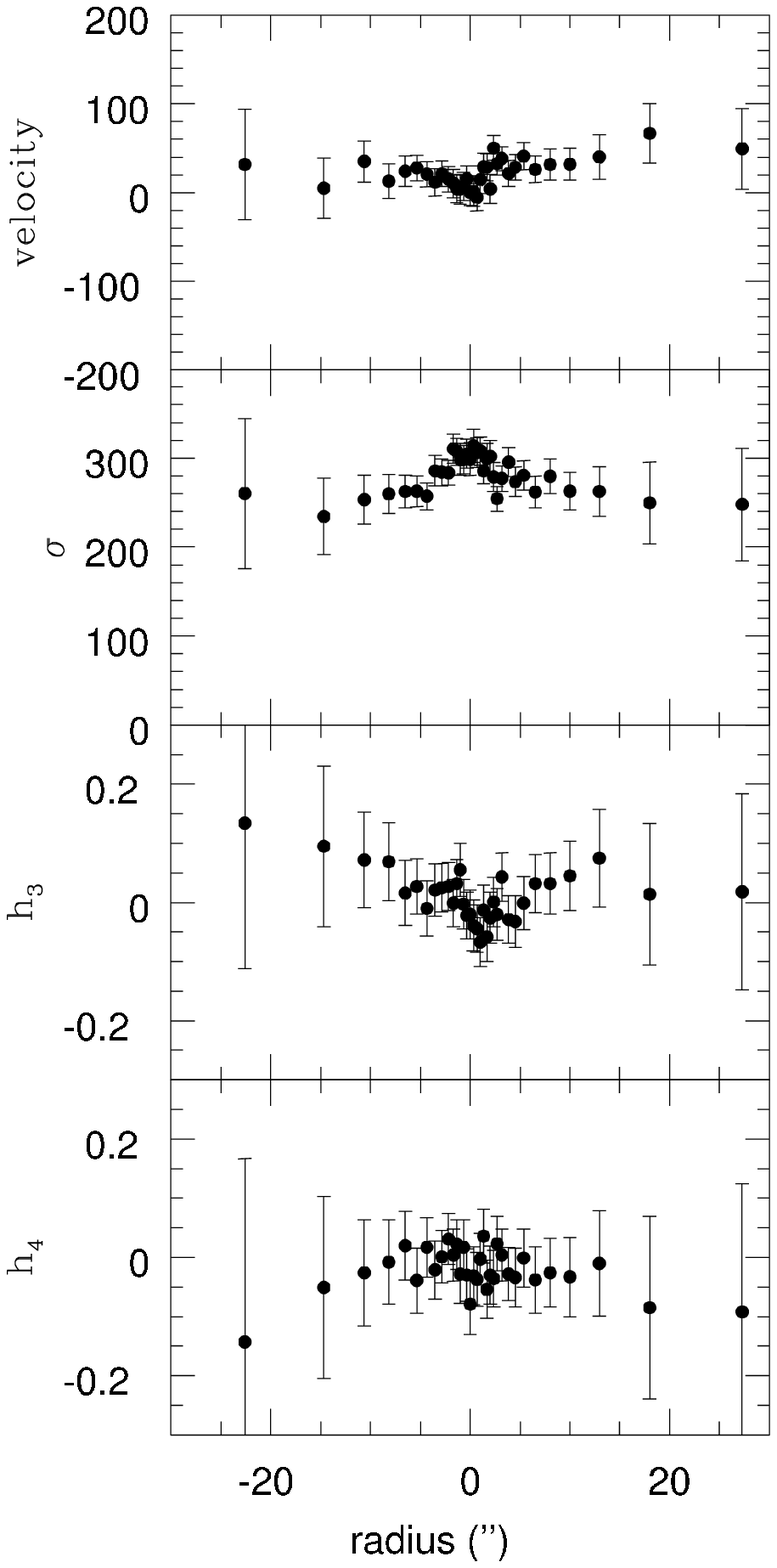}
\includegraphics{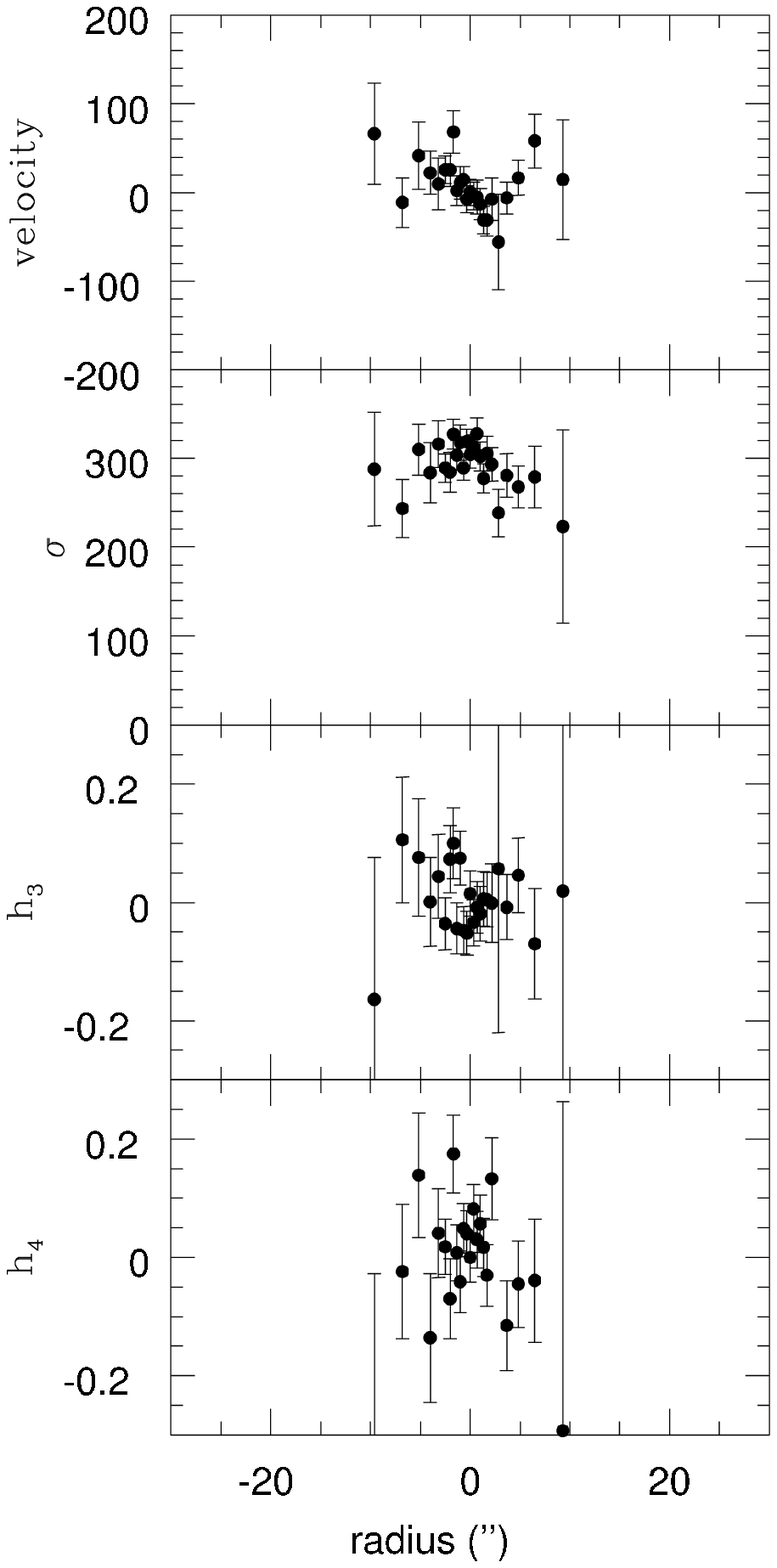}} 
\caption{Stellar kinematics 
of NGC~4105 
for major (left) and minor (right) axis.
From top to bottom: velocity, velocity dispersion, $h_3$ and $h_4$ parameters.  
}
\label{fig:NGC4105_maj_min}
\end{figure*}

\subsection{NGC~4105}

Long slit spectra of NGC~4105 were obtained on  March 9-13, 1994
using the ESO 2.2 m telescope with EFOSC.
The total exposure time for the  major axis (P.A.=150$^\circ$)  
was 27,900 s. The total exposure time for the minor axis (P.A.=60$^\circ$) was 14,400 s.  The scale was 0.336 arcsec pixel $^{-1}$. The wavelength calibration was made using a Helium--Argon lamp. The  template star was HR~5582 (type K3$^-$). The instrumental dispersion was $\sim 4.2$ \ang ($\sim$ 280 \kms) and was determined using Helium--Argon spectrum in the region near 5000 \ang. 
Slit width was 1.5 arcsec.

On the major axis this galaxy   shows a maximum value of the velocity $\sim$ 60 \kms (see  Fig. \ref{fig:NGC4105_maj_min}, left). Note that there is a hint of a counterrotating stellar core in the inner 3 arcsec. In general, there is a lack of symmetry about the galaxy centre.  The central value of the velocity dispersion is large: $\sim$ 320 \kms. It declines in the inner $\sim$ 5 arcsec after which  there is a tendency to remain constant (out to  $0.7 R_e$). 
The velocities are not antisymmetric: the reason for this could be the vicinity of NGC~4106 and interaction with it.
$h_3$ also shows a hint of the effects of the counterrotating stellar core in the inner 3 arcsec. At the larger radii the value of $h_3$ is consistent with zero. The $h_4$ parameter remains small (slightly negative, but consistent with zero) throughout the whole observed galaxy.
On the minor axis NGC~4105 shows rather complex behaviour and   again a lack of symmetry is evident   (see  Fig. \ref{fig:NGC4105_maj_min}, right). The velocity dispersion decreases from the central value of $\sim$ 320 \kms\ to $\sim$ 200 \kms.  Not much can be said about $h_3$ and $h_4$ parameters, except that they show asymmetries.

\section{Two--integral (2I) modelling}

In all the dynamical modelling performed below we folded the major axis data about the $y$-axis and the minor axis data were folded about $x$-axis. We took the mean value of the observed kinematical parameters in all the cases taking into account that the  velocity and the $h_3$ parameter are odd functions and the  velocity dispersion and the $h_4$ parameter are even functions of the radius.
In the modelling which follows, we tested different inclinations for all the galaxies and we present here only the best--case inclinations.

\subsection{IC~1459}

This galaxy has a counterrotating core, and therefore, two--integral axisymmetric modelling conceived by BDI based on the photometric profiles will necessarily fail in the inner regions. We tested the inclination angles from 
50$^\circ$ to 90$^\circ$ and we decided to use the inclination angle of 65$^\circ$ in all the cases because it provided the best fit to the data (although very far from  perfect). This inclination angle implies intrinsic axis ratio of $\sim 0.7$.

\begin{figure*}
\centering
\resizebox{\hsize}{!}{ 
 \includegraphics{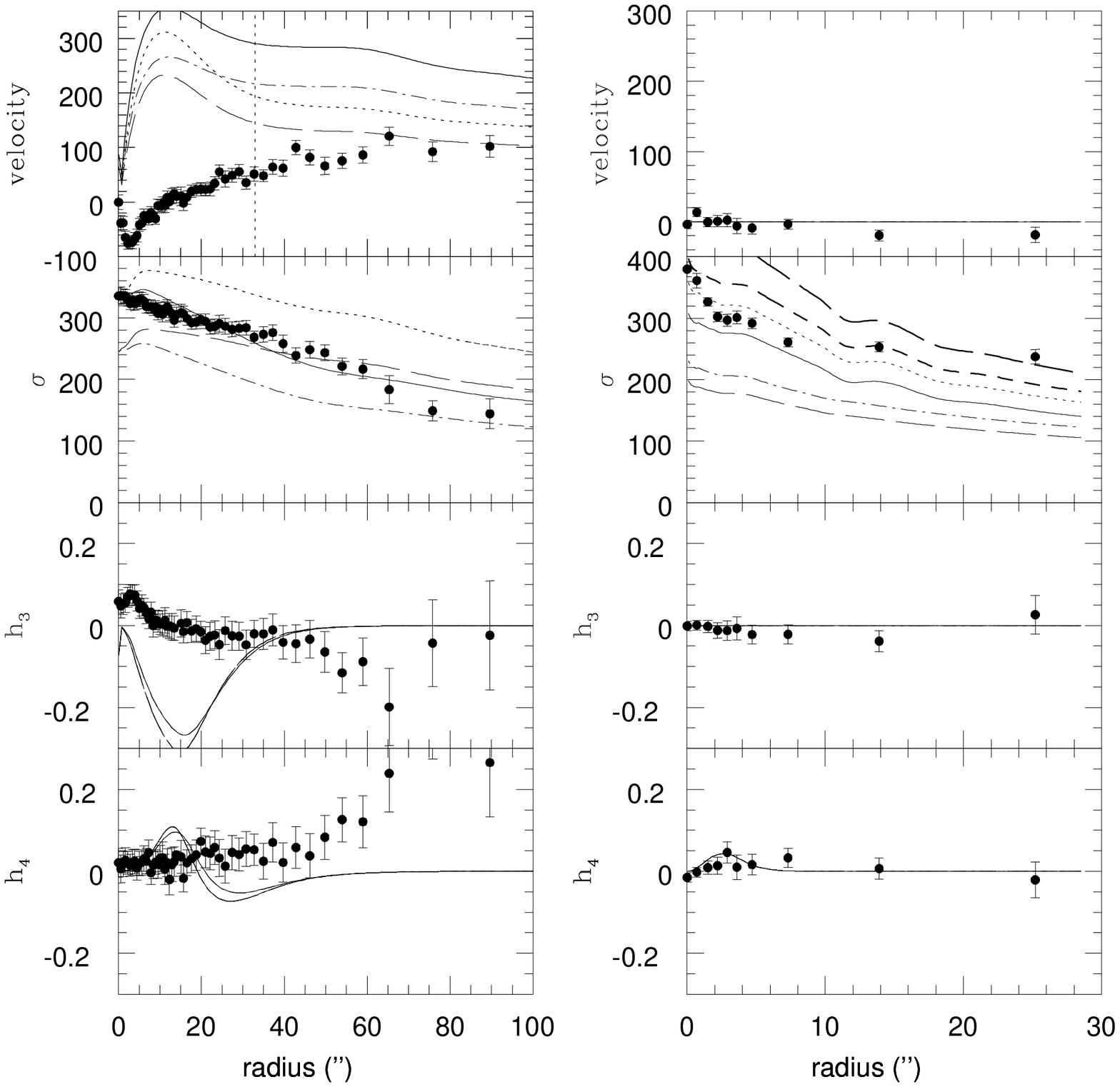}
 \includegraphics{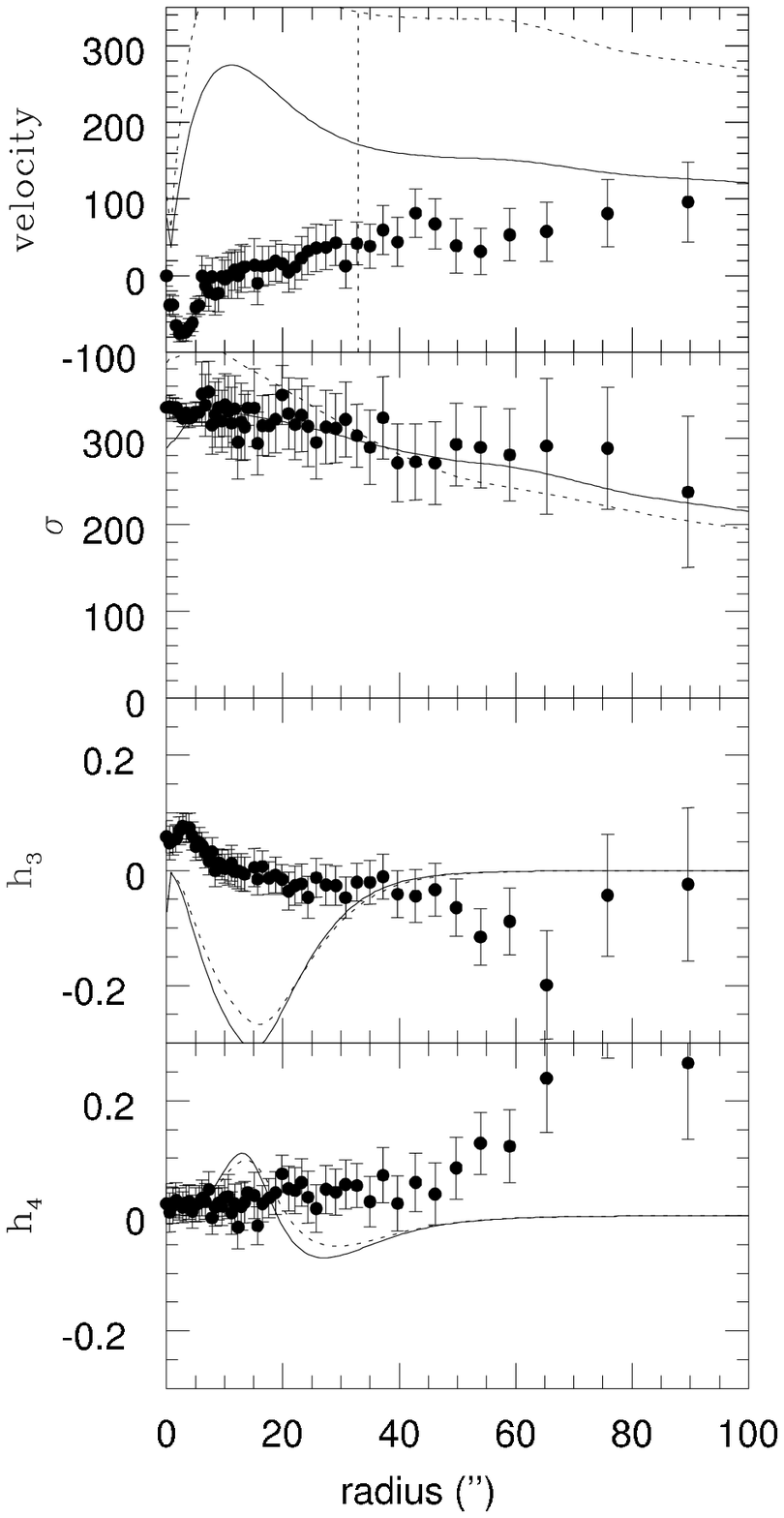}}
 \caption{Predictions of the two--integral models for IC~1459. 
 {\sl Left}: major  axis predictions.
Solid lines:  $k=1$, $M/L_B=6.83$, no dark matter included,
no embedded disc. Dashed lines: $k=0.6$, $M/L_B=3.81$, no dark matter included,
without embedded disc. Dotted lines: $k=0.6$, $M/L_B=6.83$, no dark matter included, without embedded disc.  Dot--dashed lines:   $k=1$, $M/L_B=3.81$, no dark matter included, without embedded disc . 
{\sl Middle}: minor  axis predictions.
Dotted line:  $k=1$, $M/L_B=8.54$, no dark matter included,
without embedded disc. Dashed line: $k=0.6$, $M/L_B=3.05$, no dark matter included,
without embedded disc. Solid line: $k=0.6$, $M/L_B=7.33$, no dark matter included, without embedded disc.  Dot--dashed line:   $k=1$, $M/L_B=3.05$, no dark matter included, without embedded disc. Thick long dashed line: $k=1$, $M/L_B=14.23$, no dark matter included, without embedded disc. Thick short dashed line: $k=0.6$, $M/L_B=12.21$, no dark matter included, without embedded disc.  
Vertical dotted line indicates one effective radius; in case of the minor axis it is out of scale. 
{\sl Right}: Predictions of the two--integral models for the major axis of IC~1459
for {\sl corrected} values of the observed velocity and velocity dispersion  (see text for details).
Dotted line:  $k=1$, $M/L_B=9.56$, no dark matter included,
no embedded disc.  
Solid line: $k=0.6$, $M/L_B=5.31$, no dark matter included, without embedded disc.}

\label{fig:IC1459_model}
\end{figure*}

{\sl Major axis (Fig. \ref{fig:IC1459_model}(left))}: In the case of the major axis we tested $k=0.6$ value: first, 
it gave marginally good fit for the velocity in the outer region of the galaxy 
($M/L_B=3.81$), and a marginally good fit in the region slightly beyond $1 R_e$ for the velocity dispersion (dashed line), and, second, a case of larger $M/L_B=6.83$
did not fit the velocity,  nor the velocity dispersion (dotted line).
In both of these cases no dark matter halo was included, and no embedded disc was assumed. 
If one takes $k=1$, there are two cases that we decided to present: first,
  $M/L_B=6.83$ (no dark halo, and no embedded disc) (solid lines) the velocity is extremely large (it declines from $\sim$ 350 ${\rm km \ s^{-1}}$ at 20 arcsec to $\sim$ 220
  ${\rm km \ s^{-1}}$ at 100 arcsec); the velocity dispersion can be fitted, very closely, throughout the whole observed galaxy, and second, the case when   $M/L_B=3.81$ (no dark halo, and no embedded disc) (dot--dashed lines) for which the fitted velocity has smaller 
values (although still much larger than the observed ones): in a region
between 20 arcsec and 100 arcsec the velocity decreases  from 260 ${\rm km \ s^{-1}}$  to 170 ${\rm km \ s^{-1}}$; the velocity dispersion is much lower, and the successful fit is attained only in the outer parts. The $h_3$ parameter, because of the fact that there is a counterrotating core, cannot be fitted. For the $h_4$ parameter this modelling did not give a successful fit in the outer parts where there possibly exists a radial anisotropy  (judging from the observed non--zero value of the  $h_4$ parameter).
Therefore, one can   state that only the test with $k=1$  ($M/L_B=6.83\pm 0.13$) can provide a fit to the velocity dispersion. The fact that the predicted velocity is much larger is of a crucial importance and will be addressed below.

It was argued that   the Gauss--Hermite estimates are not the best approximation of the mean line--of--sight velocity and   velocity dispersion of the LOSVD (cf. van der Marel \& Franx 1993;
Statler et al. 1996; De Rijcke et al. 2003), because their real values depend on the $h_3$ and $h_4$ parameters.
In the case of IC~1459  there are  significant departures in the Gauss--Hermite parameters from zero for a major axis so we  applied the correction for the velocity and velocity dispersion. 
 Following van der Marel \& Franx (1993) we used the following formulas to get the corrected values  that are then compared with the modelling results. The corrected values are, for the velocity:
\begin{equation}
v_{\rm corr} = v_{\rm GH} + \sqrt{3}(h_3)_{\rm GH} \sigma_{\rm GH},
\end{equation}
and for the velocity dispersion:
\begin{equation}
\sigma_{\rm corr} = \sigma_{\rm GH} (1 + \sqrt{6}(h_4)_{\rm GH}),
\label{eqn:sig_corr}
\end{equation}
where the index ``GH'' is related to the Gauss--Hermite estimates.
This correction is done only in the case of the major axis, since in the case of minor axis the departures from zero in $h_3$ and $h_4$ are minimal.
Using the Cinzano \& van der Marel modelling technique we did not correct the 
observed data in the inner regions (interior to $\sim 6$ arcsec) where the disc may be present and we corrected velocity and velocity dispersion beyond $\sim 6$ arcsec in order to compare with the models. In this inner region the model LOSVD is a sum of 2 Gaussians, and the modelled velocity dispersion is the physical velocity dispersion which is then  compared to the data. We note that the non--zero values of the $h_3$ and $h_4$ parameters are not fitted because in the outer regions the models assume Gaussian LOSVDs.
After the correction is done one can note the following change with respect to the uncorrected values (see Fig. \ref{fig:IC1459_model} (right panel)): 
the velocity dispersion values in the outer part have increased ($h_4$ is positive), but the general trend of decline remains.
When one now examines the modelling results given in Fig. \ref{fig:IC1459_model} (right panel) one can see that a better fit to the observations is obtained using $k=0.6$, and the constant mass--to--light ratio $M/L_B=5.31\pm 0.10$ that is somewhat lower than the value estimated for the best--fitting in the uncorrected case. This however does not alter the main conclusion: IC~1459 can be successfully fitted without invoking dark matter. Note however, that the error bars for $h_4$ are rather large in the outer parts. New observations of IC~1459 made (but still unpublished) by Bridges et al. (2003) should hopefully clarify the mass at $\sim 3 R_e$.

{\sl Minor axis (Fig. \ref{fig:IC1459_model} (middle panel))}:  Three $k=0.6$ cases are plotted: one for $M/L_B=3.05$ (dashed line) which does not provide a good fit  for the velocity dispersion, and the other one for which $M/L_B=7.33$ that provides a better agreement (solid line). 
Finally, a   thick  short dashed line gives a prediction of the velocity dispersion for  $M/L_B=12.21$ and obviously does not provide a  good fit (except marginally at $\sim$ 30 arcsec).   
A better fit was obtained using $k=1$: with the dotted lines we present a case with  $M/L_B=8.54$ (no dark matter, no embedded disc). However,  a fit with    $k=1$, but with a lower value of mass--to--light ($M/L_B=3.05$, dot--dashed line) predicts a velocity dispersion that is too low.
Finally, if one increases mass--to--light ratio to $M/L_B=14.23$ 
(thick long dashed line)  one can get a prediction that seems valid at   $\sim$ 25 arcsec.  
Values of both velocity and $h_3$ parameter are consistent with zero for the minor axis, and $h_4$ is fitted properly for these cases.

We have shown that in the case of the major axis the best--fitting for the velocity dispersion can be obtained
using $k=1$ and  $M/L_B=6.83\pm 0.13$ (or $5.31\pm 0.10$ obtained using corrected values of $v$ and $\sigma$) . However, with these assumptions the velocity is enormously high. This means that one is faced with  the same situation that BDI described in the case of NGC~720. Therefore, as in BDI, one can conclude that IC~1459 {\it cannot have\/} a distribution function of the form $f(E,L_z)$, and that three--integral modelling is needed.

The results for the minor axis modelling are inconclusive because the observations go out only to $\sim 1 R_e$. The slight tendency for the velocity dispersion to flatten in the outer parts of the minor axis could be a result of a predominance of tangential orbits possibly suggested by the trend seen in the $h_4$ parameter.

Because of the counterrotating core there is a strong hint that IC~1459 is the result of a merger. That is why we compared the results of Bendo \& Barnes  (2000) who used an N--body code to study the LOSVD  of simulated merger remnants with the stellar kinematics that we extracted. A reasonable agreement is seen in fig. 9 by Bendo \& Barnes (2000) which shows
the dependence of the Gauss--Hermite parameters as functions of position along the major axis for a typical 3:1 merger (merger between disc galaxies with mass ratios of 3:1). Although in the central parts there is a small discrepancy 
between the observations and the simulation, in the outer parts  there is an obvious trend for an increase in the $h_4$ parameter.
A further detailed comparison is difficult to perform because
we do not know how to scale exactly 
 Bendo \& Barnes simulation to our observed data because we do not know the effective radius of the simulated merger.
In the case of IC~1459  we have no way of knowing what the mass ratio of two disc galaxies might have been as well as other parameters involved in the simulation (such as inclination angles of the merging galaxies). For the core region we find $v/\sigma\approx 0.29$. 
  Note, that for a quoted  Bendo \& Barnes simulation
this ratio is about one. This is only a rough comparison but we hope that future studies of the projected kinematics of simulated merger remnants 
  will bring new insights to the problem of the formation of the counterrotating cores and the formation of early--type galaxies in general, as the number of observed kinematical profiles increases.

\subsection{IC~3370}

\begin{figure*}
\centering
\resizebox{\hsize}{!}{ 
 \includegraphics{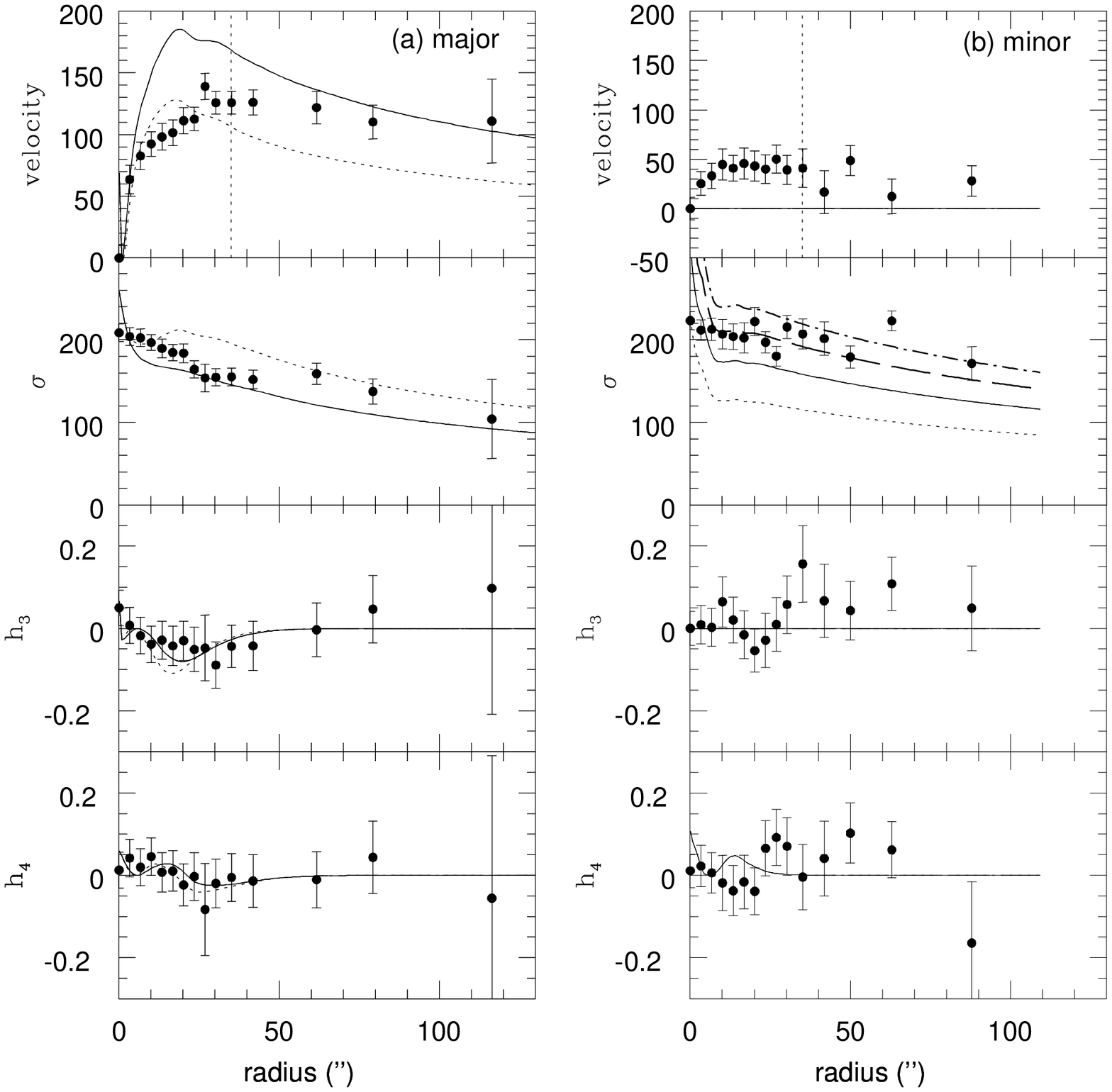}
 \includegraphics{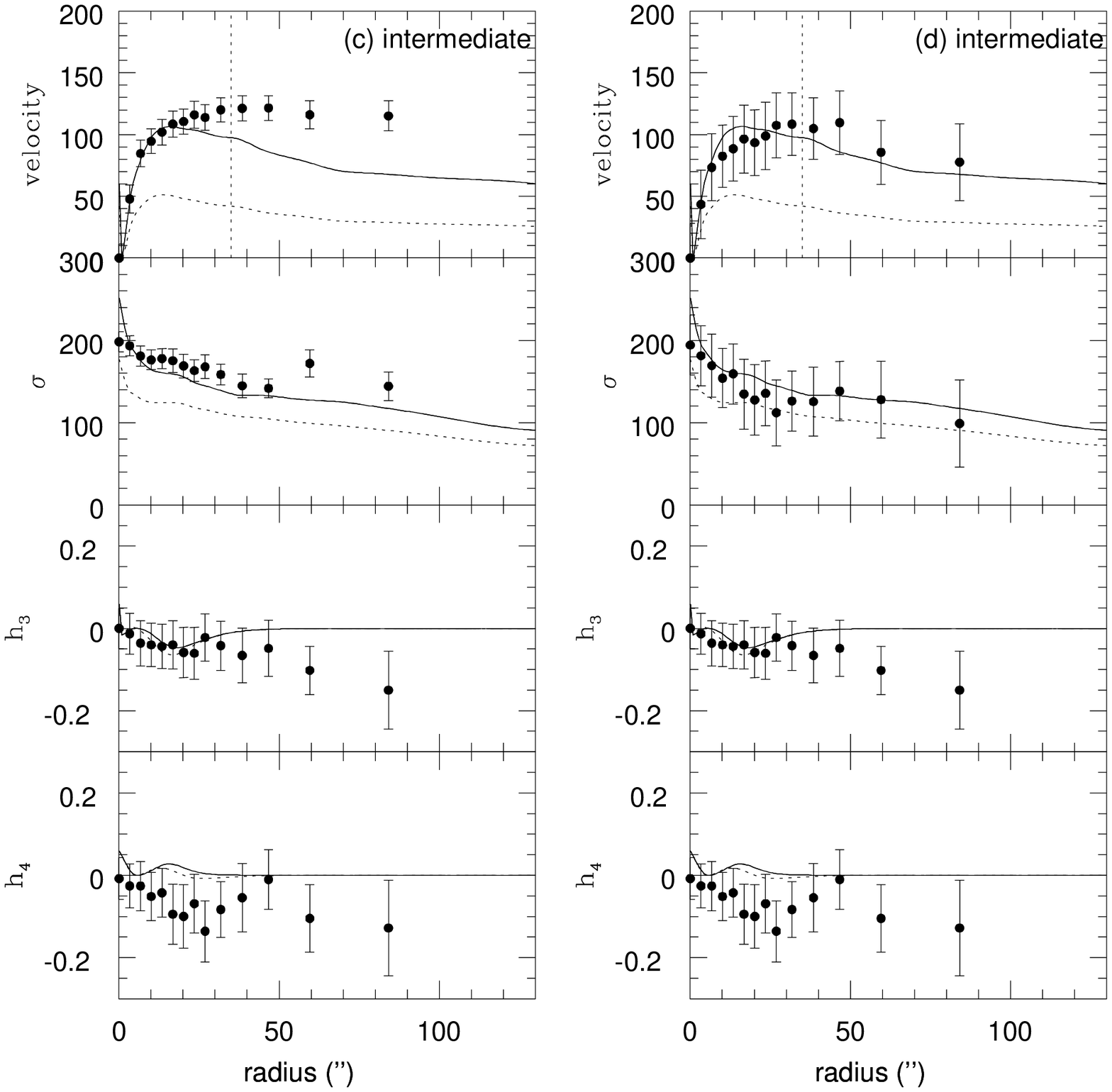}}
\caption{Predictions of the two--integral models for IC~3370. 
{\sl Panel (a)}: major axis predictions.
Dotted lines:  $k=0.6$, $M/L_B=5.42$, no dark matter included,
no embedded disc. 
Solid lines: $k=1$, $M/L_B=5.28$, no dark matter included, no embedded disc.  
{\sl Panel (b)}: minor axis predictions. Dotted line:  $k=0.6$, $M/L_B=4.80$, 
no dark matter included, no embedded disc. 
Solid line: $k=1$, $M/L_B=6.59$, no dark matter included, no embedded disc.  
 Thick    dashed line: $k=1$, $M/L_B=9.68$,  no dark matter, embedded disc. Thick    dot--dashed line: $k=1$, $M/L_B=12.64$,  no dark matter, embedded disc. {\sl Panel (c)}: Intermediate  axis predictions for the {\it uncorrected} values of the observed velocity and velocity dispersion.
Dotted lines:  $k=0.6$, $M/L_B=3.64$, no dark matter included,
no embedded disc. Solid lines: $k=1$, $M/L_B=5.19$, no dark matter included, no embedded disc
included.  
{\sl Panel (d)}: Intermediate  axis predictions for the values corrected for non--zero values of $h_3$ and $h_4$ of the observed velocity and velocity dispersion. The meaning  of the curves is the same as in plot (c). Vertical dotted line indicates one effective radius.
}

\label{fig:IC3370_model}
\end{figure*}

As might be expected the axisymmetric modelling of IC~3370 did not give a good fit to the observed data given the strong isophotal twisting present in this galaxy. We used the inclination angle of $50^\circ$ that gave the best (but far from perfect) results. This inclination angle implies intrinsic axis ratio of $\sim 0.7$.
In Fig. \ref{fig:IC3370_model} we present our modelling results for major, minor and intermediate axis. 

{\sl Major axis (Fig. \ref{fig:IC3370_model} (panel (a)))}: for the major axis $k=0.6$ (dotted lines)  gives a good fit in the inner regions ($\sim 25$ arcsec) for the velocity.  A good fit is  obtained in the outer regions ($> 50$ arcsec) for the velocity dispersion. On the contrary, $k=1$ provides a good fit for the velocity in the outer region ($>1R_e$); the velocity dispersion seems to be fitted well throughout the whole galaxy with $k=1$. We experimented with the inclusion of the inner embedded disc of 6 arcsec, 
but this does not change much the results.
The decrease of the velocity dispersion follows very closely the constant mass--to--light prediction (out to $\sim 3R_e$).  Both $h_3$ and $h_4$ are fitted reasonably in all the given cases. The mass--to--light ratio   found in all the cases at $\sim 3 R_e$ is $\sim 5.4$. Since in this case, for the major axis, both the $h_3$ and $h_4$ parameters are consistent with zero we did not
apply the correction of the  velocity and velocity dispersion as we did in the case of IC~1459. 

{\sl Minor axis (Fig. \ref{fig:IC3370_model} (panel (b)))}: because of the  fact that the axisymmetric modelling predicts zero velocity for the minor axis, a successful fit   could not be achieved (the same is true for the $h_3$ parameter). 
Modelling of the velocity dispersion therefore provided a possibility for several interesting tests.
One can see  that a $k=0.6$ fit (dotted line, $M/L_B=4.80$, no dark matter, no embedded disc) cannot produce a successful fit for the velocity dispersion.
Therefore, in all other tests in the case of the minor axis we used $k=1$.
With the solid line we present the case of $M/L_B=6.59$ without 
the embedded disc which provided a better, but  still unsatisfactory fit to the data. Again, the inner embedded disc of 6 arcsec was included.
Therefore, we  increased the mass--to--light ratio to 9.68 (case without the dark matter and with the disc represented with the thick dashed line) to achieve a better agreement. Still better agreement is obtained when one increases further
the mass--to--light ratio to 12.65: this is the case without the dark matter and the included disc represented by the thick dot--dashed line. Note, however, the discrepancy in the inner parts of the galaxy.  

{\sl Intermediate axis (Fig. \ref{fig:IC3370_model} (panels (c) and (d)))}: Several tests were done using lower values of the constant mass--to--light ratio: the modelling of the {\sl uncorrected} values of the observed velocity and velocity dispersion are given in panel (c) and the modelling of the {\sl corrected} points is given in panel (d).  Successful fits for the velocity dispersion are obtained for $k=1$ (again the $k=0.6$ case can be ruled out). All the models with $k=1$ give a good fit for the velocity in the inner parts of the galaxy ($\sim 20$ arcsec) and they all fail in the outer parts.
In a similar manner they all reproduce well the velocity dispersion profile.
 $h_3$ and $h_4$ parameters are fitted reasonably throughout the whole galaxy (modelled $h_3$ shows departures in the outer region and $h_4$ shows small departures from the data in the inner part). We note the improvement of the modelling when the points corrected for non--zero values of $h_3$ and $h_4$ are used (see Fig. \ref{fig:IC3370_model}, panel d).

Strictly speaking IC~3370 should not be modelled using the axisymmetric modelling technique. However, this technique permits the following  conclusion.
 In IC~3370 up to $\sim 3R_e$ the dark matter halo is not needed for the successful modelling: the  mass--to--light ratio varies between $\sim 5$  (based upon the major axis data) and $\sim 13$ (based upon the minor axis data). Note however, that  $M/L_B\sim 13$ is the   upper limit, 
because it must be stressed that this kind of modelling of the observed minor axis dispersions tends to {\it overestimate} the mass--to--light ratio (as given in BDI): this modelling, for a given  $M/L_B$ underestimates the minor axis dispersions since  the model will be flattened by enhanced $\overline{v_\phi^2}$ which does not contribute to the minor axis profile. The real galaxy is flattened by enhanced $\overline{v_r^2}$
which contributes on the minor axis.

\begin{figure*}
\centering
\resizebox{\hsize}{!}{ 
\includegraphics{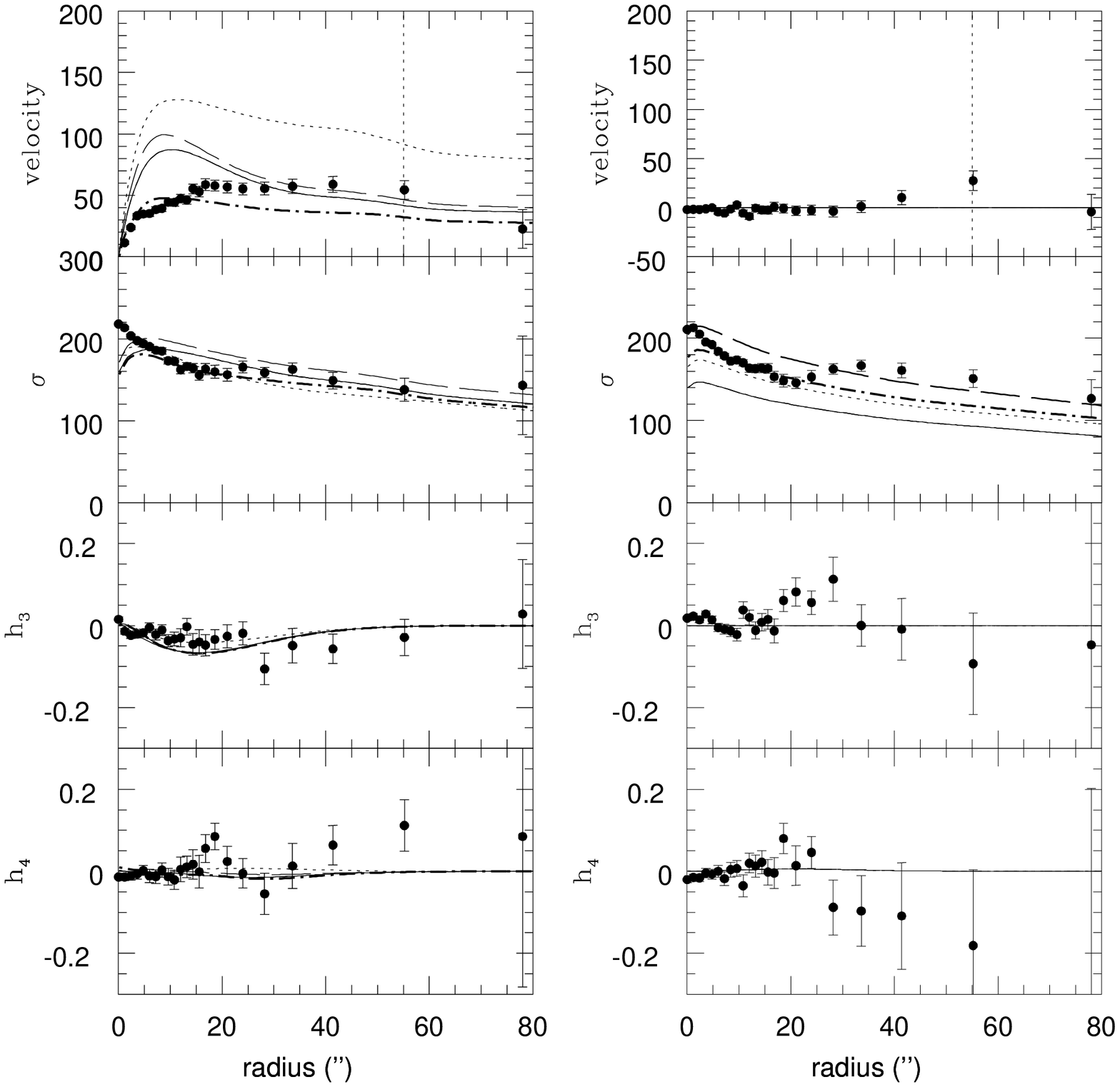}}
\caption{Predictions of the two--integral models for NGC~3379 compared to data. 
{\sl Left}: major  axis predictions.
Dotted lines:  $k=1$, $M/L_B=5.44$, no dark matter included,
no embedded disc. Dashed lines: $k=0.5$, $M/L_B=5.44$, no dark matter included,
without embedded disc. Solid lines: $k=0.5$, $M/L_B=4.75$, no dark matter included, embedded disc
included.  
Thick dot-dashed line: $k=0.4$,  $M/L_B=4.58$ no dark matter included, with embedded disc.
{\sl Right}: minor  axis predictions.
Dotted line:  $k=1$, $M/L_B=4.18$, no dark matter included,
without embedded disc. 
Solid line: $k=0.5$, $M/L_B=4.18$, no dark matter included, embedded disc
included.  
Thick long dashed line: $k=0.5$, $M/L_B=8.91$, no dark matter included, without embedded disc. Thick dot-dashed line: $k=0.5$, $M/L_B=6.69$, no dark matter included, with embedded disc.  Vertical dotted line indicates one effective radius. } 
\label{fig:NGC3379_2I}
\end{figure*}

\subsection{NGC~3379} 

NGC~3379 is a galaxy for which the evidence for dark matter is scarce (Ciardullo et al. 1993, Romanowsky et al. 2003). We present   our results for the two--integral axisymmetric modelling for the major and minor axis in Fig. \ref{fig:NGC3379_2I}. The inclination angle that we used in all the cases was 40$^\circ$ because it gave the best fit to the observed data. 

{\sl Major axis (Fig. \ref{fig:NGC3379_2I} (left))}: When one takes $k=1$ and does not include either a dark matter halo or an internal embedded disc, using   $M/L_B=5.44$ (dotted lines) one gets an exaggerated value of the velocity but a rather reasonable fit for the velocity dispersion
(especially in the inner part).
In all other cases for the major axis we will use $k=0.5$ which provides a better
fit to the data in the outer part of NGC~3379. 
Using a mass--to--light ratio, 
$M/L_B=4.75$, without an embedded disc and without a  dark halo combination gives a good fit for the velocity dispersion in the outer regions.
Also, a case with $M/L_B=5.44$ (dashed line), without the embedded disc and without a  dark halo gives a good fit in the outer part of the galaxy.  
Both $h_3$ and $h_4$ parameters are fitted similarly in all the models and the fit is very close to the observed values. Note that since the outermost points for the velocity dispersion and $h_4$ parameter appear to be discordant, we followed the advice of the referee and put a greatly increased error bar in order to provide a real uncertainty of these quantities.
To get a better fit in the inner regions (interior to $\sim 15$ arcsec) we performed a test using low $k=0.4$, inclination angle of $50^\circ$ and 
 $M/L_B=4.58$: this is shown in Fig. \ref{fig:NGC3379_2I} using a thick dot--dashed line.

{\sl Minor axis (Fig. \ref{fig:NGC3379_2I} (right))}:
Because this galaxy does not show a strong rotation on the minor axis, the velocity was fitted properly in this approach which assumes axisymmetry.
In the case of the minor axis we consider various tests related to the velocity dispersion. 
With a dotted line we present the $k=1$ case with $M/L_B=4.18$ without the dark matter, and without a disc: this does not provide a good fit.
Neither can the case with $k=0.5$ without a dark matter halo and a disc,   with $M/L_B=4.18$ which is represented with a dashed line.
The thick dot-dashed line represents a case for which $M/L_B=6.69$ ($k=0.5$, without dark matter, and with embedded  disc) and which  provides a good fit in the inner region (out to $\sim 1R_e$), but fails in the outer regions. On the contrary, a test made with  $M/L_B=8.91$  and $k=0.5$ (thick long dashed line) provides a good fit in the outer regions (beyond $\sim 1R_e$).

Our conclusion based upon the two--integral modelling that we performed 
is that in NGC~3379 there is no evidence for dark matter out to $\sim 1.46 R_e$ and that
 this galaxy can be fitted with a constant mass--to--light ratio that is 
between $\sim$ 5 and $\sim$ 9. 
The minor axis modelling suggests a mass--to--light increasing with radius, while the major axis does not. This discrepancy could be due to the third--integral effects.
These results are in agreement with the papers  by other authors using entirely different techniques.
Ciardullo et al. (1993)  found that NGC~3379 does not possess a dark halo, and that mass--to--light ratio is $\sim$ 7 (their observations of PNe extend out to $\sim 200$ arcsec). 
Also, they used distance of 10.1 Mpc to NGC~3379; if we apply the value of 13 Mpc used in our calculations the mass--to-light ratio in the $B$--band will become equal to $5.9\pm 0.9$.
Romanowsky et al. (2003) obtained the value $M/L_B=7.1\pm 0.6$ at $\sim$ 200 arcsec taking the distance of 10.3 Mpc; again if we apply the value of 13 Mpc used in our applications we calculate the mass--to-light ratio in the $B$--band
of $5.8 \pm 0.5$. Their results are  similar to ours at smaller radius and taken together they fail to demonstrate the  presence of DM over this range of radius.

\begin{figure*}
\centering
\resizebox{\hsize}{!}{ 
\includegraphics{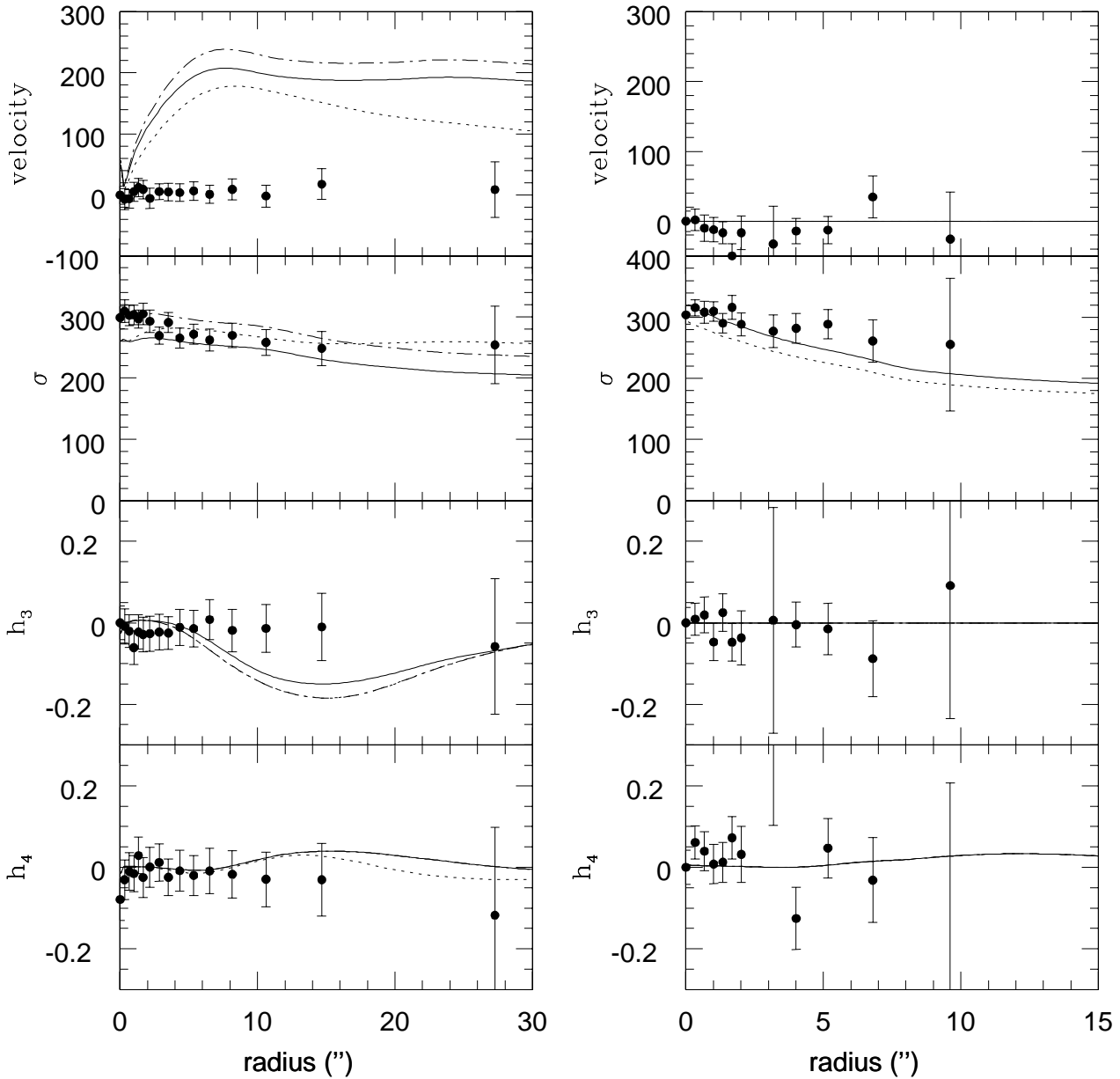}}
%
\caption{Predictions of the two--integral models for NGC4105 compared to data. 
{\sl Left}: major  axis predictions.
Dotted lines:  $k=0.5$, $M/L_B=4.50$, no dark matter included,
with embedded disc.  Solid lines: $k=1.0$, $M/L_B=4.50$, no dark matter included, embedded disc included.  Dot-dashed lines:   $k=1$, $M/L_B=5.94$,  without dark matter, with embedded disc. 
{\sl Right}: minor  axis predictions.
Dotted line:  $k=0.5$, $M/L_B=5.86$, no dark matter included,
without embedded disc. Dashed line: $k=1$, $M/L_B=6.42$, no dark matter included, without embedded disc (overlapped by solid  line). 
Solid line: $k=1$, $M/L_B=6.42$, no dark matter included, embedded disc
included.  
Dashed line: $k=0.5$, $M/L_B=5.86$, without  dark matter, without embedded disc.  Note the difference in scales for two different axes.}
\label{fig:NGC4105_2I}
\end{figure*}

 \subsection{NGC~4105}
We present our results for the two--integral axisymmetric modelling for the major and minor axis in Fig. \ref{fig:NGC4105_2I}. The inclination angle that we used in all the cases was 50$^\circ$ because it gave the best fit to the observed data. 
 
{\sl Major axis  Fig. \ref{fig:NGC4105_2I} (left))}: The case of $k=0.5$  and  $M/L_B=4.50$ (no dark matter halo, no disc included)
 provided the best-fitting to the velocity dispersion for the distance $>$ 2 arcsec (dotted line). However, this case (as well as all the others mentioned below) predicts a grossly excessive velocity. The solid lines show the predictions of the case with $k=1$, $M/L_B=4.50$ (no dark matter halo,   disc included),
 for which the velocity is again exaggerated, and velocity dispersion is lower than observed (although beyond $\sim$ 2 arcsec within the error bars). 
We experimented with the embedded discs of 6 arcsec (radius inside which the P.A. changes): their inclusion did not change much the final results. 
 Finally, with a dot-dashed line we present the case of $k=1$,  $M/L_B=5.94$ (no dark matter halo,   disc included), that predicts an even larger velocity, but  a good fit to the velocity dispersion. The $h_3$ parameter predictions provide a rather good agreement with the observations (apart from the region between 10 and 20 arcsec related to to the large rotational velocity curve which is not seen in the data and is probably related to the problem of the existence of the third integral).
The $h_4$ parameter is fitted properly in all the cases.
A good fit for the velocity could not be obtained; various mass distributions could not solve the problem of the fit to this quantity.
 
{Minor axis Fig. \ref{fig:NGC4105_2I} (right))}: The case of $k=0.5$  and  $M/L_B=5.86$ (no dark matter halo, no disc included) did not provide a successful fit to the  
velocity dispersion for a radius $<$ 10 arcsec (dotted line).
The solid line is for the case of	$k=1$ and  $M/L_B=6.42$
(without dark matter, and without a disc) and this represents the best
fit in all cases. Since $h_3$ and $h_4$ do not show large departures from zero, they are fitted properly.

Our conclusion for NGC~4105 is that this galaxy should be modelled using a three--integral approach because the rotation and velocity dispersion cannot be fitted simultaneously (the modelling results are very similar to these obtained in the case of IC~1459 above, and NGC~720 from BDI). However, judging by two--integral modelling one can see that the dark matter is not needed (out to $\sim$ 1 $R_e$) and that a successful fit (only for dispersion, for reasons given above) can be obtained for a constant mass to light ratio   $M/L_B\sim 6$.

\subsection{Schwarzschild modelling}

Because of numerous problems with fitting our 4 galaxies using 2I technique
we have also performed several tests using the three--integral  Schwarzschild (1979) orbit superposition modelling which included spherical, axisymmetric and triaxial potentials. 
Our procedures were based on the Rix et al. (1997) paper. We have implemented
a method for extraction of the velocity profiles from the orbit libraries
basd on the so--called self--organizing maps (SOMs) (Kohonen 1997; Murtagh 1995)
and the results obtained so far are only indicative: we found that the constant mass--to--light ratio potentials could provide satisfactory fits to the observed data, but more testing is needed. When performing the orbits superposition in the triaxial potential special attention must be given to selection of the representative orbit libraries. Our results so far did not differ much when we used the libraries which contained exclusively box or tube orbits.
Nevertheless, it is difficult to justify {\it a priori\/} a given selection of orbits. Development  
of this methodology is proceeding.

\section{X--ray properties of  IC~1459, NGC~3379 and NGC~4105}

Of the four galaxies presented in this paper, IC~1459, NGC~3379 and NGC~4105  are known to possess an X--ray halo.

X--rays are important for the early--type galaxies because they can provide independent constraints on the masses and mass--to--light ratios  
out to large radii  (for a review see Danziger 1997;
Mathews \& Brighenti 2003).
The basic assumptions and formulae are as follows. 
One assumes that   spherical symmetry  holds, and that the condition of hydrostatic equilibrium  is valid:
\begin{equation}
{{\rm d}P_{\rm gas}\over {\rm d}r}= -{GM(r)\rho_{\rm gas}\over r^2},
\label{eqn:DP1}
\end{equation}
where $M(r)$ is the mass interior   to the radius $r$,
and the gas obeys the perfect gas law:
\begin{equation}
P_{\rm gas}={\rho_{\rm gas} kT_{\rm gas}\over \mu m_H},
\end{equation}
where $\mu$ is the mean molecular weight for full ionization (taken to be 0.61), and
$m_H$ is the  mass of the hydrogen atom. 
From these two equations one can give the expression for the gravitating mass 
interior to radius $r$:
\begin{equation}
M(r) =-{kT_{\rm gas} r\over G\mu m_p} \left ( {{\rm d}\ln \rho\over {\rm d}\ln r} +
{{\rm d}\ln T_{\rm gas}\over {\rm d}\ln r} \right ). 
\end{equation}

If one wants to calculate the mass and mass--to--light ratio of an elliptical
galaxy based upon X--ray observations one can use the following approach
(which was used in Kim \& Fabbiano (1995), hereafter KF95, for NGC~507 \& NGC~499):
one assumes circular symmetry and derives a radial profile of the X--ray surface brightness measured in concentric rings centered on the X--ray centroid. In a given range one then fits the analytic King approximation model:
\begin{equation}
\Sigma_{\rm X}\sim \left [ 1+\left ({r\over a}\right )^2\right ]^{-3\beta +0.5}
\end{equation}
(for details see KF95). Here $a$ is the core radius (the radius where the surface brightness falls to half of its central value), and slope $\beta$. 
If the temperature of the X--ray emitting gas does not change much as a function of radius one can assume isothermality; we assumed that this holds in all the cases below. Now, using this assumption, 
one can estimate the total
gravitational mass at a given radius $r$ (assuming hydrostatic equilibrium)
in a convenient form (KF95):
\begin{equation}
M_T=1.8\times 10^{12}(3\beta +\alpha)
\left ( {T\over {\rm 1 keV}} \right ) 
\left ( {r\over 10^3{}^{\prime\prime} }\right )
\left ({d\over  {\rm 10 Mpc}}\right ) \Msun,
\label{eqn:TOT1}
\end{equation} 
here the exponent $\alpha$ is related to the temperature ($T\sim r^{-\alpha}$)
and is taken to be zero, and $\beta = 0.5$  (for IC~1459 and NGC~4105) 
and $\beta = 0.64$ (for NGC~3379) (Brown \& Bregman 2001). This formula is valid outside the core region. 

The mass--to--light ratio (in the $B$--band) can be expressed as a function of radius $r$:
\begin{equation}
{M_T\over L_B}=1.16\times 10^{-2}
10^{B\over 2.5} (3\beta +\alpha)
\left ( {T\over {\rm 1 keV}} \right ) 
\left ( {r\over 10^3{}^{\prime\prime} }\right )
\left ({d\over  {\rm 10 Mpc}}\right )^{-1} ,
\label{eqn:ML1}
\end{equation}
where $B$ is the  B magnitude of galaxy inside radius $r$ (Kim \& Fabbiano 1995).

The temperature of the X--ray halo can be estimated using a simple formula which 
connects the stellar velocity dispersion, $\sigma$ and the virial temperature (see Mathews \& Brighenti 2003):
\begin{equation}
T_{\rm vir}\approx
\mu m_p \sigma^2 /k \sim 10^7~{\rm K} \sim 1 \ {\rm keV},
\label{eqn:tvir}
\end{equation}
where $\mu$ is the mean molecular weight.  After insertion of the  value for $\mu = 0.61$ this equation can be rewritten in a convenient form:
\begin{equation}
kT_{\sigma {[\rm keV]}} = 6.367 \times 10^{-6} \sigma_{\rm [km /s]} ^2
\label{eqn:tsigma}
\end{equation}
where the temperature is given in keV and the  stellar velocity dispersion, $\sigma$,  is given in ${\rm km \ s^{-1}}$.

\begin{figure}
\centering
\resizebox{\hsize}{!}{ 
\includegraphics{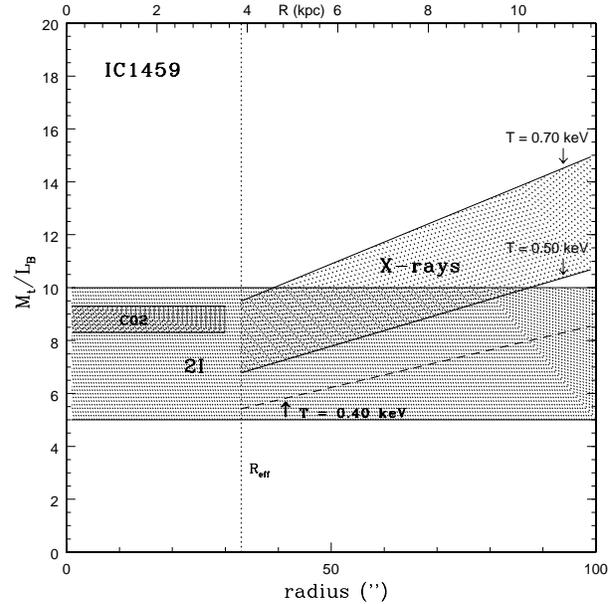}} 
\caption{Cumulative mass--to--light ratio of 
IC~1459 in the $B$--band as a function of radius.
The scale of the lower x--axis is given in arc seconds and the upper x--axis is  in kiloparsecs.
The effective radius is plotted as a dashed vertical line.
Limits on the mass based on the X--rays are given using lower and upper limits obtained using
Davis \& White (1996): $T=0.60\pm 0.10$ keV.
``2I'' refers to two--integral modelling:  the stripe in the 2I case provides the limits within which the kinematics can be fitted (see text for details).
One additional line (see text for details)  was also added  to the plot: dashed line is for the case for which $T=$0.4 keV. The stripe labelled with C02 corresponds to the estimate obtained by Cappellari et al. (2002) 
of the $M/L$ ratio which is, after the conversion to the $B$--band and the distance used in our paper, $M/L_B=8.8\pm 0.5$.}
\label{fig:IC1459_xray}
\end{figure}

\begin{figure}
\centering
\resizebox{\hsize}{!}{ 
\includegraphics{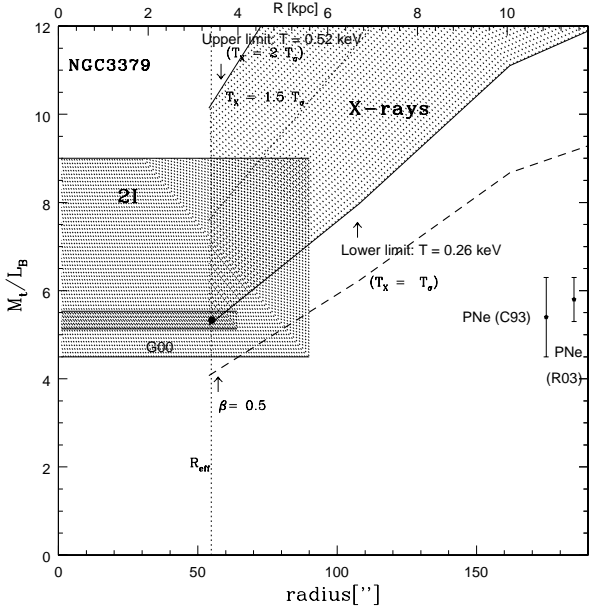}} 
\caption{Cumulative mass--to--light ratio of 
NGC~3379 in the $B$--band as a function of radius.
The meaning of the symbols is the same as in Fig. \ref{fig:IC1459_xray}
Two additional points are given in the plot: R03 refers to Romanowsky et al. (2003), C93 to Ciardullo et al. (1993).  
Two upper limits are given: one for $T_{\rm X}=1.5T_\sigma$ represented by a thick dotted line, and one for which 
 $T_{\rm X}=2T_\sigma$ as a solid line.
 The dashed line is for the case for which $T_{\rm X}=0.26$ keV and $\beta=0.5$.  The point labelled  G00 and corresponding stripe are based on the result from the 3I modelling of Gebhardt et al. (2000); it has been obtained for the $B$--band and at the distance of 13 Mpc. 
}
\label{fig:NGC3379_xray}
\end{figure}

\begin{figure}
\centering
\resizebox{\hsize}{!}{ 
\includegraphics{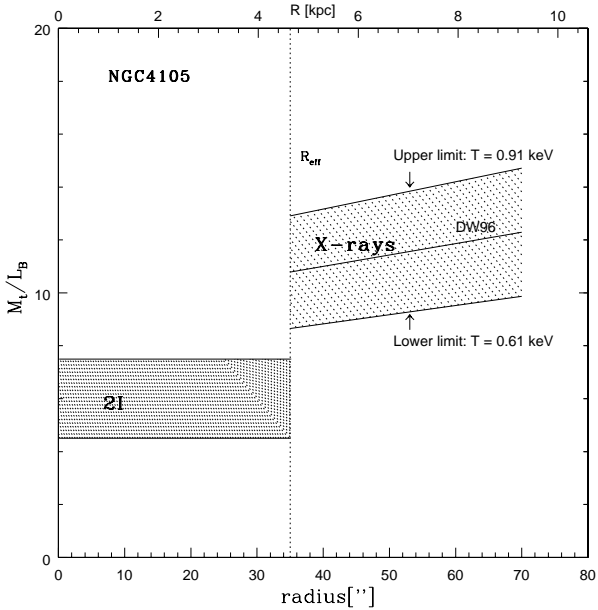}} 
\caption{The same as in Fig. \ref{fig:NGC3379_xray} but for NGC~4105.
DW96 refers to the temperature taken from the paper by Davis \& White (1996).
}
\label{fig:NGC4105_xray}
\end{figure}

In Figs. \ref{fig:IC1459_xray},  \ref{fig:NGC3379_xray} and \ref{fig:NGC4105_xray}  we sum graphically the results of the 2I modelling of IC~1459, NGC~3379 and NGC~4105 and estimates based on the X--rays calculations (beyond one effective radius; this limit is taken because we want to avoid the problems due to cooling flows in the central region and because we are interested in the comparison of different methodologies in the outer regions of the galaxies where dark matter is expected to play a significant role).
2I  modelling is represented with a
stripe which roughly indicates the uncertainty within which the kinematics of these galaxies can be fitted (see discussion on 2I modelling above).
The range of values $M/L_B=5-10$ encompasses the full range of plausible possibilities shown in Fig. \ref{fig:IC1459_model}.

In the case of IC~1459 (Fig. \ref{fig:IC1459_xray}), 
the stripe related to the X--rays is determined using the paper by Davis \& White (1996) who found $T=0.60^{+0.12}_{-0.13}$. This is in agreement with Fabbiano et al. (2003)  who estimated $T=0.5$--$0.6$ keV.
We have also added a line which corresponds to $T=0.4$ keV and which provides the best fit to the 2I modelling. Using virial assumption (Eq. \ref{eqn:tsigma}) one gets $T_\sigma = 0.73$ keV (see Fig. \ref{fig:1459_nodm1_dm1_corr}). 
Only in the   case
of $T=0.50-0.60$ keV do we have a marginal agreement (in region $1.0 < r < 2 R_e$) with results obtained using 2I modelling techniques.

This may mean that:

(i) the 2I integral models systematically underestimate the mass--to--light ratio value. This is possible but not probable since the estimated values of the mass--to--light ratio ($M/L_B$ was found to be between $\sim 6$ and $\sim 10$,
see the stripe in Fig. \ref{fig:IC1459_xray}) are in a good agreement with the mean value found in the sample of van der Marel (1991) which after rescaling to the Hubble constant of 70 ${\rm km \ s^{-1} \ Mpc^{-1}}$ becomes $M/L_B=8.33\pm 0.35$. This type of modelling precludes the possibility that with an observed decreased velocity dispersion one could have a dark matter halo with a consequent increase of a mass--to--light ratio;

(ii) the temperature estimates used above  are not correct -- note, however, that if we adopt a lower temperature, e.g. $T\sim 0.5$ keV (still allowed by the measurements of Fabbiano et al. 2003 and Davis \& White 1996)  a better agreement can be obtained (dotted line in Fig. \ref{fig:IC1459_xray}); the strong rising trend of the mass--to--light ratio will persist making this new estimate again  larger in the outer part of the galaxy. Only $T\sim 0.4$ keV  (as calculated in the paper by Brown \& Bregman 1998) (dashed line in Fig. \ref{fig:IC1459_xray}) would provide an agreement within the whole region (beyond $\sim \ 1 R_e$) for which we have the 2I mass--to--light ratio estimate; 

(iii) the  assumption of the hydrostatic equilibrium for IC~1459 is not correct, so the usage of Eq. (\ref{eqn:DP1}) is inappropriate which makes Eq. (\ref{eqn:ML1}) inapplicable in this case. We will address the problem of the departures from the hydrostatic equilibrium in a forthcoming paper (Samurovi\'c \& Danziger 2005, in preparation) and we here  refer a reader to  a very recent important work by Ciotti \& Pellegrini (2004) which   investigates the effects of deviations from equilibrium on the mass of ellipticals.

In Figures  \ref{fig:NGC3379_xray} and \ref{fig:NGC4105_xray}   we display graphically the results of the 2I modelling  and estimates based on the X-rays calculations (beyond one effective radius) for the mass--to--light ratio in the $B$--band of NGC~3379 and NGC~4105.  
The X-ray estimate for the mass--to--light ratio in the $B$--band is again given as a shaded area 
with the lower limit given as $T_\sigma$   and the upper limit
 $T_{\rm X}=1.5 T_\sigma$, because Brown \& Bregman 
(1998) for objects with few counts assumed this latter limit. 
For several galaxies  Brown \& Bregman 
(1998) found that $T_{\rm X} \approx 2 T_\sigma$: this relation as our choice was used only in the case of NGC~3379   thus giving  an additional corresponding upper limit.
The case for which  $T_{\rm X}=1.5 T_\sigma$ is also plotted as a thick dotted line in Fig. \ref{fig:NGC3379_xray}.
We have chosen to look at the temperature as virial because of the weak signal from X--rays.
In the case of NGC~3379 we  plotted available results on the mass--to--light ratio that were obtained using PNe (Ciardullo et al. 1993 (C93), and Romanowsky et al. 2003 (R03)).
Here we note a caveat that the X--ray and PN models are merely spherical.

Our conclusions regarding comparison of the results for the mass--to--light ratio obtained using 2I modelling techniques and X-ray approach are the following:

{\it {NGC~3379} (Fig. (\ref{fig:NGC3379_xray}))}: the estimates from 2I modelling are in an  agreement with the X-ray estimates. 
Our $M/L$ ratios are similar to those obtained for PNe albeit at larger radii.
We note that beyond 120 arcsec ($\sim 2.2 R_e$) a discrepancy between PNe estimates and X-ray estimate occurs.
If we reduce the value of the $\beta$ parameter from 0.64 to 0.5 we obtain a good agreement with the data between $\sim 1R_e$ and $\sim 1.5R_e$;
between $\sim 1.5$ and $\sim 3.5 R_e$  the mass--to-light ratio in the $B$--band varies between 5.5 and 9.5.
In Fig. (\ref{fig:NGC3379_xray}) we also plotted a point (at $1R_e$) and a stripe based on the paper by Gebhardt et al. (2000) which is calculated using their 3I modelling procedure.

{\it NGC~4105 (Fig. (\ref{fig:NGC4105_xray}))}: 
since in the case of this galaxy our data extend only out to $\sim 1R_e$ the comparison of the X--ray predictions with the results of the 2I modelling  
interior to this point could not be done. It is however obvious that  if one assumes that the constant mass--to--light ratio inferred from the 2I modelling is valid beyond 1 $R_e$ there is a clear discrepancy between the predictions of the two methodologies: the value of the mass-to--light ratio from the X--rays is higher (at $\sim 2R_e$ $M/L_B\sim 12$ ) than that obtained using the 2I modelling (assumed to be between $\sim 5$ and $\sim 7$).
We did not have data related to other mass tracers, so further comparisons were not possible.

Using these findings, together with the recent result of Peng et al. (2004) who have recently found that for NGC~5128 at 15 $R_e$  the mass--to--light ratio is only 13 (in the $B$--band), we can infer that for both NGC~3379 and NGC~5128 there is a discrepancy between the mass--to--light ratios calculated using X-rays and  PNe techniques at the large radii (beyond 3$R_e$) from the centre.
For the sake of comparison of the X--ray and PNe we use the recent analysis of
Kraft et al. (2003) who studied NGC~5128: using their X--ray data out to 15 kpc ($\sim 3R_e$) they found that the total mass $\sim 2\times 10^{11} M_\odot$. This is in very good agreement with the value obtained using data, technique and assumptions by  Peng et al. (2004) (for $\beta_*=0$).
At the larger distance ($\sim 15R_e$) using this approach the total cumulative mass becomes equal to $\sim 5\times 10^{11}M_\odot$ (Peng et al. 2004).

Accepting now to a first order approximation that the X--ray results discussed above are realistic we solve the Jeans equation which provides the connection between the anisotropy and the temperature of the hot interstellar 
gas through which the stars move. 
First we write the Jeans equation for the radial stellar velocity 
dispersion $\sigma_r$ (Binney \& Tremaine 1987 Sec 4.2):
\begin{equation}
{1 \over \ell_*}{{\rm d}(\ell_* \sigma_r^2) \over {\rm d}r} 
+ {2 \beta_* \sigma_r^2 \over r} = - {G M(r)\over r^2}
\label{eqn:J1}
\end{equation}
where $\ell_*$ is the stellar luminosity density 
which corresponds to the radial ($\sigma_r$)
and transverse ($\sigma_\theta$) stellar velocity dispersions.
Spherical symmetry is assumed and the equation is valid for a non--rotating system. 
By introducing a parameter $\beta_*$ 
\footnote{Note that here we will use $\beta_*$ instead of the usual $\beta$ symbol to avoid confusion with the previously defined  parameter which is related to the slope  used in the analytic King approximation model (equation \ref{eqn:TOT1}).}
one can express the nonspherical nature of  the stellar velocity dispersion:
\begin{equation}
\beta_* = 1 - \overline{v_\theta^2} / \sigma_r^2
\label{eqn:B1}
\end{equation}
where $\overline{v_\theta^2}=\overline{v_\theta}^2+\sigma_\theta^2$.
If $0 < \beta_* < 1$ the orbits are predominantly radial, which means
that the line of sight velocity profile becomes 
more strongly peaked than a Gaussian profile ($h_4$ positive), and
for  $-\infty \le \beta_* < 0$ the orbits are mostly 
tangential, which means that
the profile is more flat--topped than a Gaussian ($h_4$ negative) (Gerhard 1993, van der Marel \& Franx 1993).
The $\beta_*$ parameter can be determined from the observations: we used the estimates of the $h_4$ parameter to calculate $\beta_*$ (as given in Gerhard 1993 and van der Marel \& Franx 1993)
in the case of IC~1459.
In our study we use the estimates by Kronawitter et al. (2000) for NGC~3379 obtained using 2I modelling including $h_4$ fitting.
For NGC~4105 we will assume that $\beta_* =0$ (spherical isotropic case)
given the small observed $h_4$ parameter consistent with zero throughout the galaxy (but, nevertheless, we will also test some anisotropic models).
We solve the Jeans equation (\ref{eqn:J1}) using equation (\ref{eqn:TOT1}) to express the total mass of a given galaxy at the position $r$. For the stellar luminosity density we adopt the Hernquist (1990) profile:
\begin{equation}
\ell_* = {L\over 2\pi}{a\over r} \left ({1\over r +a}\right )^3
\label{eqn:H1}
\end{equation}
where $R_e=1.8153 a$.
The  projected line--of--sight velocity dispersion is calculated as (e.g. Binney \& Mamon 1982; Mathews \& Brighenti 2003):
\begin{equation}
\sigma^2(R) = { \int_R^{r_t}  \sigma_r^2(r)
\left[ 1 - (R/r)^2 \beta_* \right] 
\ell_*(r) (r^2 - R^2)^{-1/2} r {\rm d}r 
\over
\int_R^{r_t} \ell_*(r) (r^2 - R^2)^{-1/2} r {\rm d}r }
\label{eqn:SIG1}
\end{equation}
where the truncation radius, $r_t$,  extends well beyond the last observed kinematical point.
We took for NGC~3379 $r_t = 2R_e$,  for NGC~4105  $r_t = 1.5R_e$ and for IC~1459 $r_t = 4R_e$. Solving  equation (\ref{eqn:SIG1}) using a given  value of $T$ (known from X-ray observations, or calculated using stellar velocity dispersion) one can find the  $\beta_*$ parameter which provides the best agreement with the observed data (it is important to underline,
in the case of the hydrostatic equilibrium). 

 As a check we tested the well known case of NGC~4472 studied by Mathews \& Brighenti (2003). We confirm that indeed only with $\beta_* \sim 0.7$ can one obtain a good agreement with the observed velocity dispersion for this galaxy. As noted recently by Ciotti \& Pellegrini (2004) this value of the $\beta_*$ parameter is unrealistically high which may lead to the important conclusion that the assumed condition  of the hydrostatic equilibrium is not valid.

Using the aforementioned approach we reached the following conclusions for the three early-type galaxies with the X--ray haloes given in this paper:

\begin{figure}
\centering
\resizebox{\hsize}{!}{ 
\includegraphics{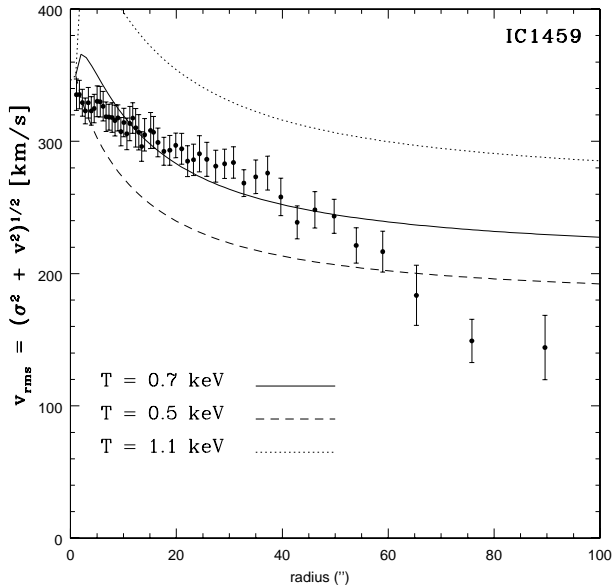}} 
\caption{Dynamical modelling of the {\it uncorrected} velocity dispersion of IC~1459 using X-ray data.  
The modelling lines based on different masses which correspond to different temperatures are the following: the case of $T=0.7$ keV is given with a solid line, the case of $T=0.5$ keV is given with a dashed line and the case of $T=1.1$ keV is given with a dotted line. In all cases spherical isotropy ($\beta_*=0$) and hydrostatic equilibrium are assumed. 
}
\label{fig:1459_nodm1_dm1}
\end{figure}

\begin{figure}
\centering
\resizebox{\hsize}{!}{ 
\includegraphics{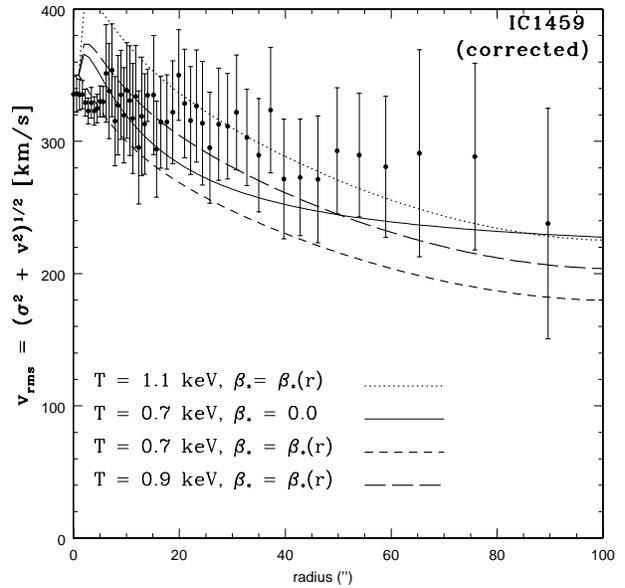}} 
\caption{Dynamical modelling of the velocity dispersion corrected for non--zero values of the $h_4$ parameter of IC~1459 using X-ray data.  
The modelling lines based on different masses which correspond to different temperatures are the following: the case of $T=0.7$ keV is again given with a  solid line ($\beta_*=0$) and a short dashed line  is for the case of $T=0.7$ keV ($\beta_*=\beta_*(r)$). The case of $T=0.9$ keV and $\beta=\beta_*(r)$ is given with a long dashed line and the case of $T=1.1$ keV and $\beta=\beta_*(r)$ is given with a dotted line. Again, in all cases  hydrostatic equilibrium is assumed. 
}
\label{fig:1459_nodm1_dm1_corr}
\end{figure}

(i) IC~1459 (Fig. (\ref{fig:1459_nodm1_dm1})). First, we tested a spherical isotropic model ($\beta_*=0$) using different estimates for the temperature: $T=0.7$ keV (solid line), 
$T=0.5$ keV (dashed line) and $T=1.1$ keV (dotted line). In all three cases spherical isotropy ($\beta_*=0$) and validity of hydrostatic equilibrium were assumed and the values of the velocity dispersion were uncorrected (in the sense of Eq. \ref{eqn:sig_corr}).
Only the case for which $T=0.7$ keV can provide a good fit to the observed data out to $\sim 50$ arcsec. 
In the outer regions the {\it observed} velocity dispersion tends to decrease and none of the fits is successful. This best--fitting value of $T=0.7$ keV 
is, as expected, in an excellent agreement with the value found using a virial assumption (see  Eq. \ref{eqn:tsigma}). This value, however,
is somewhat higher than our preferred value of $T=0.4-0.6$ keV found with the 2I modelling, thus implying a higher X--ray based mass--to--light ratio than one inferred from stellar dynamics (as shown in Fig. \ref{fig:IC1459_xray}).

Second, since for IC~1459 (Fig. \ref{fig:1459_nodm1_dm1_corr}) we have departures from zero of the $h_4$ which imply that radial orbits dominate ($h_4> 0$ beyond $\sim 20$ arcsec) we have also modelled the corrected (again, in the sense of Eq. \ref{eqn:sig_corr}) values of the velocity dispersion: the measured points are the same as in Fig. \ref{fig:IC1459_model} (right panel). The fit which uses $T=0.7$ keV and $\beta_*=0$ (the same as in Fig. \ref{fig:1459_nodm1_dm1} repeated here again with solid line) can  produce a reasonable agreement with the data.
If we now allow anisotropies ($\beta_*=\beta(r)$) based on our observed $h_4(r)$ values as given above we can see that $T=0.7$ keV, except for the central parts
(interior to $\sim 20$ arcsec)
this fit given with the short dashed line fails to fit the corrected velocity dispersion. We assume that $h_4 \sim 0$ $(\beta_*\sim 0)$ in the inner regions of the galaxy and in the outer parts at $\sim 3 R_e$ $h_4$ is $\sim 0.2$ $(\beta \sim 1$). A slightly higher value of the temperature of $T=0.9$ KeV, taken together with the same anisotropies ($\beta_*=\beta(r)$) provides a better fit especially in the intermediate region (between $\sim 10$ and $\sim 50$ arcsec).
Finally, a fit which used $T=1.1$ KeV and $\beta_*=\beta(r)$ provides the best fit (among the ones tested) in the region between $\sim 20$ and $\sim 100$ arcsec.
Again, in the case of the modelling of the corrected velocity dispersion
data the value of the temperature based on the X--rays is higher than that inferred from stellar dynamics ($T\sim 0.4-0.5$ keV): the preferred value of $T=1.1$ keV is higher than that in the uncorrected case ($T=0.7$ keV).
The reason for this discrepancy 
could be due to the inherent assumptions of the two methodologies (such as the validity of hydrostatic equilibrium and constant value of the temperature in the case of the X--rays and mass profile used in the case of the BDI approach; see also the discussion about the mass--to--light ratio of IC~1459 earlier in this Section). It is important to stress that before we can use the X--ray temperatures to model the velocity dispersion we need to understand the reasons for the differences which might be due to different types of mass profiles and/or problems with the BDI models.

\begin{figure}
\centering
\resizebox{\hsize}{!}{ 
\includegraphics{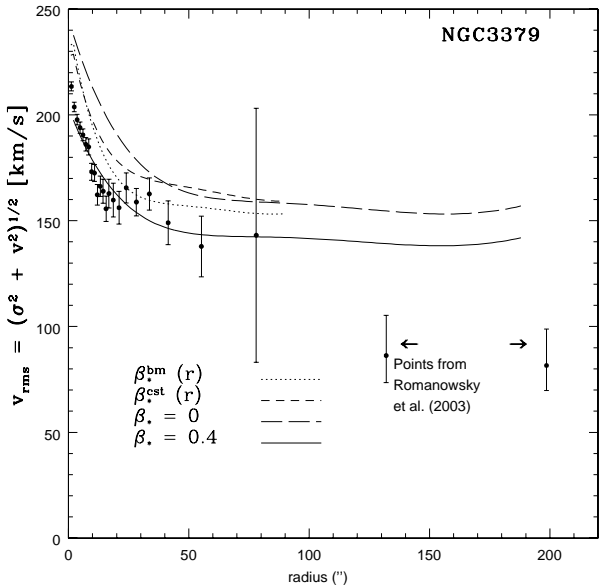}} 
\caption{Dynamical modelling of the velocity dispersion of NGC~3379 using X-ray data.  Observational points are folded about $y$-axis and are taken from Statler \& Smecker-Hane (1999).
Long dashed line is the
spherical isotropic case ($\beta_*=0$),
 when hydrostatic equilibrium is assumed.
Thin solid line is for the case when hydrostatic equilibrium is assumed for which radial orbits dominate ($\beta_*=0.4$). 
The short dashed line is based on the estimate of the $\beta_*$-parameter
obtained for the constant  mass--to--light ratio model ("cst" model) of Kronawitter et al. (2000) and the dotted line is based on the estimate of the $\beta_*$-parameter obtained for the "best model" ("bm") of Kronawitter et al. (2000): they extend only interior to $\sim 100$ arcsec. Two outermost points given as filled circles are taken from Romanowsky et al. (2003).
}
\label{fig:3379_nodm1_dm1}
\end{figure}

(ii) NGC~3379 (Fig. (\ref{fig:3379_nodm1_dm1})). 
In our modelling we have used the temperature $T=0.26$ keV, as given in Table \ref{tab:table_1}.
A spherical isotropic model ($\beta_*=0$) is excluded: in this case the model is completely inconsistent with the observed data. Two models based on the paper by Kronawitter et al. (2000) do not provide a good fit to the data either:   the case of the "best model" (dark matter implied: $0<\beta_*<0.3$) provides marginally better fit than that of the constant $M/L$ ratio   ($0<\beta_*<0.2$). After a failure
to obtain a successful fit using $\beta_*$ values taken from the literature we experimented with different (positive and constant) values of $\beta_*$: the case of $\beta_*=0.4$ provides a very good fit to the data.
Thus we can conclude that using the approach based on the X-rays (two--integral approach) we can say that for NGC~3379   the dark matter is not dynamically dominant to $\sim 1.5R_e$ because at $\sim 1.5R_e$ one can get the same cumulative mass--to--light ratio as it is obtained using stellar dynamical modelling.
Note that the gradient of the mass seen in the X--rays is not consistent with the constant mass--to--light ratio.  
However, as in Mathews \& Brighenti (2003) the value of the $\beta$ parameter is larger than that obtained using stellar dynamics although not much. Note that in our 2I dynamical modelling we have used the $h_4$ parameter to describe the anisotropies and not $\beta_*$.
We find that the positive value of the $\beta_*$ parameter ($\beta\sim 0.4$) is in agreement with
the positive values of the $h_4$ parameters ($h_4\sim 0.05$) extracted from the observations and indicating radial anisotropies in the outer parts of NGC~3379
(for the relations between $h_4$ and $\beta$ see van der Marel \& Franx 1993, Gerhard 1993).

\begin{figure}
\centering
\resizebox{\hsize}{!}{ 
\includegraphics{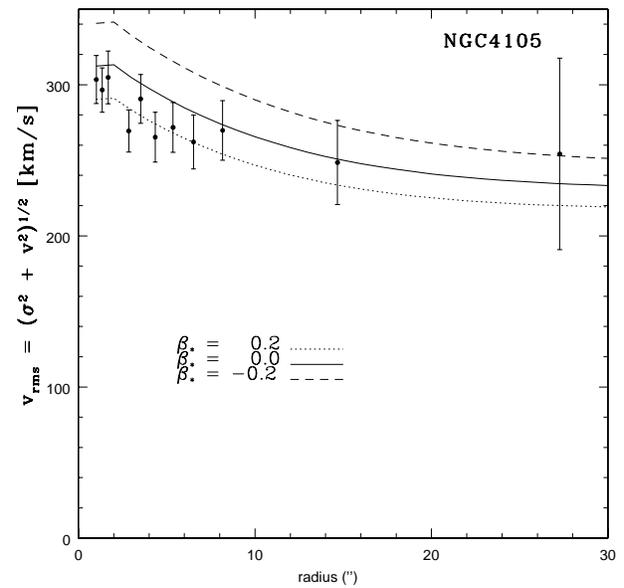}} 
\caption{Dynamical modelling of the velocity dispersion NGC~4105 using X-ray data. Solid line is for the spherical isotropic case ($\beta_*=0$),
the dashed line is for the case of $\beta_*=-0.2$ (predominantly tangential orbits)
and the dotted line is for the case of $\beta_*=0.2$ (predominantly radial orbits).}
\label{fig:4105_beta0_1}
\end{figure}

(iii) NGC~4105 (Fig. \ref{fig:4105_beta0_1}). 
In the modelling which we performed we have used the temperature $T=0.76$ keV, as given in Table \ref{tab:table_1}.
A spherical isotropic model ($\beta_*=0$) provides a good fit to the observed data (note that we again neglected possible influence of NGC~4106). We have also tested a case with $\beta_*=0.2$ (predominantly radial orbits) which provides a reasonable fit to the observed velocity dispersion and a case with ($\beta_*=-0.2$) (predominantly tangential orbits) which does not provide a good fit to the data. 
Note that in the case of NGC~4105 the error bars for the temperature of the X-ray halo  given by Davis \& White (1996) are huge (especially the upper one): $T_X = 0.76 ^{+4.40}_{-0.69}$ keV. However, as can be seen in  Fig. \ref{fig:NGC4105_xray}
the value $T_X=0.76$ keV falls just in the middle between the upper and lower limit as obtained from the stellar velocity dispersion which justifies its usage in the dynamical modelling which we performed. As a side note, we stress that with the upper limit of $T_X=0.76 + 4.40 = 5.16$ keV no fit could be obtained to the velocity dispersion even in the case when the maximum allowed value of $\beta_* =1$ was used: velocity dispersion values were always grossly exaggerated: $\sigma > 450$ \kms \ throughout the whole galaxy.

\section{Discussion and conclusions}   
   
We presented the long--slit spectra of four early--type galaxies (IC~1459, IC~3370,
NGC~3379 and NGC~4105) that extend out to $\sim 1-3 R_e$ for which we extracted full LOSVDs. For NGC~3379 we had the observations out to $\sim 0.55R_e$ so for the modelling purposes we have used the data from the literature.
The presented photometry of these four galaxies was necessary for the dynamical modelling which we performed. In this paper we have made two--integral (2I) modelling of these four galaxies and compared the results with the those obtained using the spherical X--ray modelling (in the case of IC~1459, NGC~3379 and NGC~4105). 

(i) IC~1459 has a counter--rotating stellar core and its $h_4$ has in the outer parts values that are significantly different from zero implying the existence of radial anisotropies. We could not obtain the simultaneous fit to the velocity and velocity dispersion (for the major axis) which means that we are faced with a situation where the 2I approach is not completely satisfactory and that the motion of stars in this galaxy requires three integrals. However, study of the 2I modelling of the velocity dispersion  permits us to draw conclusions about the constant mass--to--light ratio in this galaxy: we infer that in the $B$--band it is 
$5 < M/L_B < 10$ (in solar units, for the Hubble constant of 
70 ${\rm km\ s^{-1}  Mpc^{-1}}$). Because of the fact that the velocity dispersion shows a decreasing trend compatible with a given constant mass--to--light ratio we conclude that up to $\sim 3 R_e$ ($\sim 100$ arcsec $\approx 11.7$ kpc) there is no need for the dark matter halo, or at least, the contribution of the dark matter within this radius is
dynamically unimportant. Of course, we cannot exclude the existence of the dark halo that is significant at the larger radii and we hope that the ongoing studies (which use globular clusters as mass tracers) will clarify this.
Using the X--ray modelling we found that the temperatures of $T\sim 0.7$ keV (obtained in the uncorrected case) and $T=1.1$ keV (obtained in the corrected case) can provide a good fit to the dynamics of IC~1459 interior to $\sim$ 50 arcsec and between $\sim 20$ and $\sim 100$ arcsec respectively. This is higher than the value obtained using the stellar dynamics 2I modelling which lies between 0.4 and 0.6 keV. The difference between the BDI and Jeans approaches may be due to the fact that both approaches provide a different {\it prediction\/} of the X--ray temperature irrespective of the observed temperature ($T_X$) and irrespective of the validity of hydrostatic equilibrium: in this case 
the difference could be attributed to the different types of mass profiles used, and/or to the problems of the BDI modelling.

(ii) IC~3370 may be even a more complex galaxy to model with the 2I approach.
This galaxy shows a large isophotal twist and it has a significant non--zero velocity on the minor axis which strongly suggests its triaxiality. Therefore,
although 2I modelling necessarily provides a poor fit for some quantities (for example, velocity and $h_3$ parameter for the minor axis) it  nevertheless 
can give some insight into the dark matter content based on the study of the velocity dispersion. Again, as in the case of IC~1459 we can conclude that the dark matter within $\sim 3 R_e$ ($\sim 100$ arcsec $\approx 20.3$ kpc) is not needed (or it is not dynamically important) because we can obtain a good fit to the velocity dispersion using a constant mass--to--light ratio: $5 < M/L_B < 13$ (in solar units, for the Hubble constant of 70 ${\rm km\ s^{-1}  Mpc^{-1}}$).  In the case of this galaxy an X--ray halo was not detected, so we did not perform the X--ray modelling.

(iii) For {\sl NGC~3379} based on the 2I modelling procedures  we find no evidence for dark matter which dominates inside $\sim 1.5 R_e$ and found that the constant mass--to--light ratio of this galaxy in the $B$--band for $h_0=0.7$ is between 5 and 9.
This result is in agreement with previous studies of this galaxy (Ciardullo et al. 1993; Romanowsky et al. 2003). Analysis based on the X-rays leads to the same conclusion concerning the dark matter within $2R_e$ and the mass--to--light ratio of this galaxy if we adopt the parameters which are in agreement with those found in the literature: $\beta=0.64$ (where $\beta$ is the slope used in the analytic King approximation model) and the temperature of the X-ray halo is between $\sim$ 0.26 (=1$T_\sigma$) and 0.52(=2$T_\sigma$) keV where $T_\sigma$ is the temperature based on the stellar velocity dispersion. 
(If we use $T_{\rm X}=0.25$ keV and $\beta=0.5$, we can get a fit between $\sim 1$ and $\sim 1.5$ $R_e$
and between $\sim 1.5$ and $\sim 3.5 R_e$  the mass--to-light ratio in the $B$--band varies between 5.5 and 9.5.)
Solving Jeans equation we find that a spherical isotropic model ($\beta_*=0$) is excluded. The observations of the stellar velocity dispersion (interior to $\sim 1.5R_e$) favour $\beta_*=0.4$ (predominantly radial orbits) under the assumption of the hydrostatic equilibrium. Therefore, based on an X-ray analysis we conclude that interior to $\sim 1.5R_e$ the cumulative mass--to--light ratio is in agreement with that based on the dynamical modelling and 
therefore dark matter in NGC~3379 is not dynamically dominant in this region, 
although we note that the gradient of the mass profile is not consistent with the results from stellar dynamical modelling. Beyond  $\sim 2R_e$ we note the discrepancies regarding the mass--to--light ratio based on the X--rays and that based on the PNe.

Therefore, regarding NGC~3379, we conclude that, until better estimates for both stellar velocity dispersions and the $h_4$ parameter are available, it remains difficult to reconcile predictions of the X-ray methodology with the observations of the stellar velocity dispersions at radii $>2R_e$, 
without rather strong radial anisotropies for which strong hints are already available. The problems related to the X-ray methodology such as assumptions of hydrostatic equilibrium and the absolute value of the temperature $T_X$ and its radial dependence together with the difficulties in X--ray binary source subtraction may also play their part in this complex situation. 

(iv) {\sl NGC~4105}: based on the 2I modelling procedures which we performed we find no evidence for dark matter inside $\sim 1 R_e$ and found that the constant mass--to--light ratio of this galaxy in the $B$--band for $h_0=0.7$ is $\sim$ 6. Since in the case of this galaxy we had observational data only out to $\sim 1R_e$ we did not do the comparison with the X--rays out to this radius.
If we assume that the constant mass--to--light ratio of 5--7 (obtained using stellar dynamics) beyond $1R_e$  is still valid we notice a  discrepancy between this value and X-ray predictions which give at $2 R_e$ the value of $\sim 12$.
Since we did not have other tracers for the mass we could not make further comparisons as we did in the case of NGC~3379.
Solving the Jeans equation we find that a spherical isotropic model ($\beta_*=0$) in this case provides a good fit for a stellar velocity dispersion  out to $\sim 1R_e$, although we cannot exclude a case for which $\beta_*\sim 0.2$. Dark matter out to $\sim 1R_e$ in both of these cases is not needed. We note that in our modelling (both 2I and X--ray) we neglected possible interaction with the companion galaxy NGC~4106.

To summarize, based on the dynamical modelling we could not unambiguously detect the significant amounts of dark matter in our four galaxies out to $\sim 1-3R_e$ within the uncertainty of the modelling.
In the case of IC~1459, NGC~3379 and NGC~4105 we had an additional independent way to determine a mass--to--light ratio. We used the measurements of the temperature of their X--ray haloes and found that for IC~1459 there exists a discrepancy between the temperatures based on the 2I modelling ($T\sim 0.4-0.6$ keV) and that based on the X--ray modelling ($T\sim 0.7-1.1$ keV). For NGC~3379 the agreement between different methodologies is good: in its case between $\sim 1$ and $\sim 1.5R_e$ stellar dynamical modelling and X--ray modelling are in agreement. 
Unfortunately, for IC~3370 an X--ray halo was not detected so such an independent study was not possible.
For IC~1459 and NGC~3379 beyond $\sim 3R_e$ and for NGC~4105 beyond $\sim 1 R_e$
X--rays show the need for dark matter, but in these regions we may face orbit anisotropies (for example, radial, as hinted by positive values of the $h_4$ parameter in the cases of IC~1459 and NGC~3379). These anisotropies together with the inherent problems of the X--ray methodology (such as the assumption of spherical symmetry and lack of hydrostatic equilibrium) make the complex problem of dark matter in the early--type galaxies  even more difficult. This is currently under investigation.

\section*{acknowledgements}
We thank K. Freeman and M. Carollo for providing the observational material for NGC~3379 and NGC~4105 presented in this work.
We thank Francesca Matteucci for useful discussions and Piercarlo Bonifacio for the help with the {\sc MIDAS} system. We thank Olivier Hainaut for providing the photometry of IC~3370. We acknowledge  the use of the Gauss--Hermite Fourier
Fitting Software developed by R.P. van der Marel and M. Franx and the use of the Two--integral Jeans modelling Software developed by R.P. van der Marel and J.J. Binney.  This research has made use of the NASA/IPAC Extragalactic Database (NED) which is operated by the Jet Propulsion Laboratory, California Institute of Technology, under contract with the National Aeronautics and Space Administration.  
One of the authors (SS) has been partially supported by
the Ministry of Science and Environmental Protection of the
Republic of Serbia through the projects: no.~1468, ``Structure
Kinematics, and Dynamics of the Milky Way'' and
P1196  ``Astrophysical Spectroscopy of Extragalactic Objects''.
S.S. was supported with grants from MIUR COFIN 1998 Prot. No. 9802909231$\_$001, the ``Regione Friuli--Venezia--Giulia'' L.R. 3/98, CNAA Prot. No. 14/a  and ASI Prot. No. I/R/043/02. S.S. expresses his gratitude to the TRIL (Training and Research in Italian Laboratories) programme of the Abdus Salam International Centre for Theoretical Physics.
All unpublished observational material was obtained at ESO, La Silla
and Mt. Stromlo and Siding Springs Observatory.
We thank the referee, A. Romanowsky, for providing helpful comments.

\vfill\eject

\appendix{APPENDIX A: PHOTOMETIC DATA}

\begin{table*}
\caption{Photometric data of IC~1459}
\begin{center}
\begin{tabular}{rrr}
\hline
\noalign{\smallskip}
\multicolumn{1}{c}{Radius} &
\multicolumn{1}{c}{$B_{\rm maj}$} &
\multicolumn{1}{c}{$B_{\rm min}$}   \\
\noalign{\smallskip}
\hline
\noalign{\smallskip}  
 
1.12 & 17.54 & 17.53 \\
2.24 & 17.69 & 17.57 \\
3.36 & 17.98 & 17.93 \\
4.48 & 18.26 & 18.31 \\
5.60 & 18.53 & 18.64 \\
6.72 & 18.75 & 18.94 \\
7.84 & 18.97 & 19.23 \\
8.96 & 19.17 & 19.44 \\
10.08 & 19.35 & 19.65 \\
11.20 & 19.52 & 19.84 \\
12.32 & 19.63 & 20.01 \\
13.44 & 19.77 & 20.13 \\
14.56 & 19.87 & 20.27 \\
15.68 & 19.97 & 20.41 \\
16.80 & 20.07 & 20.54 \\
17.92 & 20.18 & 20.62 \\
19.04 & 20.26 & 20.77 \\
20.16 & 20.33 & 20.82 \\
21.28 & 20.38 & 20.94 \\
22.40 & 20.51 & 21.01 \\
23.52 & 20.60 & 21.13 \\
24.64 & 20.72 & 21.20 \\
25.76 & 20.78 & 21.28 \\
26.88 & 20.85 & 21.35 \\
28.00 & 20.93 & 21.41 \\
29.12 & 20.99 & 21.48 \\
30.24 & 21.06 & 21.55 \\
31.36 & 21.09 & 21.56 \\
32.48 & 21.14 & 21.69 \\
33.60 & 21.26 & 21.79 \\
34.72 & 21.29 & 21.81 \\
35.84 & 21.32 & 21.81 \\
36.96 & 21.36 & 21.92 \\
38.08 & 21.45 & 22.01 \\
39.20 & 21.46 & 22.05 \\
40.32 & 21.54 & 22.07 \\
41.44 & 21.57 & 22.15 \\
42.56 & 21.68 & 22.23 \\
43.68 & 21.68 & 22.26 \\
44.80 & 21.71 & 22.28 \\
45.92 & 21.79 & 22.38 \\
47.04 & 21.83 & 22.48 \\
48.16 & 21.84 & 22.54 \\
49.28 & 21.91 & 22.56 \\
50.40 & 21.99 & 22.66 \\
51.52 & 22.07 & 22.71 \\
52.64 & 22.06 & 22.78 \\
53.76 & 22.10 & 22.84 \\
54.88 & 22.13 & 22.82 \\
56.00 & 22.25 & 22.83 \\
57.12 & 22.26 & 22.95 \\
58.24 & 22.32 & 22.98 \\
59.36 & 22.38 & 22.99 \\
60.48 & 22.40 & 23.13 \\
61.60 & 22.39 & 23.06 \\
62.72 & 22.45 & 23.08 \\
63.84 & 22.53 & 23.33 \\
64.96 & 22.51 & 23.22 \\
66.08 & 22.56 & 23.27 \\
67.20 & 22.63 & 23.28 \\
68.32 & 22.65 & 23.16 \\
69.44 & 22.64 & 23.45 \\
70.56 & 22.76 & 23.62 \\

\noalign{\smallskip}
\hline
\noalign{\medskip}
\end{tabular}
\end{center}
\end{table*}
\vfill\eject

\begin{table*}
\caption{Photometric data of IC~1459 (cont.)}
\begin{center}
\begin{tabular}{rrr}
\hline
\noalign{\smallskip}
\multicolumn{1}{c}{Radius} &
\multicolumn{1}{c}{$B_{\rm maj}$} &
\multicolumn{1}{c}{$B_{\rm min}$}   \\
\noalign{\smallskip}
\hline
\noalign{\smallskip}

71.68 & 22.77 & 23.57 \\
72.80 & 22.79 & 23.51 \\
73.92 & 22.82 & 23.53 \\
75.04 & 22.81 & 23.68 \\
76.16 & 22.98 & 23.62 \\
77.28 & 22.89 & 23.60 \\
78.40 & 23.04 & 23.74 \\
79.52 & 23.03 & 23.88 \\
80.64 & 23.00 & 23.87 \\
81.76 & 23.15 & 23.92 \\
82.88 & 23.07 & 24.58 \\
84.00 & 23.17 & 24.18 \\
85.12 & 23.00 & 24.17 \\
86.24 & 23.25 & 24.54 \\
87.36 & 23.21 & 24.24 \\
88.48 & 23.16 & 24.21 \\
89.60 & 23.34 & 24.72 \\
90.72 & 23.34 & 24.38 \\
91.84 & 23.34 & 24.04 \\
92.96 & 23.34 & 23.59 \\
94.08 & 23.36 & 23.82 \\
95.20 & 23.32 & 24.70 \\
96.32 & 23.44 & 24.38 \\
97.44 & 23.56 & 24.62 \\
98.56 & 23.48 & 24.51 \\
99.68 & 23.60 & 24.77 \\
  
\noalign{\smallskip}
\hline
\noalign{\medskip}
\end{tabular}
\end{center}
\end{table*}

\begin{table*}
\caption{Photometric data of IC~1459}
\begin{center}
\begin{tabular}{rrrrrrrr}
\hline
\noalign{\smallskip}
\multicolumn{1}{c}{Radius} &
\multicolumn{1}{c}{ell} &
\multicolumn{1}{c}{e${}\_$ell} &
\multicolumn{1}{c}{$a_4$} &
\multicolumn{1}{c}{e${}\_{a_4}$} &
\multicolumn{1}{c}{PA} & 
\multicolumn{1}{c}{e${}\_{\rm PA}$}  \\
\noalign{\smallskip}
\hline
\noalign{\smallskip}  
 
2.37 & 0.0789 & 0.0222 & 0.0098 & 0.0112 & 25.0340 & 8.5371 \\ 
2.61 & 0.1555 & 0.0155 & 0.0014 & 0.0081 & 25.0340 & 3.1138 \\ 
2.87 & 0.2305 & 0.0116 & -0.0010 & 0.0064 & 29.3903 & 1.6299 \\ 
3.16 & 0.2619 & 0.0070 & -0.0036 & 0.0043 & 36.6932 & 0.9039 \\ 
3.48 & 0.2469 & 0.0050 & 0.0048 & 0.0030 & 36.2416 & 0.6791 \\ 
3.82 & 0.2217 & 0.0024 & 0.0040 & 0.0012 & 34.3339 & 0.3531 \\ 
4.21 & 0.2208 & 0.0022 & 0.0048 & 0.0009 & 33.9219 & 0.3290 \\ 
4.63 & 0.2181 & 0.0018 & 0.0043 & 0.0008 & 33.9717 & 0.2642 \\ 
5.09 & 0.2207 & 0.0018 & 0.0041 & 0.0008 & 34.7649 & 0.2591 \\ 
5.60 & 0.2239 & 0.0019 & 0.0043 & 0.0007 & 34.2680 & 0.2707 \\ 
6.16 & 0.2237 & 0.0019 & 0.0045 & 0.0009 & 34.1239 & 0.2799 \\ 
6.78 & 0.2259 & 0.0021 & 0.0073 & 0.0008 & 34.4263 & 0.3025 \\ 
7.45 & 0.2295 & 0.0020 & 0.0080 & 0.0009 & 34.9283 & 0.2867 \\ 
8.20 & 0.2357 & 0.0015 & 0.0065 & 0.0006 & 34.5964 & 0.2107 \\ 
9.02 & 0.2385 & 0.0016 & 0.0077 & 0.0006 & 34.4031 & 0.2149 \\ 
9.92 & 0.2416 & 0.0016 & 0.0062 & 0.0008 & 34.8394 & 0.2195 \\ 
10.91 & 0.2413 & 0.0014 & 0.0062 & 0.0007 & 35.0805 & 0.1927 \\ 
12.00 & 0.2481 & 0.0017 & 0.0068 & 0.0009 & 35.4974 & 0.2205 \\ 
13.20 & 0.2576 & 0.0016 & 0.0081 & 0.0008 & 35.7242 & 0.2098 \\ 
14.52 & 0.2631 & 0.0015 & 0.0054 & 0.0008 & 36.2231 & 0.1831 \\ 
15.98 & 0.2670 & 0.0014 & 0.0043 & 0.0009 & 36.3186 & 0.1769 \\ 
17.58 & 0.2721 & 0.0015 & 0.0015 & 0.0010 & 36.9959 & 0.1890 \\ 
19.33 & 0.2781 & 0.0016 & 0.0055 & 0.0010 & 37.4972 & 0.1963 \\ 
21.27 & 0.2816 & 0.0020 & 0.0037 & 0.0013 & 37.9748 & 0.2338 \\ 
23.39 & 0.2743 & 0.0015 & 0.0009 & 0.0010 & 38.7111 & 0.1853 \\ 
25.73 & 0.2669 & 0.0015 & 0.0036 & 0.0010 & 39.2994 & 0.1859 \\ 
28.31 & 0.2743 & 0.0019 & 0.0031 & 0.0013 & 40.3620 & 0.2370 \\ 
31.14 & 0.2746 & 0.0016 & 0.0072 & 0.0010 & 40.2397 & 0.1912 \\ 
34.25 & 0.2735 & 0.0017 & 0.0062 & 0.0011 & 40.5417 & 0.2108 \\ 
37.67 & 0.2799 & 0.0017 & 0.0037 & 0.0011 & 40.6800 & 0.2014 \\ 
41.44 & 0.2793 & 0.0019 & 0.0025 & 0.0013 & 41.5781 & 0.2319 \\ 
45.59 & 0.2734 & 0.0019 & -0.0060 & 0.0012 & 41.6269 & 0.2277 \\ 
50.14 & 0.2605 & 0.0021 & 0.0044 & 0.0013 & 42.3462 & 0.2649 \\ 
55.16 & 0.2633 & 0.0017 & -0.0022 & 0.0011 & 42.8055 & 0.2126 \\ 
60.67 & 0.2633 & 0.0036 & -0.0046 & 0.0032 & 42.8055 & 0.4574 \\ 
66.74 & 0.3460 & 0.0166 & 0.0246 & 0.0128 & 42.8055 & 1.6807 \\ 
73.42 & 0.2916 & 0.0121 & 0.0232 & 0.0085 & 44.9324 & 1.3949 \\ 
80.76 & 0.2655 & 0.0040 & 0.0021 & 0.0027 & 42.7246 & 0.5053 \\ 
88.83 & 0.2655 & 0.0039 & -0.0096 & 0.0026 & 42.7246 & 0.4818 \\ 
97.72 & 0.2866 & 0.0028 & -0.0042 & 0.0020 & 43.5526 & 0.3290 \\

\noalign{\smallskip}
\hline
\noalign{\medskip}
\end{tabular}
\end{center}
\end{table*}

\begin{table*}
\caption{Photometric data of IC~3370}
\begin{center}
\begin{tabular}{rrrrrrrrrr}
\hline
\noalign{\smallskip}
\multicolumn{1}{c}{Radius} &
\multicolumn{1}{c}{$B_{\rm maj}$} &
\multicolumn{1}{c}{$B_{\rm min}$} &
\multicolumn{1}{c}{ell} &
\multicolumn{1}{c}{e${}\_$ell} &
\multicolumn{1}{c}{$a_4$} &
\multicolumn{1}{c}{e${}\_{a_4}$} &
\multicolumn{1}{c}{PA} & 
\multicolumn{1}{c}{e${}\_{\rm PA}$}  \\
\noalign{\smallskip}
\hline
\noalign{\smallskip}  
 
1.01 & 20.32 & 20.35 & 0.1779 & 0.0345 & 0.0023 & 0.0083 & 28.4463 & 6.3102 \\ 
1.11 & 20.33 & 20.37 & 0.1532 & 0.0270 & -0.0077 & 0.0051 & 30.3158 & 5.6492 \\ 
1.22 & 20.35 & 20.39 & 0.1532 & 0.0272 & -0.0076 & 0.0047 & 35.2232 & 5.5920 \\ 
1.34 & 20.36 & 20.42 & 0.1671 & 0.0252 & -0.0150 & 0.0116 & 35.5865 & 4.8310 \\ 
1.47 & 20.38 & 20.44 & 0.1868 & 0.0315 & -0.0204 & 0.0144 & 35.2309 & 5.3989 \\ 
1.62 & 20.40 & 20.47 & 0.1999 & 0.0180 & -0.0103 & 0.0087 & 34.9541 & 3.0065 \\ 
1.78 & 20.43 & 20.50 & 0.2081 & 0.0180 & -0.0136 & 0.0094 & 33.7971 & 2.8176 \\ 
1.96 & 20.45 & 20.53 & 0.2071 & 0.0145 & -0.0127 & 0.0072 & 34.3572 & 2.3044 \\ 
2.16 & 20.48 & 20.56 & 0.2176 & 0.0107 & -0.0070 & 0.0052 & 31.5482 & 1.6367 \\ 
2.37 & 20.51 & 20.60 & 0.2250 & 0.0088 & -0.0015 & 0.0034 & 30.4501 & 1.3044 \\ 
2.61 & 20.54 & 20.65 & 0.2346 & 0.0095 & -0.0002 & 0.0046 & 28.8893 & 1.3446 \\ 
2.87 & 20.57 & 20.69 & 0.2461 & 0.0096 & -0.0027 & 0.0047 & 28.1640 & 1.2951 \\ 
3.16 & 20.61 & 20.74 & 0.2523 & 0.0087 & -0.0076 & 0.0038 & 28.5882 & 1.1625 \\ 
3.48 & 20.65 & 20.80 & 0.2578 & 0.0086 & -0.0042 & 0.0040 & 28.2886 & 1.1202 \\ 
3.82 & 20.70 & 20.85 & 0.2666 & 0.0089 & 0.0015 & 0.0036 & 28.5728 & 1.1267 \\ 
4.21 & 20.75 & 20.92 & 0.2803 & 0.0080 & 0.0050 & 0.0034 & 28.5190 & 0.9652 \\ 
4.63 & 20.80 & 20.99 & 0.3009 & 0.0089 & 0.0075 & 0.0040 & 28.6105 & 1.0058 \\ 
5.09 & 20.86 & 21.06 & 0.3293 & 0.0104 & 0.0135 & 0.0056 & 28.0826 & 1.0915 \\ 
5.60 & 20.93 & 21.14 & 0.3748 & 0.0127 & 0.0216 & 0.0068 & 27.5783 & 1.2197 \\ 
6.16 & 21.00 & 21.23 & 0.4074 & 0.0115 & 0.0050 & 0.0069 & 27.2899 & 1.0344 \\ 
6.78 & 21.07 & 21.33 & 0.4178 & 0.0099 & -0.0096 & 0.0068 & 27.2899 & 0.8805 \\ 
7.45 & 21.15 & 21.43 & 0.4241 & 0.0106 & -0.0203 & 0.0073 & 24.5118 & 0.9281 \\ 
8.20 & 21.24 & 21.54 & 0.3938 & 0.0146 & -0.0264 & 0.0081 & 21.2977 & 1.3407 \\ 
9.02 & 21.34 & 21.65 & 0.3673 & 0.0151 & -0.0222 & 0.0082 & 19.2960 & 1.4664 \\ 
9.92 & 21.44 & 21.78 & 0.3194 & 0.0107 & 0.0023 & 0.0050 & 20.5110 & 1.1535 \\ 
10.91 & 21.55 & 21.91 & 0.2797 & 0.0113 & 0.0231 & 0.0060 & 19.6155 & 1.3511 \\ 
12.00 & 21.66 & 22.04 & 0.2586 & 0.0079 & 0.0204 & 0.0047 & 22.4270 & 1.0151 \\ 
13.20 & 21.79 & 22.19 & 0.2597 & 0.0054 & 0.0071 & 0.0034 & 26.4154 & 0.6823 \\ 
14.53 & 21.92 & 22.34 & 0.2534 & 0.0049 & 0.0074 & 0.0032 & 29.5984 & 0.6348 \\ 
15.98 & 22.06 & 22.50 & 0.2539 & 0.0057 & 0.0022 & 0.0037 & 32.0069 & 0.7313 \\ 
17.58 & 22.21 & 22.66 & 0.2313 & 0.0055 & 0.0119 & 0.0035 & 32.3434 & 0.7686 \\ 
19.33 & 22.37 & 22.83 & 0.2261 & 0.0059 & 0.0096 & 0.0036 & 32.9040 & 0.8371 \\ 
21.27 & 22.53 & 23.01 & 0.2176 & 0.0073 & 0.0195 & 0.0045 & 33.7490 & 1.0765 \\ 
23.39 & 22.70 & 23.18 & 0.2003 & 0.0063 & 0.0106 & 0.0038 & 32.9594 & 1.0006 \\ 
25.73 & 22.88 & 23.36 & 0.1740 & 0.0063 & 0.0014 & 0.0038 & 37.2009 & 1.1426 \\ 
28.30 & 23.06 & 23.53 & 0.1634 & 0.0073 & -0.0059 & 0.0043 & 35.9724 & 1.3946 \\ 
31.14 & 23.24 & 23.71 & 0.1631 & 0.0066 & -0.0140 & 0.0039 & 36.6420 & 1.2666 \\ 
34.25 & 23.43 & 23.88 & 0.1507 & 0.0088 & -0.0254 & 0.0053 & 39.0821 & 1.8091 \\ 
37.67 & 23.61 & 24.05 & 0.1439 & 0.0093 & -0.0346 & 0.0056 & 38.3693 & 2.0001 \\ 
41.44 & 23.80 & 24.21 & 0.1629 & 0.0070 & -0.0267 & 0.0043 & 41.7802 & 1.3302 \\ 
45.59 & 23.98 & 24.37 & 0.1481 & 0.0129 & -0.0351 & 0.0090 & 48.1979 & 2.6974 \\ 
50.14 & 24.16 & 24.53 & 0.2027 & 0.0100 & -0.0275 & 0.0066 & 48.1979 & 1.5722 \\ 
55.16 & 24.33 & 24.69 & 0.1798 & 0.0109 & -0.0502 & 0.0080 & 48.8121 & 1.9028 \\ 
60.67 & 24.49 & 24.86 & 0.1956 & 0.0102 & -0.0325 & 0.0069 & 49.0052 & 1.6644 \\ 
66.74 & 24.65 & 25.05 & 0.1956 & 0.0122 & -0.0237 & 0.0078 & 47.0718 & 1.9807 \\ 
73.42 & 24.80 & 25.25 & 0.1591 & 0.0144 & -0.0433 & 0.0103 & 47.0718 & 2.8120 \\ 
80.76 & 24.96 & 25.48 & 0.2944 & 0.0167 & 0.0311 & 0.0117 & 62.2400 & 1.9216 \\ 
88.83 & 25.11 & 25.74 & 0.2116 & 0.0324 & 0.0656 & 0.0260 & 63.7530 & 4.7387 \\ 
97.72 & 25.28 & 26.01 & 0.1937 & 0.0285 & 0.0158 & 0.0166 & 70.8523 & 4.2561 \\

\noalign{\smallskip}
\hline
\noalign{\medskip}
\end{tabular}
\end{center}
\end{table*}

\begin{table*}
\caption{Photometric data of NGC~3379}
\begin{center}
\begin{tabular}{rrrrrrrrrr}
\hline
\noalign{\smallskip}
\multicolumn{1}{c}{Radius} &
\multicolumn{1}{c}{ell} &
\multicolumn{1}{c}{e${}\_$ell} &
\multicolumn{1}{c}{$a_4$} &
\multicolumn{1}{c}{e${}\_{a_4}$} &
\multicolumn{1}{c}{PA} & 
\multicolumn{1}{c}{e${}\_{\rm PA}$}  \\
\noalign{\smallskip}
\hline
\noalign{\smallskip}  

1.04 & 0.0420 &  0.0179 & -0.0199 & 0.0077 & 31.7308 & 12.6662 \\ 
1.15 & 0.0411 &  0.0144 & -0.0182 & 0.0061 & 31.4050 & 10.1941 \\ 
1.26 & 0.0364 &  0.0109 & -0.0120 & 0.0048 & 30.9507 & 8.8246 \\ 
1.39 & 0.0364 &  0.0090 & -0.0100 & 0.0040 & 38.9286 & 7.2954 \\ 
1.52 & 0.0468 &  0.0061 & -0.0033 & 0.0029 & 38.9286 & 3.8845 \\ 
1.68 & 0.0560 &  0.0065 & -0.0033 & 0.0033 & 35.8077 & 3.4417 \\ 
1.84 & 0.0497 &  0.0045 & -0.0047 & 0.0021 & 35.2196 & 2.6751 \\ 
2.03 & 0.0477 &  0.0040 & 0.0006 & 0.0019 & 42.1754 & 2.4513 \\ 
2.23 & 0.0513 &  0.0029 & -0.0018 & 0.0012 & 48.6264 & 1.6554 \\ 
2.45 & 0.0484 &  0.0032 & -0.0009 & 0.0015 & 46.5158 & 1.9447 \\ 
2.70 & 0.0472 &  0.0024 & 0.0003 & 0.0011 & 49.1606 & 1.5022 \\ 
2.97 & 0.0545 &  0.0020 & 0.0009 & 0.0010 & 52.3800 & 1.0774 \\ 
3.27 & 0.0570 &  0.0016 & 0.0014 & 0.0007 & 53.0786 & 0.8062 \\ 
3.59 & 0.0598 &  0.0018 & 0.0009 & 0.0009 & 58.6417 & 0.8868 \\ 
3.95 & 0.0601 &  0.0014 & 0.0004 & 0.0006 & 59.2378 & 0.6864 \\ 
4.35 & 0.0638 &  0.0016 & 0.0005 & 0.0008 & 61.9261 & 0.7383 \\ 
4.78 & 0.0712 &  0.0012 & 0.0003 & 0.0006 & 64.9966 & 0.5020 \\ 
5.26 & 0.0707 &  0.0013 & 0.0012 & 0.0007 & 66.1964 & 0.5561 \\ 
5.79 & 0.0757 &  0.0013 & -0.0000 & 0.0006 & 69.1280 & 0.5120 \\ 
6.37 & 0.0801 &  0.0011 & -0.0006 & 0.0005 & 69.6258 & 0.4070 \\ 
7.00 & 0.0829 &  0.0009 & 0.0004 & 0.0005 & 70.0371 & 0.3291 \\ 
7.70 & 0.0851 &  0.0009 & 0.0010 & 0.0005 & 71.3429 & 0.3188 \\ 
8.47 & 0.0857 &  0.0009 & 0.0004 & 0.0005 & 71.5087 & 0.3147 \\ 
9.32 & 0.0857 &  0.0008 & 0.0020 & 0.0004 & 72.4738 & 0.2620 \\ 
10.25 & 0.0842 &  0.0007 & 0.0000 & 0.0004 & 73.0124 & 0.2546 \\ 
11.28 & 0.0846 &  0.0007 & 0.0005 & 0.0004 & 73.2315 & 0.2527 \\ 
12.41 & 0.0809 &  0.0008 & 0.0002 & 0.0004 & 73.2120 & 0.2892 \\ 
13.65 & 0.0794 &  0.0008 & 0.0005 & 0.0004 & 73.5209 & 0.2851 \\ 
15.01 & 0.0789 &  0.0008 & -0.0001 & 0.0004 & 73.1663 & 0.2998 \\ 
16.51 & 0.0809 &  0.0007 & -0.0015 & 0.0004 & 73.6499 & 0.2755 \\ 
18.16 & 0.0825 &  0.0008 & -0.0011 & 0.0004 & 73.2797 & 0.2816 \\ 
19.98 & 0.0823 &  0.0008 & -0.0003 & 0.0004 & 73.3162 & 0.2944 \\ 
21.98 & 0.0844 &  0.0008 & -0.0008 & 0.0004 & 73.6756 & 0.2757 \\ 
24.18 & 0.0842 &  0.0008 & -0.0009 & 0.0004 & 73.1237 & 0.2663 \\ 
26.59 & 0.0851 &  0.0008 & -0.0010 & 0.0004 & 72.7601 & 0.2656 \\ 
29.25 & 0.0837 &  0.0008 & -0.0014 & 0.0005 & 72.6128 & 0.3025 \\ 
32.18 & 0.0867 &  0.0010 & -0.0006 & 0.0005 & 72.6128 & 0.3348 \\ 
35.40 & 0.0909 &  0.0008 & -0.0016 & 0.0005 & 72.5209 & 0.2780 \\ 
38.94 & 0.0984 &  0.0019 & -0.0017 & 0.0010 & 72.1395 & 0.5751 \\ 
42.83 & 0.1018 &  0.0011 & -0.0004 & 0.0006 & 72.3862 & 0.3325 \\ 
47.11 & 0.1018 &  0.0038 & 0.0171 & 0.0047 & 72.3862 & 1.1097 \\ 
51.82 & 0.1497 &  0.0050 & 0.0222 & 0.0028 & 72.3862 & 1.0235 \\ 
57.01 & 0.1314 &  0.0024 & -0.0002 & 0.0014 & 71.0137 & 0.5648 \\ 
62.71 & 0.1296 &  0.0017 & -0.0007 & 0.0009 & 69.7442 & 0.3917 \\ 
68.98 & 0.1335 &  0.0017 & 0.0023 & 0.0009 & 69.7442 & 0.3833 \\ 
75.88 & 0.1282 &  0.0021 & -0.0016 & 0.0012 & 69.8391 & 0.5120 \\ 
83.46 & 0.1352 &  0.0020 & 0.0002 & 0.0011 & 68.4398 & 0.4645 \\

\noalign{\smallskip}
\hline
\noalign{\medskip}
\end{tabular}
\end{center}
\end{table*}

\begin{table*}
\caption{Photometric data of NGC~4105}
\begin{center}
\begin{tabular}{rrr}
\hline
\noalign{\smallskip}
\multicolumn{1}{c}{Radius} &
\multicolumn{1}{c}{$B_{\rm maj}$} &
\multicolumn{1}{c}{$B_{\rm min}$}   \\
\noalign{\smallskip}
\hline
\noalign{\smallskip}

1.01 & 16.60 &    16.44 \\ 
1.34 & 16.79 &    16.68 \\ 
1.68 & 16.99 &    16.95 \\ 
2.02 & 17.17 &    17.19 \\ 
2.35 & 17.34 &    17.40 \\ 
2.69 & 17.48 &    17.57 \\ 
3.02 & 17.58 &    17.75 \\ 
3.36 & 17.70 &    17.90 \\ 
3.70 & 17.78 &    18.02 \\ 
4.03 & 17.90 &    18.13 \\ 
4.37 & 17.97 &    18.25 \\ 
4.70 & 18.09 &    18.38 \\ 
5.04 & 18.17 &    18.45 \\ 
5.38 & 18.25 &    18.53 \\ 
5.71 & 18.36 &    18.63 \\ 
6.05 & 18.41 &    18.74 \\ 
6.38 & 18.51 &    18.82 \\ 
6.72 & 18.57 &    18.87 \\ 
7.06 & 18.66 &    18.94 \\ 
7.39 & 18.71 &    19.05 \\ 
7.73 & 18.76 &    19.09 \\ 
8.06 & 18.83 &    19.14 \\ 
8.40 & 18.90 &    19.27 \\ 
8.74 & 18.97 &    19.27 \\ 
9.07 & 19.00 &    19.31 \\ 
9.41 & 19.04 &    19.39 \\ 
9.74 & 19.09 &    19.48 \\ 
10.08 & 19.13 &    19.45 \\ 
10.42 & 19.19 &    19.48 \\ 
10.75 & 19.24 &    19.61 \\ 
11.09 & 19.29 &    19.57 \\ 
11.42 & 19.38 &    19.76 \\ 
11.76 & 19.37 &    19.69 \\ 
12.10 & 19.41 &    19.80 \\ 
12.43 & 19.44 &    19.76 \\ 
12.77 & 19.55 &    19.86 \\ 
13.10 & 19.52 &    19.94 \\ 
13.44 & 19.51 &    19.91 \\ 
13.78 & 19.57 &    19.93 \\ 
14.11 & 19.67 &    19.99 \\ 
14.45 & 19.70 &    20.02 \\ 
14.78 & 19.64 &    19.97 \\ 
15.12 & 19.67 &    20.05 \\ 
15.46 & 19.67 &    20.12 \\ 
15.79 & 19.74 &    20.12 \\ 
16.13 & 19.83 &    20.10 \\ 
16.46 & 19.82 &    20.15 \\ 
16.80 & 19.91 &    20.18 \\ 
17.14 & 19.90 &    20.24 \\ 
17.47 & 19.85 &    20.23 \\ 
17.81 & 19.91 &    20.22 \\ 
18.14 & 19.97 &    20.37 \\ 
18.48 & 19.93 &    20.26 \\ 
18.82 & 20.03 &    20.33 \\ 
19.15 & 19.93 &    20.36 \\ 
19.49 & 19.90 &    20.33 \\ 
19.82 & 20.03 &    20.39 \\ 
20.16 & 19.98 &    20.30 \\ 
20.50 & 20.10 &    20.43 \\ 
20.83 & 20.14 &    20.44 \\ 

\noalign{\smallskip}
\hline
\noalign{\medskip}
\end{tabular}
\end{center}
\end{table*}

\begin{table*}
\caption{Photometric data of NGC~4105 (cont.)}
\begin{center}
\begin{tabular}{rrr}
\hline
\noalign{\smallskip}
\multicolumn{1}{c}{Radius} &
\multicolumn{1}{c}{$B_{\rm maj}$} &
\multicolumn{1}{c}{$B_{\rm min}$}   \\
\noalign{\smallskip}
\hline
\noalign{\smallskip}

21.17 & 20.03 &    20.40 \\ 
21.50 & 20.06 &    20.41 \\ 
21.84 & 20.19 &    20.45 \\ 
22.18 & 20.20 &    20.51 \\ 
22.51 & 20.17 &    20.41 \\ 
22.85 & 20.18 &    20.51 \\ 
23.18 & 20.26 &    20.48 \\ 
23.52 & 20.15 &    20.47 \\ 
23.86 & 20.28 &    20.64 \\ 
24.19 & 20.24 &    20.56 \\ 
24.53 & 20.18 &    20.73 \\ 
24.86 & 20.23 &    20.51 \\ 
25.20 & 20.25 &    20.67 \\ 
25.54 & 20.40 &    20.53 \\ 
25.87 & 20.38 &    20.55 \\ 
26.21 & 20.33 &    20.67 \\ 
26.54 & 20.39 &    20.70 \\ 
26.88 & 20.49 &    20.59 \\ 
27.22 & 20.44 &    20.58 \\ 
27.55 & 20.32 &    20.68 \\ 
27.89 & 20.38 &    20.68 \\ 
28.22 & 20.45 &    20.71 \\ 
28.56 & 20.43 &    20.74 \\ 
28.90 & 20.47 &    20.75 \\ 
29.23 & 20.58 &    20.81 \\ 
29.57 & 20.52 &    20.76 \\ 
29.90 & 20.58 &    21.03 \\

\noalign{\smallskip}
\hline
\noalign{\medskip}
\end{tabular}
\end{center}
\end{table*}

\begin{table*}
\caption{Photometric data of NGC~4105}
\begin{center}
\begin{tabular}{rrrrrrrrrr}
\hline
\noalign{\smallskip}
\multicolumn{1}{c}{Radius} &
\multicolumn{1}{c}{ell} &
\multicolumn{1}{c}{e${}\_$ell} &
\multicolumn{1}{c}{$a_4$} &
\multicolumn{1}{c}{e${}\_{a_4}$} &
\multicolumn{1}{c}{PA} & 
\multicolumn{1}{c}{e${}\_{\rm PA}$}  \\
\noalign{\smallskip}
\hline
\noalign{\smallskip}  

1.07 & 0.0309 &  0.0073 & -0.0008 & 0.0029 & 198.5218 & 6.9764 \\ 
1.18 & 0.0390 &  0.0064 & -0.0034 & 0.0024 & 187.2130 & 4.8666 \\ 
1.30 & 0.0461 &  0.0049 & -0.0030 & 0.0016 & 179.1972 & 3.1360 \\ 
1.42 & 0.0606 &  0.0037 & -0.0022 & 0.0014 & 174.7109 & 1.7980 \\ 
1.57 & 0.0718 &  0.0046 & -0.0003 & 0.0017 & 169.7080 & 1.9388 \\ 
1.72 & 0.0908 &  0.0034 & 0.0024 & 0.0014 & 164.3650 & 1.1180 \\ 
1.90 & 0.1077 &  0.0031 & 0.0026 & 0.0015 & 160.6236 & 0.8842 \\ 
2.09 & 0.1286 &  0.0041 & 0.0025 & 0.0021 & 157.5560 & 0.9839 \\ 
2.29 & 0.1464 &  0.0037 & 0.0004 & 0.0021 & 156.5570 & 0.7885 \\ 
2.52 & 0.1585 &  0.0033 & -0.0042 & 0.0017 & 155.4411 & 0.6422 \\ 
2.78 & 0.1635 &  0.0038 & -0.0069 & 0.0015 & 154.9087 & 0.7275 \\ 
3.05 & 0.1719 &  0.0055 & -0.0097 & 0.0024 & 154.0606 & 1.0044 \\ 
3.36 & 0.1802 &  0.0049 & -0.0114 & 0.0022 & 151.7974 & 0.8679 \\ 
3.70 & 0.1935 &  0.0045 & -0.0098 & 0.0019 & 150.6190 & 0.7382 \\ 
4.07 & 0.1966 &  0.0049 & -0.0082 & 0.0024 & 151.6672 & 0.7954 \\ 
4.47 & 0.2071 &  0.0043 & -0.0048 & 0.0020 & 151.8860 & 0.6633 \\ 
4.92 & 0.2068 &  0.0042 & -0.0054 & 0.0018 & 152.9017 & 0.6469 \\ 
5.41 & 0.2089 &  0.0044 & -0.0055 & 0.0019 & 154.2281 & 0.6783 \\ 
5.95 & 0.2071 &  0.0036 & -0.0073 & 0.0018 & 153.8673 & 0.5592 \\ 
6.55 & 0.2053 &  0.0033 & -0.0042 & 0.0017 & 154.6262 & 0.5140 \\ 
7.20 & 0.2149 &  0.0036 & -0.0022 & 0.0017 & 153.3746 & 0.5293 \\ 
7.92 & 0.2160 &  0.0033 & -0.0038 & 0.0018 & 153.4159 & 0.4974 \\ 
8.71 & 0.2126 &  0.0029 & -0.0047 & 0.0017 & 153.2645 & 0.4434 \\ 
9.59 & 0.2139 &  0.0032 & -0.0067 & 0.0018 & 153.3385 & 0.4807 \\ 
10.55 & 0.2095 &  0.0028 & -0.0052 & 0.0017 & 153.6522 & 0.4293 \\ 
11.60 & 0.2113 &  0.0031 & -0.0044 & 0.0018 & 152.5015 & 0.4725 \\ 
12.76 & 0.2143 &  0.0029 & -0.0020 & 0.0018 & 153.5367 & 0.4342 \\ 
14.04 & 0.2217 &  0.0030 & -0.0003 & 0.0019 & 153.2547 & 0.4416 \\ 
15.44 & 0.2249 &  0.0034 & 0.0009 & 0.0022 & 153.2996 & 0.4909 \\ 
16.98 & 0.2336 &  0.0034 & 0.0007 & 0.0022 & 152.9260 & 0.4679 \\ 
18.68 & 0.2324 &  0.0035 & 0.0004 & 0.0022 & 151.6540 & 0.4890 \\ 
20.55 & 0.2559 &  0.0037 & 0.0047 & 0.0024 & 151.6540 & 0.4760 \\ 
22.60 & 0.2545 &  0.0063 & -0.0122 & 0.0041 & 149.2091 & 0.8134 \\ 
24.86 & 0.2545 &  0.0037 & -0.0055 & 0.0024 & 150.8761 & 0.4807 \\ 
27.35 & 0.2737 &  0.0091 & 0.0067 & 0.0062 & 150.8761 & 1.1101 \\ 

\noalign{\smallskip}
\hline
\noalign{\medskip}
\end{tabular}
\end{center}
\end{table*}

\appendix{APPENDIX B: KINEMATIC DATA}

\begin{table*}
\caption{Kinematic data of IC~1459 (major axis)}
\begin{center}
\begin{tabular}{rrrrrrrrr}
\hline
\noalign{\smallskip}
\multicolumn{1}{c}{Radius} &
\multicolumn{1}{c}{$v$ (km/s)} &
\multicolumn{1}{c}{e${}\_{}v$} &
\multicolumn{1}{c}{$\sigma$ (km/s)} &
\multicolumn{1}{c}{e${}\_\sigma$} &
\multicolumn{1}{c}{$h_3$} &
\multicolumn{1}{c}{e${}\_{h_3}$} &
\multicolumn{1}{c}{$h_4$} & 
\multicolumn{1}{c}{e${}\_h_4$}  \\
\noalign{\smallskip}
\hline
\noalign{\smallskip}

 -89.608 & 95.698 & 25.135 & 169.480 & 24.391 & -0.068 & 0.133 & 0.133 & 0.133 \\
-75.794 & 102.597 & 19.720 & 163.257 & 16.297 & -0.079 & 0.106 & 0.241 & 0.107 \\
-65.327 & 121.165 & 17.883 & 183.567 & 22.703 & -0.199 & 0.094 & 0.239 & 0.094 \\
-58.974 & 76.006 & 16.481 & 234.049 & 15.407 & -0.019 & 0.058 & 0.069 & 0.063 \\
-53.963 & 61.948 & 14.635 & 241.966 & 13.329 & 0.007 & 0.049 & 0.075 & 0.054 \\
-49.787 & 81.218 & 13.897 & 231.859 & 12.874 & -0.032 & 0.050 & 0.075 & 0.054 \\
-46.163 & 110.24 & 15.757 & 262.035 & 13.837 & 0.018 & 0.046 & 0.080 & 0.054 \\
-42.810 & 124.364 & 13.676 & 242.944 & 12.458 & -0.057 & 0.046 & 0.071 & 0.051 \\
-39.725 & 66.984 & 13.232 & 255.728 & 14.135 & 0.024 & 0.041 & -0.015 & 0.048 \\
-37.223 & 67.771 & 14.594 & 278.077 & 12.821 & 0.028 & 0.039 & 0.076 & 0.048 \\
-34.981 & 48.423 & 13.359 & 270.989 & 12.674 & 0.065 & 0.038 & 0.030 & 0.044 \\
-32.739 & 56.906 & 10.967 & 258.893 & 10.091 & 0.005 & 0.033 & 0.058 & 0.038 \\
-30.785 & 38.417 & 12.847 & 267.527 & 11.879 & 0.002 & 0.036 & 0.051 & 0.043 \\
-29.112 & 69.605 & 12.108 & 268.088 & 10.983 & 0.024 & 0.034 & 0.060 & 0.040 \\
-27.428 & 61.859 & 13.102 & 275.242 & 12.066 & -0.010 & 0.035 & 0.050 & 0.042 \\
-25.748 & 52.271 & 12.670 & 274.420 & 12.825 & 0.014 & 0.034 & -0.001 & 0.041 \\
-24.355 & 70.152 & 14.619 & 287.086 & 13.682 & -0.003 & 0.037 & 0.040 & 0.045 \\
-23.234 & 35.261 & 12.820 & 279.566 & 11.885 & 0.038 & 0.034 & 0.044 & 0.041 \\
-22.114 & 20.275 & 11.958 & 266.581 & 11.656 & 0.003 & 0.034 & 0.024 & 0.040 \\
-20.994 & 19.439 & 12.491 & 280.641 & 11.685 & -0.004 & 0.033 & 0.041 & 0.039 \\
-19.873 & 21.448 & 10.453 & 291.712 & 9.357 & 0.024 & 0.026 & 0.066 & 0.032 \\
-18.753 & 30.122 & 12.012 & 286.600 & 11.272 & 0.031 & 0.031 & 0.037 & 0.037 \\
-17.634 & 36.696 & 10.995 & 283.593 & 10.608 & 0.017 & 0.028 & 0.022 & 0.034 \\
-16.513 & 20.862 & 11.180 & 301.964 & 10.874 & 0.059 & 0.027 & 0.011 & 0.032 \\
-15.680 & 12.527 & 12.161 & 308.807 & 12.062 & 0.020 & 0.027 & 0.008 & 0.033 \\
-15.120 & 18.488 & 13.908 & 302.048 & 13.638 & 0.063 & 0.033 & 0.005 & 0.040 \\
-14.560 & 0.628 & 16.895 & 305.528 & 17.641 & 0.054 & 0.040 & -0.037 & 0.048 \\
-14.000 & 20.766 & 12.372 & 293.623 & 12.266 & 0.020 & 0.030 & 0.005 & 0.036 \\
-13.440 & 22.538 & 11.843 & 289.379 & 11.235 & 0.027 & 0.030 & 0.031 & 0.036 \\
-12.880 & 18.442 & 14.275 & 312.678 & 13.395 & 0.040 & 0.032 & 0.045 & 0.041 \\
-12.320 & 13.073 & 13.877 & 309.492 & 14.580 & 0.026 & 0.031 & -0.034 & 0.038 \\
-11.760 & 20.24 & 12.140 & 318.476 & 11.560 & 0.042 & 0.026 & 0.039 & 0.034 \\
-11.200 & 7.132 & 13.393 & 310.274 & 12.989 & 0.058 & 0.031 & 0.016 & 0.037 \\
-10.640 & 11.204 & 12.971 & 308.312 & 11.974 & 0.075 & 0.031 & 0.044 & 0.038 \\
-10.080 & 1.964 & 11.344 & 316.647 & 10.728 & 0.050 & 0.025 & 0.040 & 0.032 \\
-9.520 & 14.763 & 11.508 & 308.410 & 10.749 & 0.042 & 0.026 & 0.045 & 0.033 \\
-8.960 & -11.319 & 10.450 & 307.970 & 9.865 & 0.065 & 0.024 & 0.032 & 0.030 \\
-8.400 & 2.195 & 11.973 & 315.360 & 11.706 & 0.052 & 0.027 & 0.015 & 0.033 \\
-7.840 & -0.192 & 11.165 & 326.330 & 11.215 & 0.081 & 0.024 & -0.011 & 0.029 \\
-7.280 & -11.324 & 10.741 & 315.678 & 9.805 & 0.053 & 0.024 & 0.063 & 0.031 \\
-6.720 & -14.223 & 11.613 & 322.202 & 11.122 & 0.093 & 0.026 & 0.018 & 0.031 \\
-6.160 & -18.614 & 11.714 & 341.184 & 11.177 & 0.067 & 0.024 & 0.052 & 0.031 \\
-5.600 & -19.66 & 12.040 & 338.357 & 11.596 & 0.091 & 0.026 & 0.029 & 0.031 \\
-5.040 & -32.463 & 11.824 & 341.004 & 11.608 & 0.077 & 0.024 & 0.025 & 0.030 \\
-4.480 & -44.091 & 11.259 & 338.169 & 10.803 & 0.104 & 0.025 & 0.023 & 0.029 \\
-3.920 & -50.476 & 11.266 & 334.689 & 10.751 & 0.097 & 0.025 & 0.029 & 0.030 \\
-3.360 & -51.924 & 11.881 & 338.544 & 11.498 & 0.089 & 0.025 & 0.027 & 0.031 \\
-2.800 & -59.817 & 10.187 & 330.681 & 9.706 & 0.104 & 0.023 & 0.022 & 0.027 \\
-2.240 & -56.934 & 10.723 & 341.247 & 10.165 & 0.117 & 0.024 & 0.027 & 0.028 \\
-1.680 & -56.246 & 11.498 & 346.636 & 10.854 & 0.101 & 0.024 & 0.048 & 0.030 \\
-1.120 & -34.563 & 12.155 & 343.103 & 11.845 & 0.076 & 0.025 & 0.034 & 0.031 \\
-0.560 & -11.207 & 13.910 & 341.779 & 13.939 & 0.077 & 0.029 & 0.009 & 0.034 \\

\noalign{\smallskip}
\hline
\noalign{\medskip}
\end{tabular}
\end{center}
\end{table*}

\begin{table*}
\caption{Kinematic data of IC~1459 (major axis, continued)}
\begin{center}
\begin{tabular}{rrrrrrrrr}
\hline
\noalign{\smallskip}
\multicolumn{1}{c}{Radius} &
\multicolumn{1}{c}{$v$ (km/s)} &
\multicolumn{1}{c}{e${}\_{}v$} &
\multicolumn{1}{c}{$\sigma$ (km/s)} &
\multicolumn{1}{c}{e${}\_\sigma$} &
\multicolumn{1}{c}{$h_3$} &
\multicolumn{1}{c}{e${}\_{h_3}$} &
\multicolumn{1}{c}{$h_4$} & 
\multicolumn{1}{c}{e${}\_h_4$}  \\
\noalign{\smallskip}
\hline
\noalign{\smallskip}

0.000 & 0.000 & 13.584 & 335.607 & 13.451 & 0.059 & 0.028 & 0.021 & 0.035 \\
0.560 & 64.773 & 13.550 & 330.543 & 13.831 & -0.019 & 0.027 & 0.003 & 0.034 \\
1.120 & 41.025 & 12.329 & 327.415 & 12.543 & -0.029 & 0.025 & 0.001 & 0.031 \\
1.680 & 72.978 & 12.977 & 324.131 & 13.110 & -0.010 & 0.027 & 0.006 & 0.033 \\
2.240 & 93.515 & 13.384 & 317.287 & 13.243 & -0.022 & 0.029 & 0.014 & 0.036 \\
2.800 & 88.161 & 12.975 & 315.420 & 12.818 & -0.051 & 0.029 & 0.008 & 0.035 \\
3.360 & 94.115 & 13.326 & 319.895 & 13.020 & -0.059 & 0.029 & 0.017 & 0.036 \\
3.920 & 83.247 & 13.626 & 311.288 & 13.196 & -0.052 & 0.031 & 0.020 & 0.038 \\
4.480 & 77.888 & 12.533 & 311.528 & 12.755 & -0.016 & 0.028 & -0.008 & 0.034 \\
5.040 & 50.191 & 13.194 & 319.595 & 13.135 & -0.006 & 0.028 & 0.013 & 0.035 \\
5.600 & 57.395 & 13.182 & 321.551 & 13.124 & -0.007 & 0.028 & 0.015 & 0.035 \\
6.160 & 30.099 & 12.318 & 311.903 & 12.207 & -0.018 & 0.027 & 0.010 & 0.034 \\
6.720 & 45.402 & 11.682 & 315.057 & 11.229 & 0.031 & 0.026 & 0.032 & 0.032 \\
7.280 & 49.54 & 11.678 & 320.937 & 11.378 & 0.020 & 0.025 & 0.028 & 0.031 \\
7.840 & 38.143 & 11.796 & 309.733 & 11.779 & 0.016 & 0.026 & 0.004 & 0.032 \\
8.400 & 50.055 & 13.290 & 316.146 & 12.994 & 0.053 & 0.029 & 0.015 & 0.036 \\
8.960 & 48.897 & 13.160 & 327.410 & 13.156 & 0.037 & 0.027 & 0.013 & 0.034 \\
9.520 & 27.645 & 12.004 & 306.298 & 12.232 & 0.022 & 0.027 & -0.012 & 0.033 \\
10.080 & 14.984 & 11.952 & 311.881 & 11.556 & 0.043 & 0.027 & 0.023 & 0.033 \\
10.640 & 17.1 & 11.947 & 303.106 & 11.402 & 0.063 & 0.028 & 0.023 & 0.034 \\
11.200 & 18.861 & 11.859 & 316.752 & 12.039 & 0.033 & 0.026 & -0.006 & 0.031 \\
11.760 & 2.951 & 11.570 & 316.747 & 11.553 & 0.046 & 0.025 & 0.003 & 0.031 \\
12.320 & 10.752 & 12.174 & 311.036 & 12.311 & 0.028 & 0.027 & -0.005 & 0.033 \\
12.880 & -4.117 & 10.728 & 301.099 & 10.872 & 0.046 & 0.025 & -0.013 & 0.031 \\
13.440 & -7.849 & 10.265 & 302.861 & 9.980 & 0.040 & 0.024 & 0.016 & 0.029 \\
14.000 & -1.452 & 12.259 & 316.277 & 12.168 & 0.020 & 0.026 & 0.012 & 0.033 \\
14.560 & -0.432 & 11.647 & 305.249 & 11.583 & 0.016 & 0.026 & 0.005 & 0.032 \\
15.120 & -3.782 & 13.528 & 314.441 & 12.288 & 0.052 & 0.031 & 0.066 & 0.040 \\
15.680 & 15.954 & 10.125 & 304.898 & 10.658 & 0.050 & 0.024 & -0.042 & 0.029 \\
16.240 & 2.923 & 12.771 & 296.385 & 12.075 & 0.045 & 0.031 & 0.031 & 0.038 \\
17.073 & -3.435 & 10.380 & 301.319 & 9.704 & 0.043 & 0.025 & 0.040 & 0.030 \\
18.194 & -16.026 & 10.976 & 299.967 & 10.196 & 0.046 & 0.026 & 0.043 & 0.032 \\
19.314 & -26.115 & 10.609 & 302.258 & 9.316 & 0.055 & 0.026 & 0.080 & 0.033 \\
20.436 & -27.453 & 12.201 & 308.059 & 11.141 & 0.068 & 0.029 & 0.054 & 0.036 \\
21.551 & -28.393 & 12.007 & 303.493 & 10.798 & 0.055 & 0.029 & 0.064 & 0.036 \\
22.674 & -33.819 & 10.082 & 292.630 & 8.810 & 0.084 & 0.026 & 0.072 & 0.032 \\
23.795 & -41.113 & 11.382 & 294.164 & 10.656 & 0.090 & 0.029 & 0.025 & 0.034 \\
24.915 & -33.013 & 12.538 & 298.468 & 11.986 & 0.039 & 0.030 & 0.026 & 0.037 \\
27.988 & -36.619 & 12.789 & 287.368 & 11.879 & 0.040 & 0.033 & 0.042 & 0.039 \\
29.669 & -42.04 & 10.953 & 297.999 & 10.392 & 0.076 & 0.027 & 0.022 & 0.032 \\
31.350 & -33.386 & 11.511 & 300.428 & 10.263 & 0.096 & 0.029 & 0.058 & 0.035 \\
33.303 & -46.175 & 10.867 & 277.943 & 10.008 & 0.045 & 0.029 & 0.047 & 0.035 \\
35.541 & -48.5 & 12.146 & 275.367 & 11.475 & 0.106 & 0.036 & 0.019 & 0.040 \\
37.782 & -61.208 & 11.763 & 273.917 & 10.507 & 0.049 & 0.033 & 0.065 & 0.039 \\
40.294 & -57.907 & 12.915 & 260.229 & 11.573 & 0.106 & 0.041 & 0.058 & 0.045 \\
43.106 & -76.088 & 9.915 & 234.690 & 9.735 & 0.032 & 0.035 & 0.045 & 0.038 \\
46.136 & -54.036 & 11.510 & 234.418 & 12.348 & 0.085 & 0.043 & -0.004 & 0.047 \\
49.795 & -51.694 & 13.862 & 255.001 & 11.893 & 0.097 & 0.045 & 0.091 & 0.051 \\
53.928 & -89.869 & 14.058 & 200.652 & 26.597 & 0.238 & 0.071 & 0.177 & 0.069 \\
58.963 & -97.594 & 15.891 & 199.162 & 27.492 & 0.158 & 0.075 & 0.173 & 0.075 \\
65.327 & -120.835 & 17.883 & 183.567 & 22.703 & 0.199 & 0.094 & 0.239 & 0.094 \\
73.939 & -82.389 & 23.726 & 135.026 & 32.675 & 0.007 & 0.142 & 0.522 & 0.148 \\
86.244 & -108.638 & 37.625 & 118.830 & 69.486 & -0.020 & 0.247 & 0.398 & 0.234 \\

\noalign{\smallskip}
\hline
\noalign{\medskip}
\end{tabular}
\end{center}
\end{table*}

\begin{table*}
\caption{Kinematic data of IC~3370 (major axis)}
\begin{center}
\begin{tabular}{rrrrrrrrr}
\hline
\noalign{\smallskip}
\multicolumn{1}{c}{Radius} &
\multicolumn{1}{c}{$v$ (km/s)} &
\multicolumn{1}{c}{e${}\_{}v$} &
\multicolumn{1}{c}{$\sigma$ (km/s)} &
\multicolumn{1}{c}{e${}\_\sigma$} &
\multicolumn{1}{c}{$h_3$} &
\multicolumn{1}{c}{e${}\_{h_3}$} &
\multicolumn{1}{c}{$h_4$} & 
\multicolumn{1}{c}{e${}\_h_4$}  \\
\noalign{\smallskip}
\hline
\noalign{\smallskip}  

  -118.709 & -97.283 & 33.982 & 108.359 & 47.825 & -0.102 & 0.306 & -0.009 & 0.346\\
 -76.994 & -98.17 & 13.729 & 149.995 & 15.174 & -0.020 & 0.082 & 0.054 & 0.088 \\
 -58.261 & -108.416 & 13.138 & 172.560 & 12.817 & 0.040 & 0.065 & 0.070 & 0.068 \\
 -46.710 & -131.07 & 10.158 & 157.412 & 11.417 & 0.093 & 0.060 & 0.023 & 0.064 \\
 -38.513 & -118.186 & 9.072 & 161.628 & 10.925 & 0.077 & 0.052 & -0.015 & 0.058 \\
 -31.791 & -111.942 & 9.111 & 152.370 & 10.386 & 0.086 & 0.056 & 0.027 & 0.059 \\
 -26.880 & -148.232 & 10.553 & 138.989 & 16.823 & 0.072 & 0.080 & -0.138 & 0.112 \\
 -23.520 & -97.746 & 9.498 & 157.489 & 10.566 & 0.043 & 0.054 & 0.035 & 0.058 \\
 -20.160 & -94.81 & 10.537 & 186.111 & 11.442 & 0.047 & 0.048 & 0.000 & 0.051 \\
 -16.800 & -77.852 & 10.333 & 183.333 & 9.788 & 0.047 & 0.048 & 0.066 & 0.049 \\
 -13.440 & -82.048 & 10.792 & 193.314 & 11.130 & 0.033 & 0.046 & 0.015 & 0.048 \\
 -10.080 & -70.944 & 9.994 & 189.044 & 9.632 & 0.036 & 0.044 & 0.050 & 0.045 \\
 -6.720 & -69.167 & 11.289 & 205.146 & 10.671 & 0.007 & 0.044 & 0.041 & 0.045 \\
 -3.360 & -60.06 & 11.441 & 208.147 & 10.757 & 0.011 & 0.044 & 0.040 & 0.045 \\
 0.000 & 0.000 & 11.151 & 208.537 & 10.793 & 0.050 & 0.043 & 0.025 & 0.044 \\
 3.360 & 67.06 & 10.770 & 199.582 & 10.236 & 0.026 & 0.044 & 0.044 & 0.045 \\
 6.720 & 96.202 & 11.002 & 199.779 & 11.602 & -0.028 & 0.045 & -0.001 & 0.048 \\
 10.080 & 113.86 & 10.915 & 204.124 & 10.313 & -0.041 & 0.043 & 0.041 & 0.044 \\
 13.440 & 114.294 & 9.822 & 185.917 & 10.724 & -0.024 & 0.045 & 0.000 & 0.048 \\
 16.800 & 125.151 & 9.606 & 185.910 & 11.544 & -0.038 & 0.045 & -0.045 & 0.051 \\
 20.160 & 127.636 & 9.426 & 181.309 & 11.575 & -0.013 & 0.045 & -0.047 & 0.052 \\
 23.520 & 127.365 & 8.849 & 170.744 & 10.957 & -0.059 & 0.047 & -0.041 & 0.054 \\
 26.880 & 129.717 & 9.302 & 168.540 & 11.390 & -0.023 & 0.049 & -0.028 & 0.056 \\
 30.240 & 139.625 & 9.165 & 156.671 & 12.257 & -0.092 & 0.058 & -0.066 & 0.068 \\
 35.141 & 133.451 & 8.889 & 148.509 & 10.985 & -0.010 & 0.055 & 0.005 & 0.062 \\
 41.864 & 121.188 & 8.819 & 146.365 & 12.311 & 0.008 & 0.057 & -0.051 & 0.071 \\
 50.100 & 146.844 & 8.659 & 154.099 & 12.364 & -0.034 & 0.054 & -0.083 & 0.069 \\
 61.694 & 135.349 & 9.836 & 145.376 & 14.707 & 0.033 & 0.067 & -0.091 & 0.088 \\
 79.294 & 122.278 & 12.685 & 124.982 & 15.916 & 0.073 & 0.097 & 0.033 & 0.104 \\
 116.432 & 124.591 & 18.182 & 99.847 & 30.532 & 0.093 & 0.184 & -0.102 & 0.247 \\

\noalign{\smallskip}
\hline
\noalign{\medskip}
\end{tabular}
\end{center}
\end{table*}

\begin{table*}
\caption{Kinematic data of IC~3370 (minor axis)}
\begin{center}
\begin{tabular}{rrrrrrrrr}
\hline
\noalign{\smallskip}
\multicolumn{1}{c}{Radius} &
\multicolumn{1}{c}{$v$ (km/s)} &
\multicolumn{1}{c}{e${}\_{}v$} &
\multicolumn{1}{c}{$\sigma$ (km/s)} &
\multicolumn{1}{c}{e${}\_\sigma$} &
\multicolumn{1}{c}{$h_3$} &
\multicolumn{1}{c}{e${}\_{h_3}$} &
\multicolumn{1}{c}{$h_4$} & 
\multicolumn{1}{c}{e${}\_h_4$}  \\
\noalign{\smallskip}
\hline
\noalign{\smallskip}  

-80.156 & 41.732 & 15.472 & 185.858 & 20.348 & -0.170 & 0.103 & -0.305 & 0.149 \\
-61.574 & 30.78 & 17.496 & 233.660 & 11.910 & -0.133 & 0.065 & 0.170 & 0.068 \\
-50.043 & 54.865 & 15.371 & 183.722 & 13.351 & -0.067 & 0.071 & 0.107 & 0.073 \\
-41.662 & 1.02 & 21.828 & 201.291 & 20.062 & -0.058 & 0.089 & 0.057 & 0.091 \\
-35.373 & 29.818 & 19.630 & 232.101 & 18.222 & -0.243 & 0.093 & -0.038 & 0.080 \\
-30.240 & 34.231 & 14.851 & 186.155 & 13.909 & -0.105 & 0.070 & 0.062 & 0.070 \\
-26.880 & 21.567 & 14.278 & 182.174 & 11.897 & 0.016 & 0.065 & 0.133 & 0.068 \\
-23.520 & 40.021 & 14.302 & 183.965 & 12.714 & 0.056 & 0.066 & 0.096 & 0.067 \\
-20.160 & 48.205 & 15.211 & 230.440 & 16.886 & 0.035 & 0.052 & -0.050 & 0.057 \\
-16.800 & 37.463 & 15.702 & 216.157 & 17.973 & -0.020 & 0.058 & -0.051 & 0.065 \\
-13.440 & 35.915 & 13.037 & 205.530 & 15.336 & -0.081 & 0.056 & -0.064 & 0.061 \\
-10.080 & 36.775 & 15.706 & 220.740 & 18.126 & -0.078 & 0.061 & -0.070 & 0.067 \\
-6.720 & 34.023 & 12.548 & 216.791 & 13.353 & 0.033 & 0.046 & -0.020 & 0.049 \\
-3.360 & 35.387 & 12.093 & 206.551 & 12.634 & 0.025 & 0.047 & -0.002 & 0.050 \\
 0.000 & 0.000 & 11.312 & 223.119 & 10.956 & 0.072 & 0.041 & 0.011 & 0.041 \\
3.360 & -15.449 & 11.714 & 216.859 & 10.655 & 0.043 & 0.043 & 0.047 & 0.044 \\
6.720 & -32.512 & 11.494 & 208.624 & 11.032 & 0.038 & 0.044 & 0.032 & 0.045 \\
10.080 & -52.446 & 10.533 & 193.031 & 10.462 & 0.051 & 0.046 & 0.033 & 0.047 \\
13.440 & -45.915 & 11.340 & 202.402 & 12.168 & -0.040 & 0.046 & -0.011 & 0.049 \\
16.800 & -54.17 & 11.744 & 188.615 & 12.172 & -0.051 & 0.053 & 0.019 & 0.055 \\
20.160 & -38.137 & 14.072 & 213.722 & 15.152 & -0.073 & 0.055 & -0.028 & 0.057 \\
23.520 & -40.025 & 15.812 & 210.258 & 15.055 & -0.002 & 0.060 & 0.035 & 0.062 \\
26.880 & -78.507 & 13.215 & 178.116 & 13.214 & 0.035 & 0.063 & 0.051 & 0.066 \\
30.240 & -44.075 & 21.753 & 245.018 & 17.623 & 0.011 & 0.067 & 0.079 & 0.070 \\
35.131 & -52.099 & 13.435 & 181.799 & 13.803 & 0.070 & 0.064 & 0.030 & 0.066 \\
41.859 & -32.566 & 14.869 & 201.688 & 14.593 & 0.076 & 0.062 & 0.025 & 0.063 \\
50.024 & -42.578 & 14.372 & 174.449 & 13.211 & 0.020 & 0.070 & 0.098 & 0.072 \\
62.955 & 6.104 & 17.907 & 212.082 & 19.989 & 0.084 & 0.073 & -0.046 & 0.077 \\
87.898 & -14.423 & 17.562 & 156.723 & 22.006 & -0.072 & 0.106 & -0.024 & 0.120 \\

\noalign{\smallskip}
\hline
\noalign{\medskip}
\end{tabular}
\end{center}
\end{table*}

\begin{table*}
\caption{Kinematic data of IC~3370 (intermediate axis)}
\begin{center}
\begin{tabular}{rrrrrrrrr}
\hline
\noalign{\smallskip}
\multicolumn{1}{c}{Radius} &
\multicolumn{1}{c}{$v$ (km/s)} &
\multicolumn{1}{c}{e${}\_{}v$} &
\multicolumn{1}{c}{$\sigma$ (km/s)} &
\multicolumn{1}{c}{e${}\_\sigma$} &
\multicolumn{1}{c}{$h_3$} &
\multicolumn{1}{c}{e${}\_{h_3}$} &
\multicolumn{1}{c}{$h_4$} & 
\multicolumn{1}{c}{e${}\_h_4$}  \\
\noalign{\smallskip}
\hline
\noalign{\smallskip}

  -59.788 & -97.318 & 11.334 & 180.295 & 16.307 & 0.052 & 0.059 & -0.142 & 0.082 \\
 -46.693 & -103.027 & 9.992 & 133.376 & 11.399 & 0.006 & 0.068 & 0.069 & 0.072 \\
 -38.494 & -103.169 & 10.171 & 147.593 & 14.390 & -0.018 & 0.066 & -0.062 & 0.083 \\
 -31.754 & -100.368 & 9.884 & 157.162 & 12.581 & 0.079 & 0.060 & -0.037 & 0.068 \\
 -26.880 & -99.208 & 10.273 & 166.574 & 14.369 & 0.027 & 0.057 & -0.095 & 0.074 \\
 -23.520 & -102.625 & 10.907 & 156.696 & 13.198 & 0.038 & 0.063 & -0.004 & 0.071 \\
 -20.160 & -87.717 & 10.239 & 162.518 & 14.397 & 0.064 & 0.061 & -0.100 & 0.077 \\
 -16.800 & -84.928 & 10.636 & 174.007 & 14.551 & 0.078 & 0.059 & -0.107 & 0.073 \\
 -13.440 & -87.972 & 10.301 & 170.535 & 12.116 & 0.051 & 0.054 & -0.016 & 0.059 \\
 -10.080 & -81.43 & 9.921 & 169.085 & 11.912 & 0.061 & 0.053 & -0.025 & 0.059 \\
 -6.720 & -74.647 & 10.695 & 173.391 & 12.346 & 0.053 & 0.055 & -0.012 & 0.060 \\
 -3.360 & -47.07 & 11.336 & 190.177 & 12.093 & 0.047 & 0.050 & 0.002 & 0.053 \\
 0.000 & 0.000 & 11.413 & 198.156 & 12.091 & 0.077 & 0.049 & -0.008 & 0.051 \\
  3.360 & 48.517 & 10.996 & 196.522 & 13.149 & 0.021 & 0.047 & -0.053 & 0.054 \\
 6.720 & 94.703 & 10.941 & 188.287 & 12.965 & -0.019 & 0.049 & -0.040 & 0.056 \\
 10.080 & 107.835 & 10.712 & 183.460 & 13.909 & -0.019 & 0.051 & -0.078 & 0.063 \\
 13.440 & 115.963 & 10.342 & 185.005 & 13.115 & -0.036 & 0.049 & -0.069 & 0.058 \\
 16.800 & 132.163 & 9.937 & 176.510 & 13.251 & -0.002 & 0.050 & -0.082 & 0.063 \\
 20.160 & 133.392 & 9.804 & 175.128 & 13.410 & -0.053 & 0.052 & -0.100 & 0.066 \\
 23.520 & 129.282 & 9.888 & 169.665 & 14.142 & -0.083 & 0.058 & -0.134 & 0.076 \\
 26.880 & 128.456 & 9.650 & 169.223 & 14.866 & -0.017 & 0.056 & -0.177 & 0.084 \\
 31.771 & 139.59 & 9.432 & 159.740 & 14.169 & -0.005 & 0.057 & -0.129 & 0.080 \\
 38.494 & 139.441 & 9.034 & 142.028 & 11.807 & -0.149 & 0.067 & -0.047 & 0.074 \\
 46.683 & 140.017 & 10.226 & 150.049 & 14.515 & -0.091 & 0.069 & -0.089 & 0.086 \\
 59.581 & 134.721 & 12.377 & 163.816 & 15.429 & -0.152 & 0.079 & -0.067 & 0.085 \\
 84.130 & 115.096 & 12.181 & 144.219 & 17.536 & -0.150 & 0.094 & -0.128 & 0.116 \\

\noalign{\smallskip}
\hline
\noalign{\medskip}
\end{tabular}
\end{center}
\end{table*}

\begin{table*}
\caption{Kinematic data of NGC~3379 (major axis)}
\begin{center}
\begin{tabular}{rrrrrrrrr}
\hline
\noalign{\smallskip}
\multicolumn{1}{c}{Radius} &
\multicolumn{1}{c}{$v$ (km/s)} &
\multicolumn{1}{c}{e${}\_{}v$} &
\multicolumn{1}{c}{$\sigma$ (km/s)} &
\multicolumn{1}{c}{e${}\_\sigma$} &
\multicolumn{1}{c}{$h_3$} &
\multicolumn{1}{c}{e${}\_{h_3}$} &
\multicolumn{1}{c}{$h_4$} & 
\multicolumn{1}{c}{e${}\_h_4$}  \\
\noalign{\smallskip}
\hline
\noalign{\smallskip}  

 -25.879 & 63.194 & 8.513 & 162.073 & 7.368 & -0.090 & 0.045 & 0.109 & 0.046 \\
-14.051 & 68.302 & 7.637 & 179.704 & 7.898 & -0.073 & 0.036 & 0.028 & 0.037 \\
-8.753 & 56.908 & 4.830 & 176.622 & 5.180 & -0.012 & 0.023 & 0.022 & 0.024 \\
-6.255 & 52.234 & 5.303 & 183.786 & 5.603 & 0.015 & 0.024 & 0.024 & 0.025 \\
-4.809 & 51.106 & 5.572 & 199.728 & 5.931 & -0.015 & 0.023 & 0.014 & 0.024 \\
-3.624 & 42.337 & 5.403 & 207.183 & 5.545 & 0.002 & 0.021 & 0.026 & 0.022 \\
-2.773 & 45.139 & 5.834 & 214.089 & 5.978 & -0.011 & 0.022 & 0.023 & 0.023 \\
-2.183 & 39.724 & 5.935 & 219.934 & 5.964 & -0.019 & 0.021 & 0.029 & 0.023 \\
-1.593 & 35.234 & 5.928 & 225.600 & 6.096 & -0.010 & 0.021 & 0.016 & 0.022 \\
-1.003 & 29.393 & 5.767 & 224.304 & 5.916 & -0.003 & 0.020 & 0.018 & 0.022 \\
-0.413 & 19.156 & 5.681 & 217.610 & 5.945 & 0.016 & 0.021 & 0.012 & 0.022 \\
-0.177 & 5.516 & 5.703 & 223.417 & 6.094 & 0.013 & 0.020 & 0.001 & 0.022 \\ 
0.767 & -3.167 & 5.601 & 217.749 & 5.619 & 0.016 & 0.020 & 0.030 & 0.022 \\
1.357 & -9.204 & 5.750 & 208.990 & 5.938 & 0.024 & 0.022 & 0.022 & 0.023 \\
1.947 & -11.184 & 6.080 & 203.897 & 6.422 & 0.040 & 0.024 & 0.013 & 0.026 \\
2.798 & -18.15 & 5.505 & 201.697 & 5.764 & 0.028 & 0.022 & 0.019 & 0.023 \\
3.983 & -24.77 & 5.559 & 196.204 & 5.828 & 0.032 & 0.023 & 0.021 & 0.024 \\
5.428 & -29.387 & 5.263 & 184.628 & 5.622 & 0.010 & 0.023 & 0.020 & 0.025 \\
7.704 & -30.205 & 4.881 & 175.466 & 5.150 & 0.025 & 0.023 & 0.028 & 0.024 \\
12.261 & -45.279 & 5.224 & 171.950 & 5.111 & 0.027 & 0.025 & 0.060 & 0.026 \\
20.582 & -60.748 & 7.666 & 155.531 & 8.560 & 0.033 & 0.042 & 0.011 & 0.043 \\

\noalign{\smallskip}
\hline
\noalign{\medskip}
\end{tabular}
\end{center}
\end{table*}

\begin{table*}
\caption{Kinematic data of NGC~4105 (major axis)}
\begin{center}
\begin{tabular}{rrrrrrrrr}
\hline
\noalign{\smallskip}
\multicolumn{1}{c}{Radius} &
\multicolumn{1}{c}{$v$ (km/s)} &
\multicolumn{1}{c}{e${}\_{}v$} &
\multicolumn{1}{c}{$\sigma$ (km/s)} &
\multicolumn{1}{c}{e${}\_\sigma$} &
\multicolumn{1}{c}{$h_3$} &
\multicolumn{1}{c}{e${}\_{h_3}$} &
\multicolumn{1}{c}{$h_4$} & 
\multicolumn{1}{c}{e${}\_h_4$}  \\
\noalign{\smallskip}
\hline
\noalign{\smallskip}  

 -22.585 & 31.695 & 62.504 & 260.281 & 84.348 & 0.134 & 0.246 & -0.143 & 0.310 \\
 -14.673 & 4.87 & 33.698 & 234.433 & 42.807 & 0.095 & 0.136 & -0.051 & 0.154 \\
 -10.619 & 35.136 & 23.159 & 253.402 & 27.209 & 0.072 & 0.081 & -0.026 & 0.090 \\
 -8.165 & 12.979 & 19.660 & 259.904 & 21.912 & 0.069 & 0.066 & -0.008 & 0.071 \\
 -6.519 & 24.065 & 17.210 & 262.390 & 18.105 & 0.016 & 0.055 & 0.020 & 0.058 \\
 -5.356 & 27.889 & 14.490 & 262.859 & 17.444 & 0.027 & 0.047 & -0.039 & 0.055 \\
 -4.347 & 20.754 & 14.195 & 257.176 & 15.235 & -0.010 & 0.047 & 0.017 & 0.050 \\
 -3.520 & 11.619 & 15.405 & 285.667 & 17.093 & 0.021 & 0.044 & -0.021 & 0.049 \\
 -2.847 & 21.285 & 14.241 & 284.228 & 15.041 & 0.025 & 0.041 & 0.001 & 0.044 \\
 -2.173 & 15.245 & 14.320 & 283.497 & 14.142 & 0.027 & 0.041 & 0.031 & 0.043 \\
 -1.680 & 10.452 & 16.009 & 310.394 & 16.202 & -0.001 & 0.040 & 0.004 & 0.044 \\
 -1.344 & 4.165 & 15.589 & 307.387 & 15.176 & 0.032 & 0.040 & 0.022 & 0.042 \\
 -1.008 & 3.815 & 16.464 & 298.603 & 18.045 & 0.055 & 0.045 & -0.028 & 0.050 \\
 -0.672 & 7.031 & 16.050 & 298.183 & 16.030 & -0.003 & 0.042 & 0.017 & 0.046 \\
 -0.336 & 15.546 & 14.806 & 304.751 & 16.300 & -0.022 & 0.039 & -0.030 & 0.044 \\
 0.000 & 0.000 & 14.837 & 298.764 & 18.337 & -0.020 & 0.041 & -0.079 & 0.051 \\
 0.336 & 1.583 & 17.086 & 313.825 & 18.558 & -0.039 & 0.043 & -0.031 & 0.049 \\
 0.672 & -5.221 & 15.298 & 306.415 & 16.960 & -0.044 & 0.040 & -0.037 & 0.046 \\
 1.008 & 14.739 & 15.538 & 308.173 & 15.828 & -0.067 & 0.041 & -0.003 & 0.044 \\
 1.344 & 28.992 & 15.082 & 285.592 & 14.677 & -0.013 & 0.042 & 0.036 & 0.045 \\
 1.680 & 28.555 & 15.058 & 299.200 & 17.439 & -0.058 & 0.042 & -0.054 & 0.049 \\
 2.016 & 4.331 & 16.399 & 301.758 & 18.116 & -0.026 & 0.043 & -0.030 & 0.049 \\
 2.352 & 50.178 & 14.001 & 278.795 & 16.299 & 0.001 & 0.042 & -0.036 & 0.048 \\
 2.688 & 32.482 & 13.129 & 254.500 & 13.943 & -0.020 & 0.044 & 0.023 & 0.046 \\
 3.181 & 38.352 & 13.591 & 277.016 & 14.385 & 0.043 & 0.041 & 0.004 & 0.044 \\
 3.855 & 21.539 & 14.533 & 295.552 & 16.118 & -0.029 & 0.040 & -0.028 & 0.045 \\
 4.528 & 28.619 & 14.184 & 273.435 & 16.533 & -0.032 & 0.044 & -0.034 & 0.050 \\
 5.356 & 41.094 & 15.362 & 280.693 & 16.450 & -0.001 & 0.045 & -0.001 & 0.049 \\
 6.516 & 26.132 & 14.800 & 261.816 & 17.802 & 0.032 & 0.049 & -0.038 & 0.056 \\
 8.017 & 31.639 & 17.392 & 279.589 & 19.685 & 0.032 & 0.052 & -0.026 & 0.058 \\
 9.996 & 31.828 & 18.018 & 262.988 & 21.318 & 0.045 & 0.059 & -0.033 & 0.067 \\
 12.943 & 40.285 & 25.093 & 262.602 & 27.882 & 0.075 & 0.083 & -0.010 & 0.089 \\
 18.002 & 66.794 & 33.495 & 249.708 & 45.852 & 0.014 & 0.120 & -0.085 & 0.154 \\
 27.268 & 49.384 & 45.562 & 248.060 & 63.312 & 0.018 & 0.166 & -0.092 & 0.216 \\

\noalign{\smallskip}
\hline
\noalign{\medskip}
\end{tabular}
\end{center}
\end{table*}

\begin{table*}
\caption{Kinematic data of NGC~4105 (minor axis)}
\begin{center}
\begin{tabular}{rrrrrrrrr}
\hline
\noalign{\smallskip}
\multicolumn{1}{c}{Radius} &
\multicolumn{1}{c}{$v$ (km/s)} &
\multicolumn{1}{c}{e${}\_{}v$} &
\multicolumn{1}{c}{$\sigma$ (km/s)} &
\multicolumn{1}{c}{e${}\_\sigma$} &
\multicolumn{1}{c}{$h_3$} &
\multicolumn{1}{c}{e${}\_{h_3}$} &
\multicolumn{1}{c}{$h_4$} & 
\multicolumn{1}{c}{e${}\_h_4$}  \\
\noalign{\smallskip}
\hline
\noalign{\smallskip}

   -9.613 & 66.253 & 56.865 & 287.610 & 63.660 & -0.164 & 0.240 & -0.404 & 0.377 \\
   -6.800 & -11.076 & 27.873 & 243.361 & 32.630 & 0.106 & 0.106 & -0.024 & 0.114 \\
   -5.167 & 41.767 & 37.886 & 309.688 & 28.733 & 0.076 & 0.099 & 0.139 & 0.105 \\
   -4.009 & 22.206 & 24.338 & 283.535 & 34.077 & 0.001 & 0.075 & -0.136 & 0.109 \\
   -3.180 & 9.865 & 28.872 & 315.807 & 26.594 & 0.044 & 0.071 & 0.041 & 0.075 \\
   -2.504 & 25.53 & 15.694 & 288.956 & 15.782 & -0.036 & 0.044 & 0.018 & 0.047 \\
   -2.016 & 25.516 & 18.632 & 284.057 & 22.655 & 0.073 & 0.057 & -0.070 & 0.068 \\
   -1.680 & 68.354 & 23.839 & 327.022 & 16.758 & 0.100 & 0.060 & 0.175 & 0.066 \\
   -1.344 & 2.133 & 16.576 & 303.480 & 16.686 & -0.044 & 0.043 & 0.008 & 0.047 \\
   -1.008 & 11.083 & 18.115 & 317.555 & 19.687 & 0.075 & 0.046 & -0.041 & 0.052 \\
   -0.672 & 14.812 & 14.372 & 288.949 & 13.448 & -0.047 & 0.040 & 0.049 & 0.042 \\
   -0.336 & -7.478 & 15.040 & 318.497 & 13.816 & -0.052 & 0.037 & 0.040 & 0.039 \\
   0.000 & 0.000 & 14.757 & 303.940 & 15.209 & 0.015 & 0.038 & 0.000 & 0.042 \\
   0.336 & -3.794 & 15.515 & 312.880 & 13.146 & -0.034 & 0.039 & 0.082 & 0.041 \\
   0.672 & -5.044 & 19.103 & 327.555 & 17.912 & -0.008 & 0.044 & 0.030 & 0.048 \\
   1.008 & -13.082 & 17.626 & 301.887 & 15.988 & -0.019 & 0.046 & 0.057 & 0.048 \\
   1.344 & -31.054 & 15.749 & 277.036 & 16.262 & 0.006 & 0.046 & 0.017 & 0.049 \\
   1.680 & -30.863 & 17.706 & 305.189 & 19.507 & 0.005 & 0.046 & -0.030 & 0.052 \\
   2.169 & -7.328 & 24.048 & 292.992 & 18.893 & -0.001 & 0.066 & 0.133 & 0.069 \\
   2.844 & -55.537 & 53.998 & 238.476 & 26.755 & 0.057 & 0.277 & 0.760 & 0.297 \\
   3.668 & -5.908 & 17.948 & 280.591 & 24.505 & -0.008 & 0.055 & -0.115 & 0.076 \\
   4.829 & 16.595 & 19.588 & 267.684 & 23.580 & 0.046 & 0.063 & -0.045 & 0.073 \\
   6.466 & 58.218 & 30.208 & 278.722 & 34.763 & -0.070 & 0.093 & -0.039 & 0.104 \\
   9.289 & 14.762 & 67.459 & 223.187 & 108.529 & 0.019 & 0.327 & -0.293 & 0.556 \\

\noalign{\smallskip}
\hline
\noalign{\medskip}
\end{tabular}
\end{center}
\end{table*}

\label{lastpage}

\begin{thebibliography}{99}

\bibitem[\protect\citeauthoryear{Bendo \& Barnes}{2000}]{bb00}Bendo G.J., Barnes
 J.E., 2000, MNRAS, 316, 315

\bibitem[\protect\citeauthoryear{Binney}{2004}]{bin04}Binney J.J., 2004,  in  Ryder S.,  Pisano D.J., Walker M., Freeman K., eds., 
ASP Conf. Ser. Vol.  220, Dark Matter in Galaxies, p.3

\bibitem[\protect\citeauthoryear{Binney \& Mamon}{1982}]{bm82}Binney J.J., Mamon, 1982,   MNRAS, 200, 361

\bibitem[\protect\citeauthoryear{Binney \& Merrifield}{1998}]{BM98}Binney J.J., Merrifield M.R., 1998,  {Galactic Astronomy},  Princeton Univ. Press, Princeton, NJ

\bibitem[\protect\citeauthoryear{Binney \& Tremaine}{1987}]{BT87}Binney J.J., Tremaine S., 1987,  {Galactic Dynamics},  Princeton Univ. Press, Princeton, NJ

\bibitem[\protect\citeauthoryear{BDI}{1990}]{bdi}Binney J.J., Davies R.D., Illingworth G.D.,  1990, ApJ, {361}, 78 (BDI)

\bibitem[\protect\citeauthoryear{Bridges et al.}{2003}]{bri03}Bridges T. et al., 2003, in press, in  Bridges T.,  Forbes D., eds,  IAU General Assembly, July 2003,
Joint Discussion 6: Extragalactic Globular Clusters and their  Host Galaxies,    {preprint (astro-ph/0310324)}

\bibitem[\protect\citeauthoryear{Brown,  \& Bregman}{1998}]{bb98}Brown, B.A., Bregman J.N., 1998, ApJ, 495, L75

\bibitem[\protect\citeauthoryear{Brown,  \& Bregman}{2001}]{bb01}Brown B.A., Bregman J.N: 2001, ApJ, 547, 154

\bibitem[\protect\citeauthoryear{Cap etal}{1990}]{cap00}Capaccioli M., Held E.V., Lorenz H., Vietri M., 1990,  AJ, 99, 1813

\bibitem[\protect\citeauthoryear{Cappellari, M.,   Verolme, E.K.,    van der Marel, R.P., et al}{2002}.]{cap02}Cappellari M., Verolme, E.K., van der Marel, R.P., Kleijn, G. A. Verdoes, Illingworth, G.D., Franx, M., Carollo, C.M., de Zeeuw, P.T., 2002, ApJ, 578, 787

\bibitem[\protect\citeauthoryear{Carollo,  Danziger \& Buson}{1993}]{car93}Carollo C.M., Danziger I.J., Buson L., 1993, MNRAS, 265, 553

\bibitem[\protect\citeauthoryear{Carollo et al.}{1995}]{car95}Carollo C.M., de Zeeuw P.T., van der Marel R.P., Danziger I.J., Qian E.E., 1995, ApJ, 441, L25

\bibitem[\protect\citeauthoryear{Ciardullo etal}{1993}]{cia93}Ciardullo R., Jacoby G.H., Dejonghe H.G., 1993, ApJ, 414, 454

\bibitem[\protect\citeauthoryear{Cinzano  \& van der Marel}{1994}]{cvdm94}Cinzano P., van der Marel, R.P., 1994, MNRAS, 270, 325 

\bibitem[\protect\citeauthoryear{Ciotti \& Pellegrini}{2004}]{cp04}Ciotti L., Pellegrini S., 2004, MNRAS,  350, 609.

\bibitem[\protect\citeauthoryear{C\^ot\'e et al.}{2003}]{cot03}{C\^ot\'e} P., McLaughlin D.E., Cohen J.G.,  Blakeslee J.P., 2003, 591, 850

\bibitem[\protect\citeauthoryear{Cretton, N. \& van der Bosch}{1999}]{cre99}Cretton N., van der Bosch F.C., 1999, ApJ, 514, 704

\bibitem[\protect\citeauthoryear{Cretton et al.}{2000}]{cre00}Cretton N., Rix H-W., de Zeeuw P.T., 2000,  ApJ, 536, 319

\bibitem[\protect\citeauthoryear{Danziger}{1997}]{dan97}Danziger I.J., 1997, in    Persic M.,   Salucci P., eds., ASP Conf. Ser. Vol.  117,
{Dark and Visible Matter in Galaxies,} p. 28

\bibitem[\protect\citeauthoryear{Davis \& White}{1996}]{dw96}Davis D.S., White R.E., 1996, ApJ,  470, L35

\bibitem[\protect\citeauthoryear{De Bruyne et al.}{2001}]{deb01}De Bruyne V., Dejonghe H., Pizzella A., Bernardi M., Zeilinger W.W., 2001, ApJ, 546,  903

\bibitem[\protect\citeauthoryear{De Rijcke et al.}{2003}]{der03}De Rijcke S., Dejonghe H., Zeilinger W.W., Hau G.K.T., 2003, A\& A, 400, 119
 
\bibitem[\protect\citeauthoryear{Fabbiano et al.}{2003}]{fab03}Fabbiano G. et al., 2003, ApJ, 588, 175

\bibitem[\protect\citeauthoryear{Fasano  \& Bonoli}{1989}]{fb89}Fasano G., Bonoli C.,  1989, A\& ASS,  {79}, 291

\bibitem[\protect\citeauthoryear{Fisher}{1997}]{fish97}Fisher D., 1997, AJ, 113, 950


\bibitem[\protect\citeauthoryear{Forbes, Reizel \& Williger}{1995}]{for95}Forbes D.A., Reizel D.B., Williger G.M., 1995, AJ, 109, 1576


\bibitem[\protect\citeauthoryear{Gebhardt et al}{2000}]{geb00}Gebhardt K. et al., 2000, AJ, 119, 1157 
 

\bibitem[\protect\citeauthoryear{Gerhard}{1993}]{ger93}Gerhard O., 1993, MNRAS, 265, 213

\bibitem[\protect\citeauthoryear{Gerhard et al.}{1998}]{ger98}Gerhard O., Jeske G., Saglia R.P., Bender R., 1998, MNRAS, 295, 197

\bibitem[\protect\citeauthoryear{Gregg et al.}{2004}]{gre04}Gregg, M.D.,
 Ferguson, H.C., Minniti, D., Tanvir, N., Catchpole, R., 2004, AJ, 127, 1441


\bibitem[\protect\citeauthoryear{Hernquist}{1990}]{hern90}Hernquist L., 1990, ApJ, 356, 359


\bibitem[\protect\citeauthoryear{Jaffe}{1983}]{jaf83}Jaffe W., 1983, MNRAS, 202, 995

\bibitem[\protect\citeauthoryear{Jarvis}{1987}]{jar87}Jarvis B., 1987, AJ, {94}, 30 (J87)

\bibitem[\protect\citeauthoryear{Kim \& Fabbiano}{1995}]{kf95}Kim D.-W., Fabbiano G., 1995, ApJ, 441, 182 (KF95)

\bibitem[\protect\citeauthoryear{Kohonen}{1997}]{koh97}Kohonen T., 1997, {Self-Organizing Maps}, Springer-Verlag, Berlin and Heidelberg


\bibitem[\protect\citeauthoryear{Kraft}{2003}]{kr2003}Kraft R.P., 
V\'azquez S.E., Forman W.R., Jones C., Murray S.S., Hardcastle M.J., 
Worrall D.M., Churazov E., 2003, ApJ, 592, 129 
\bibitem[\protect\citeauthoryear{Kronawitter et al.}{2000}]{kro00}Kronawitter A, Saglia R.P., Gerhard O., Bender R.,  2000, A\& AS, 144, 53

\bibitem[\protect\citeauthoryear{Loewenstein  \& White}{1999}]{LW99}Loewenstein M., White R.E., 1999, ApJ, 518, 50

\bibitem[\protect\citeauthoryear{Longhetti et al.}{1998}]{lon98}Longhetti M., Rampazzo R., Bressan A., Chiosi C., 1998, A\& AS, 130, 267 

\bibitem[\protect\citeauthoryear{Mathews  \& Brighenti}{2003}]{bm03}Mathews W.G., Brighenti F., 2003, ARA\& A, 41, 191

\bibitem[\protect\citeauthoryear{M\'endez et al.}{2001}]{men01}M\'endez, R.H., Riffeser, A., Kudritzki, R.-P., Matthias, M., Freeman, K.C., Arnaboldi, M., Capaccioli, M., Gerhard, O. E.,  2001, ApJ, 563, 135

 
\bibitem[\protect\citeauthoryear{Murtagh}{1995}]{mur95}Murtagh F., 1995, Pattern Recognition Letters, 16, 399

\bibitem[\protect\citeauthoryear{Peng, Ford \& Freeman}{2004}]{pen04}Peng E.W., Ford H.C., Freeman K.C., 2004, ApJ, 602, 685


\bibitem[\protect\citeauthoryear{Reduzzi \& Rampazzo}{1996}]{1996}Reduzzi L., Rampazzo R., 1996, A \& AS, 116, 515


\bibitem[\protect\citeauthoryear{Richtler et al. 2004}{2004}]{Ric04}
Richtler T. et al., 2004, AJ, 127,  2094


\bibitem[\protect\citeauthoryear{Rix et al.}{1993}]{rix97}Rix H.-W., de Zeeuw P.T., Cretton N., van der Marel R.P., Carollo C.M., 1997, ApJ, {488}, 702

\bibitem[\protect\citeauthoryear{Romanowsky et al.}{2003}]{rom03}Romanowsky A.J.,  Douglas N.G., Arnaboldi M., Kuijken K., Merrifield M.R., Napolitano N.R., Capaccioli M. \&  Freeman K.C., 2003, Sci, 5640,  1696

\bibitem[\protect\citeauthoryear{Saglia et al.}{1993}]{sag93}Saglia R.P.
 et al., 1993, ApJ, 403, 567

\bibitem[\protect\citeauthoryear{Saglia etal.}{2000}]{sag00}Saglia R.P., Kronawitter A., Gerhard O. \& Bender R., 2000, AJ, 119,  153

\bibitem[\protect\citeauthoryear{Samurovi{\'c},{\'C}irkovi{\'c} \& Milo{\v s}evi{\'c}-Zdjelar}{1999}]{SCMZ}Samurovi{\'c} S.,
{\'C}irkovi{\'c} M.M., Milo{\v s}evi{\'c}-Zdjelar V., 1999,
MNRAS, 309, 63

\bibitem[\protect\citeauthoryear{Sargent et al.}{1997}]{sar77}Sargent W.L.W., Schechter P.L., Boksenberg A., Shortridg, K., 1977, AJ, {212}, 326

\bibitem[\protect\citeauthoryear{Schwarzschild}{1979}]{sch79}Schwarzschild M., 1979, ApJ, 232, 236

\bibitem[\protect\citeauthoryear{Simkin}{1974}]{sim74}Simkin S.M., 1974, A\& A, 31, 129

\bibitem[\protect\citeauthoryear{Slee et al.}{1994}]{sle94}Slee O. B., Sadler E. M., Reynolds J. E., Ekers R.D., 1994, MNRAS, 269, 928

\bibitem[\protect\citeauthoryear{Sparks}{1986}]{spa86}Sparks W. B., Hough J.H., Axon D.J., Bailey J., 1986, MNRAS, 218, 429

\bibitem[\protect\citeauthoryear{Statler}{1995}]{sta95}Statler T.S., 1995, AJ, 109, 1371

\bibitem[\protect\citeauthoryear{Statler}{2001}]{sta01}Statler T.,
2001, ApJ, 121, 244  

\bibitem[\protect\citeauthoryear{Statler etal.}{1996}]{sta96}Statler T., Smecker-Hane T., Cecil G., 1996, AJ, 111, 1512

\bibitem[\protect\citeauthoryear{Statler et al.}{1999}]{sta99}Statler T., Dejonghe H.,  Smecker-Hane T., 1999, ApJ, 117, 126

\bibitem[\protect\citeauthoryear{Statler Hane}{1999}]{stah99}Statler T., Smecker-Hane T., 1999, ApJ, 117,  839


\bibitem[\protect\citeauthoryear{Statler \& McNamara}{2002}]{stamcn}Statler T., McNamara, B.R., 2002, ApJ, 581, 1032

\bibitem[\protect\citeauthoryear{Tonry}{1983}]{ton83}Tonry J.L., 1983, ApJ, 266, 58 

\bibitem[\protect\citeauthoryear{Tonry \& Davis}{1979}]{td79}Tonry J., Davis M., 1979, AJ, {84}, 1511

\bibitem[\protect\citeauthoryear{van der Marel}{1991}]{vdm91}van der Marel R.P., 1991, MNRAS, 253, 710

\bibitem[\protect\citeauthoryear{van der Marel \& Franx}{1993}]{vdmf93}van der Marel R.P., Franx M., 1993, ApJ, 407, 525

\bibitem[\protect\citeauthoryear{van der Marel et al.}{1990}]{vdm90}van der Marel R.P., Binney J.J., Davies R.L., 1990, MNRAS, 245, 582

\bibitem[\protect\citeauthoryear{Verdoes Kleijn et al.}{2000}]{vk00}Verdoes Kleijn G. A., van der Marel R. P., Carollo C. M., de Zeeuw P.T., 2000, AJ, 120, 1221

\bibitem[\protect\citeauthoryear{Williams \& Schwarzschild}{1979}]{ws79}Williams T.B., Schwarzschild M., 1979, ApJ, 227, 56

\end{thebibliography}
\end{document}